\documentclass[a4paper, 12pt]{article}

\usepackage[utf8]{inputenc}
\usepackage{xcolor}
\usepackage[paperwidth=8.5in, paperheight=11in,top=2.5cm,bottom=2.5cm,left=2.5cm,right=2.5cm]{geometry}

\usepackage{amsmath,amsfonts,amssymb,bm}
\usepackage{mathtools}

\usepackage{tikz}
\usepackage{listofitems}      
\usetikzlibrary{arrows.meta}  
\usepackage[outline]{contour} 

\tikzset{>=latex} 
\tikzstyle{node}=[thick,circle,draw=black,minimum size=22,inner sep=0.5,outer sep=0.6]
\tikzstyle{node in}=[node,black,draw=black,fill=white]
\tikzstyle{node hidden}=[node,black,draw=black,fill=white]
\tikzstyle{node out}=[node,black,draw=black,fill=white]
\tikzstyle{connect}=[thick,black] 
\tikzstyle{connect arrow}=[-{Latex[length=4,width=3.5]},thick,black,shorten <=0.5,shorten >=1]
\tikzset{ 
  node 1/.style={node in},
  node 2/.style={node hidden},
  node 3/.style={node out},
}
\def\nstyle{int(\lay<\Nnodlen?min(2,\lay):3)} 

\usepackage{comment}
\usepackage{soul}

\usepackage{subcaption}

\usepackage{natbib}

\newcommand{\dd}{\text{d}}

\title{Deep calibration with random grids}

\author{Fabio Baschetti\thanks{Scuola Normale Superiore, Pisa, Italy. Email address: fabio.baschetti@sns.it} \and Giacomo~Bormetti\thanks{Dipartimento di Matematica, Universit\`a di Bologna, Bologna, Italy. Email address: giacomo.bormetti@unibo.it}\and Pietro~Rossi\thanks{Prometeia S.p.A., Bologna, Italy. Email address: pietro.rossi@prometeia.it}}

\date{\today}

\begin{document}

\maketitle

\begin{abstract}
    We propose a neural network-based approach to calibrating stochastic volatility models, which combines the pioneering grid approach by Horvath et al. (2021) with the pointwise two-stage calibration of Bayer et al. (2018) and Liu et al. (2019). Our methodology inherits robustness from the former while not suffering from the need for interpolation/extrapolation techniques, a clear advantage ensured by the pointwise approach. The crucial point to the entire procedure is the generation of implied volatility surfaces on random grids, which one dispenses to the network in the training phase. We support the validity of our calibration technique with several empirical and Monte Carlo experiments for the rough Bergomi and Heston models under a simple but effective parametrization of the forward variance curve. 
    The approach paves the way for valuable applications in financial engineering - for instance, pricing under local stochastic volatility models - and extensions to the fast-growing field of path-dependent volatility models. 
\end{abstract}


\textbf{Acknowledgements:} We warmly thank Vola Dynamics LLC for making their tools available for academic purposes. Many explorations in the paper would not have been possible without their support.

\section{Introduction}

Calibration of stochastic volatility models is a long-standing problem in quantitative finance. From a theoretical viewpoint, the issue is well understood and requires the implementation of standard optimization routines. However, some operational difficulties  often arise regarding the loop over the pricing function. Recent models in the rough regime -- such as rough Bergomi (rBergomi) of \cite{bayer2016pricing}, quadratic rough Heston (rHeston) of \cite{gatheral2020quadratic}, and evolutions thereof --  typically do not provide a (semi) closed-form expression for the characteristic function (CF) of the asset log-price, the only exception being the rHeston model of \cite{el2018perfect} (see also \cite{el2019characteristic}). On the contrary, these models -- not to mention path-dependent volatility (PDV) models \citep{guyon2014pdf,blanc2017qhawkes,gatheral2020quadratic,parent2023ewma,guyon2023volais} -- are purely simulative and only treated via Monte Carlo techniques which make calibration prohibitive (if not impossible) in terms of computational efforts. The same rHeston model, though amenable to treatment using Fast Fourier Transform (FFT) techniques, is more complex to calibrate than its classical counterpart. The CF requires numerical approximations that slow calibration down (see \cite{diethelm2004detailed, gatheral2019rational,callegaro2021fast}). It is not surprising, then, that calibration of rough volatility models is now usually dealt with using neural networks. Of course, nothing prevents the application of neural networks to standard stochastic volatility models, and one can readily find several examples in the literature. Using neural networks to calibrate financial models dates back to \cite{hernandez2017model}, which deals with the Hull-White model.\\ 

\cite{hernandez2017model} approximates the map from implied volatilities to model parameters via a large feed-forward neural network. The author calibrates volatility surfaces -- normalized w.r.t. strike and maturity -- to associate each with their optimal parameters and trains the network to learn the relationship between volatilities and parameter values. In this direct approach, one feeds newly coming normalized volatility surfaces to the neural network and recovers optimal parameters as an output. \cite{bayer2018deep} represents the first attempt of a two-stage calibration process. A million combinations of rBergomi parameters are drawn uniformly and randomly from the parameter space. Each is associated with a strike-expiration pair, and the corresponding option contract is valued using Monte Carlo methods. The authors use the synthetic dataset for training a large neural network (3 hidden layers of 4096 nodes each) to learn the pricing function from model parameters to implied volatilities. Offline training corresponds to step one in the two-stage process. The learned map is stored as neural network weights and carried to the second step, corresponding to the proper calibration. The idea is to replace the ``true'' pricing function with the neural network approximation and exploit the numerical advantages offered by the neural networks, such as the fast computation by automatic differentiation of the derivatives to set up gradient-based optimization. In~\cite{liu2019neural}, the authors apply the same approach of~\cite{bayer2018deep} to Heston and Bates stochastic volatility models. The CF is available in closed form for these models, and in the generation phase, Fourier integration replaces Monte Carlo pricing. The network is still very large (4 hidden layers of 200 nodes each), and the number of generated samples is again one million.\\

\noindent The entire procedure in \cite{bayer2018deep} and \cite{liu2019neural} goes under the name of the \textit{pointwise approach}: The network outputs individual points in implied volatility corresponding to a given strike and maturity. Henceforth, we will enforce the definition of a truly pointwise approach to underline that the generation phase guarantees that each sampled parameter set is associated with a unique option price. We will contrast our approach with the standard \textit{grid-based approach} by~\cite{horvath2021deep}. The latter employs (deep) neural networks to approximate the underlying pricing function from model parameters to volatility grids. The idea is to look at the volatility surface as a collection of pixels over a bi-dimensional grid in strike and time to maturity. Training can be performed on a (relatively small) number of couples $(\bm{\theta}, \sigma_{BS}(K^{¯}, T^{¯}))$, where $\bm{\theta}$ is one random set of model parameters and $\sigma_{BS}(\cdot, \cdot)$ are the associated implied volatilities over the specified grid $(K^{¯}, T^{¯})$. As before, this methodology is a two-step approach, where sample generation and training correspond to step one, and calibration represents step two. While one can perform data generation and training offline, calibration is extremely fast and well-suited for online execution. The optimizer replaces each call to the ``true'' pricing function by its neural network approximation $F^\mathcal{M}(\bm{\theta},\cdot)$ and exploits the availability of the gradient to ensure convergence to the minimum in a few milliseconds. Indeed, every network evaluation yields one whole grid, and calibration requires only a few iterations. Two points are worth noting about the grid-based approach in \cite{horvath2021deep}: i) generation of the samples is much faster, with 80,000 parameters combinations only (each of which is associated with 88 implied volatilities corresponding to an $11~\times 8$ strike-time to maturity grid) for the rBergomi model; ii) the neural network is very small (4 hidden layers of 30 nodes each) when compared with \cite{bayer2018deep} for the rBergomi model and with \cite{liu2019neural} for classic stochastic volatility models. \\

\noindent Finally, \cite{romer2022empirical} enhances the grid-based approach. The grid is much larger, including very short time to maturities, and the strike specification is conditional on the time to maturity. Such a choice of the grid points is more financially sensible and mitigates some need for more flexibility coming with the approach by~\cite{horvath2021deep}. Also, it addresses the need for higher accuracy in the wings of the smile. The author selects 64 maturities with 25 strikes each and trains six different networks to keep sizes low (3 hidden layers, 200 nodes per layer), acknowledging that ``short- and long-term volatility smiles behave rather differently, especially under rough volatility''. Numerical experiments deal with classic and rough stochastic volatility models and investigate the joint calibration to SPX and VIX smiles.\\ 

\noindent The key to every two-step approach - both on a grid or pointwise - is that the neural network is much faster to evaluate than the actual pricing function it is approximating. A significant difference, however, is immediately apparent as the grid-based approach requires interpolation/extrapolation steps that the pointwise method gets rid of. In fact, in a grid-based approach to calibration, one first needs to project market quotes on the regular and pre-specified grid used for training the neural network. After calibration, one must price quoted strikes and maturities, thus requiring a second run of interpolation/extrapolation. A natural alternative to performing this step is using the ``true'' pricing function under the optimal parameter specification. While this makes perfect sense, it could require slow pricing methods. The pointwise approach does not suffer from this drawback.\\ 

\noindent Regarding rough volatility models, a more technical aspect relates to the specification of the forward variance curve. The standard approach in all mentioned works is to represent the forward curve as a piecewise constant function with (at least) eight levels. Then, one has to calibrate their values jointly with the other model parameters. The approach significantly differs from the earlier proposal in \cite{el2019roughening}, where the forward variance curve is a state variable, estimated from the variance swaps, and exogenous to the calibration process. One first recovers variance swaps from an infinite log-strip of out-of-the-money options, following~\cite{gatheral2011volatility}, and then builds the forward variance curve by differentiation. The curve is piecewise constant, but its identification precedes the calibration step. In this paper, the estimation of the forward curve is part of the calibration procedure. However, we propose a very simple parametrization, which reduces the dimensionality of the calibration problem and guarantees excellent fits to the volatility surface, as we will show extensively. \\

\noindent Our contribution is three-fold. We re-interpret the pointwise approach in a (quasi-)random grid setting, combining the original approaches in \cite{bayer2018deep} and \cite{horvath2021deep}. The result is a small neural network that calibrates the entire market volatility surface in nearly one second and does not require any interpolation/extrapolation. We provide a neural network-based pricer to be used by traders to evaluate options with every strike-maturity pair within the domain of training of the network itself. Last but not least, we propose a simple parametrization of the forward variance curve, which is also supported by empirical evidence from the market. \\ 

\noindent Throughout the paper, we will often advocate the concept of a grid that is either fixed, adaptive, or random, depending on the context. We clarify the meaning of these names from the beginning. The fixed grid is the same as the one introduced by~\cite{horvath2021deep} regarding expirations and strikes. The adaptive grid is a slight modification of the former. We include shorter expirations and make the range of the strikes a function of time to maturity. The reasons for an adaptive grid are apparent after~\cite{romer2022empirical}, but our choice differs from R{\o}mer's. We do not increase the number of maturities in the grid and define our rule for selecting strikes. Finally, a random grid is adaptive, but grid times and strikes are drawn non-uniformly from some contract parameter space (rather than being specified a priori).\\

\noindent Learning from random smiles is also possible and very convenient. We leverage this possibility in the paper. We sample a bunch of maturities for each parameter set, and for each maturity, we price a random vector of strikes. As we will clearly show, this improves the out-of-sample performance of the network (in contrast to the pure pointwise approach) for two reasons. Computationally, generating strikes for a single smile can be performed very efficiently. Second, in the training phase, feeding the network with smiles corresponding to the same parameter set improves learning the conditional density implied by the model.\\

\noindent Grid approaches require two rounds of interpolation/extrapolation - as we noticed. We will always deal with the first one (i.e. projection of the market volatility surface to the grid) via the tools that Vola Dynamics LLC~\footnote{Interested readers can contact Vola Dynamics LCC inquiring about an agreement for academic purposes or buy the software by visiting the following webpage: https://voladynamics.com.} kindly made available to us. Such tools are also fundamental for validation of the parametric form of the forward variance curve that we come up with. \\

\noindent We organize the paper as follows. Section 2 introduces the calibration problem and recalls the (rough) models we will deal with in our numerical experiments. Section 3 reviews standard neural network approaches in the literature, underlines their pros and cons and introduces our pointwise approach. Sections 4 and 5 report numerical experiments under the rHeston and rBergomi models and describe the new parametrization of the forward variance curve. We support the importance of an adapted grid for adequately estimating the roughness parameter in the rHeston model and test our methodology against a few volatility surfaces from the market. Section 6 gives possible directions for future research and concludes.

\section{The calibration problem}

Loosely speaking, the mathematical description of the calibration problem is an optimization as below: 
\begin{equation}\label{pbm}
    \hat{\bm{\theta}} = \underset{\bm{\theta} \in \Theta}{\operatorname{argmin}} \ d(\sigma_{BS}^{\mathcal{M}}(\bm{\theta},K,T),\sigma_{BS}^{mkt}(K,T))\,, 
\end{equation}
where $d$ is a distance between two sets of points (e.g., the root mean squared error (RMSE)) and $\mathcal{M}=\mathcal{M}(\bm{\theta})_{\bm{\theta} \in \Theta}$ is a model with parameters $\bm{\theta} \in \Theta \subset \mathbb{R}^p, p \in \mathbb{N}$. The choice to minimize against the market implied volatilities $\sigma_{BS}^{mkt}(K,T)$ is disputable, but typically justified with increased sensitivity of short-dated deep out-of-the-money  (OTM) options to model parameters (as opposed to minimizing against option prices). This region is crucial for hedging purposes. One should try to achieve an excellent description of it. \\

\noindent Whenever one knows the CF of the asset log-price under model $\mathcal{M}$, FFT pricing is the obvious choice to perform calibration. This is the standard case with classical stochastic volatility models and first tested against rough models in \cite{el2019roughening}. However, the problem with the rHeston model is that the CF requires numerical procedures that are too costly to allow for fast calibration, as we are used to seeing with classical Heston, for instance. The pricing function's neural network approximations come naturally in such a setting, and this is even more so when Monte Carlo is the only way, like in the rBergomi model that we will also investigate in this paper.  

\subsection{Rough volatility models}

Rough volatility models gathered huge interest during the last few years, especially in response to improved fitting to the volatility surface compared to standard stochastic volatility models. They also naturally provide a theoretical justification of the short-time explosion of the observed ATM skew after \cite{fukasawa2017short} in terms of the roughness parameter. \\

\noindent Let $(\Omega,\mathcal{F},(\mathcal{F}_t)_{t \geq 0},\mathbb{Q})$ be a filtered probaility space as generated by independent Brownian motions $B_t$ and $B_t^\perp$ under the risk neutral measure $\mathbb{Q}$. The evolution of the spot price (the SPX here) undergoes the following dynamics:
\begin{align*}
    \dd S_t = S_t \sqrt{V_t}(\rho~\dd B_t + \sqrt{1-\rho^2}~\dd B_t^\perp)
\end{align*}
upon assumption of zero interest rates and dividends. Here $V_t$ is the variance process and $\rho$ stands for the correlation with the spot price $S_t$. Different specifications for the variance process give rise to different (rough) models.   

\subsubsection{rHeston}
The rHeston model of \cite{el2019characteristic} arises with a specification of the variance process as below
\begin{align*}
V_t = \xi_0(t) + \frac{\nu}{\Gamma(H+\frac{1}{2})} \int_0^t \frac{\sqrt{V_s}}{(t-s)^{\frac{1}{2}-H}} \dd B_s\,,
\end{align*}
where $\xi_0(t) = \mathbb{E}[V_t|\mathcal{F}_0 ]$ is the time-zero forward variance curve and $\nu>0$ the volatility of volatility parameter. The parameter $H$, ranging between zero and a half, represents the roughtness parameter of variance paths. As already noticed, the CF of the asset log-price is known up to the numerical solution of a fractional Riccati equation, which can be achieved via the Adams scheme of \cite{diethelm2004detailed}, the hybrid scheme of \cite{callegaro2021fast} or the rational approximation of \cite{gatheral2019rational}. The latter approach is the one we will chose to generate samples. Alternatively, \cite{abi2019lifting} and \cite{abi2019multifactor} discuss  Markovian approximations of the variance dynamics. Whatever the route one follows, Fourier techniques are aptly available for numerical integration and we exploit the FFT implementation presented in \citep{cherubinibook, baschetti2022sinc}, known as the SINC approach. 

\subsubsection{rBergomi}
The rBergomi model of \cite{bayer2016pricing} corresponds to the following specification of the variance process
\begin{align*}
& V_t = \xi_0(t) \exp \bigg( \eta Y_t^\alpha - \frac{\eta^2}{2} t^{2\alpha-1}\bigg)\,,
& Y_t^\alpha = \sqrt{2\alpha-1} \int_0^t (t-s)^{\alpha-1} \dd B_s\,,
\end{align*}
where $\alpha=2H-1$ and $\eta>0$. \cite{bennedsen2017hybrid} and \cite{mcCrickerd2018turbocharging} present efficient schemes for the simulation of the rBergomi dynamics.

\section{Neural networks}

Model calibration to real market data has been a target for machine learning algorithms since the pioneering work of \cite{hernandez2017model}. This work prompted a flourishing literature that reached a standard with the two-step approach of \cite{horvath2021deep}. For ease of exposition, in this section, we review the two-stage approach, return to the direct approach and finally investigate methods that avoid interpolation/extrapolation.

\subsection{The two-stage approach}

\begin{figure}[h!]
     \centering
     \includegraphics[width=\textwidth]{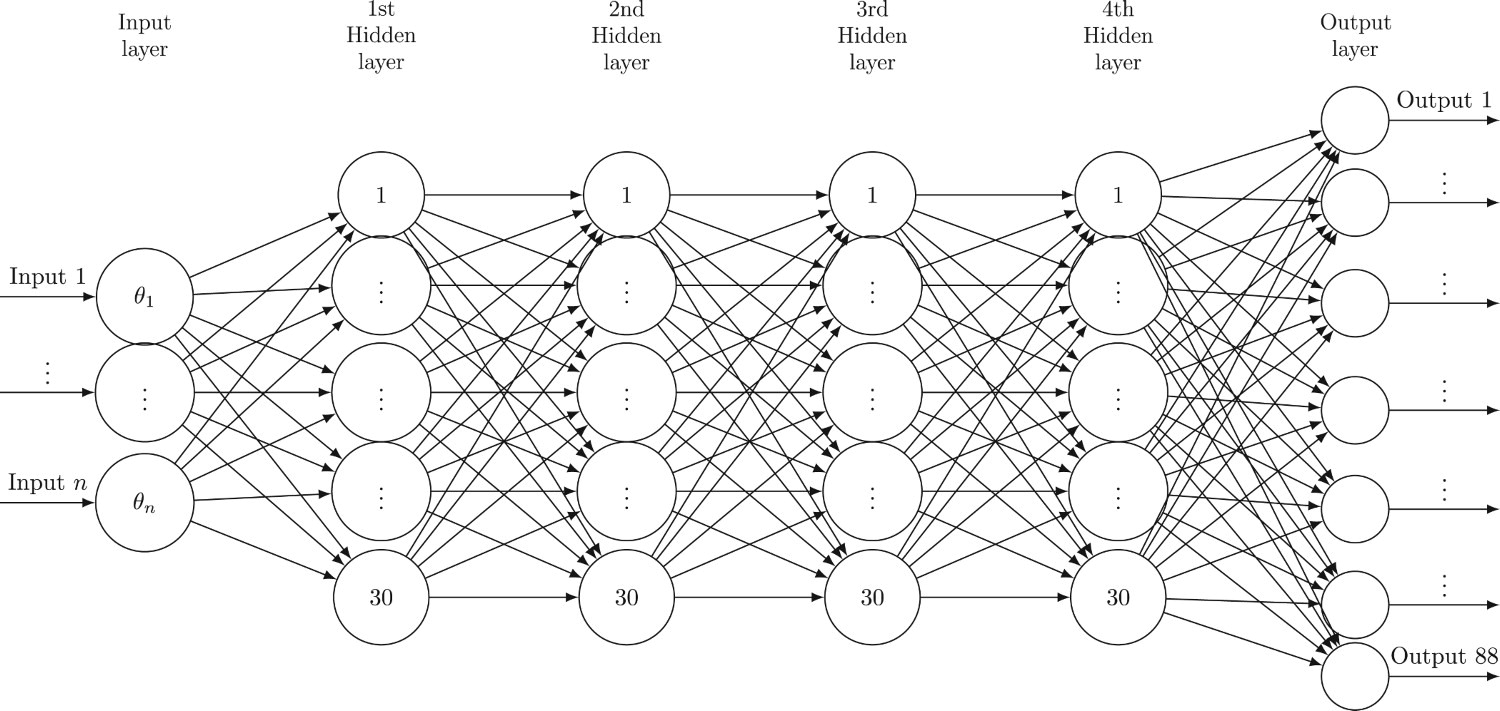}
     \caption{Neural network architecture in \cite{horvath2021deep}}
     \label{Hor_NNarchit}
\end{figure}

\cite{horvath2021deep} separate the calibration procedure into two steps: 
\begin{enumerate}
    \item Generate samples according to a model $\mathcal{M}$ and train a neural network to learn semi-parametrically the pricing map $F^{\mathcal{M}}(\bm{\theta};w)_{ij}:\bm{\theta}\mapsto \sigma_{ij}\doteq\sigma(K_i,T_j)$, from the model parameters $\bm{\theta}$ to implied volatilities $\sigma_{ij}$ on a fixed strike - time to maturity grid $\{ K_i, T_j \}_{i,j=1}^{n,m}$ in terms of the weights $w$.
    This amounts to finding neural network weights $w^\star$ that solve
    \begin{equation*}
        w^\star = \underset{w}{\operatorname{argmin}} \sum_{u=1}^{N} \sum_{i=1}^{n} \sum_{j=1}^{m} (F^{\mathcal{M}}(\bm{\theta}_u,w)_{ij} - \sigma_{BS}^{\mathcal{M}}(\bm{\theta}_u)_{ij})^2
    \end{equation*}
    where $N$ is the number of samples in the training set and $\sigma_{BS}^{\mathcal{M}}(\bm{\theta}_u)_{ij}$ denotes the grid of implied volatilities for a parameter realization $\bm{\theta}_u$.  
    \item Solve
    \begin{equation*}
        \hat{\bm{\theta}} = \underset{\bm{\theta} \in \Theta}{\operatorname{argmin}} \sum_{i=1}^{n} \sum_{j=1}^{m} (F^{\mathcal{M}}(\bm{\theta};w^\star)_{ij} - \sigma_{BS}^{mkt}(K_i,T_j))^2\,,
    \end{equation*}
    where $\sigma_{BS}^{mkt}(K_i,T_j)$ denotes the grid of market implied volatilities for the same strike-time to maturity values.
\end{enumerate}
They set $N=68,000$ and choose a relatively small grid with $n=11$ and $m=8$. As for the network architecture, they use four hidden layers of 30 nodes each. Figure \ref{Hor_NNarchit} provides a graphical representation of the network.

\noindent The  grid specification is fundamental in this kind of approach. In \cite{horvath2021deep}, the eleven strike prices are evenly spaced and range from $0.5$ to $1.5$, while the time to maturity spans the values $\{0.1,0.3,0.6,0.9,1.2,1.5,1.8,2.0\}$ years. 
Strike prices are independent of the maturity and span a constant region of the asset log-price's probability density function (PDF). We will show that starting the grid at $T=0.1$ years results in highly biased estimates of the roughness parameter $H$ in the rHeston model. Intuitively, rough models pay off the most in the shortest maturity regimes and allow for an improved fit that standard one-factor models of volatility can never achieve. We will return to this point with numerical experiments in subsection \ref{estH_rHes_fxdVSada}. \\

\begin{figure}[h!]
         \centering
         \includegraphics[width=0.6\textwidth]{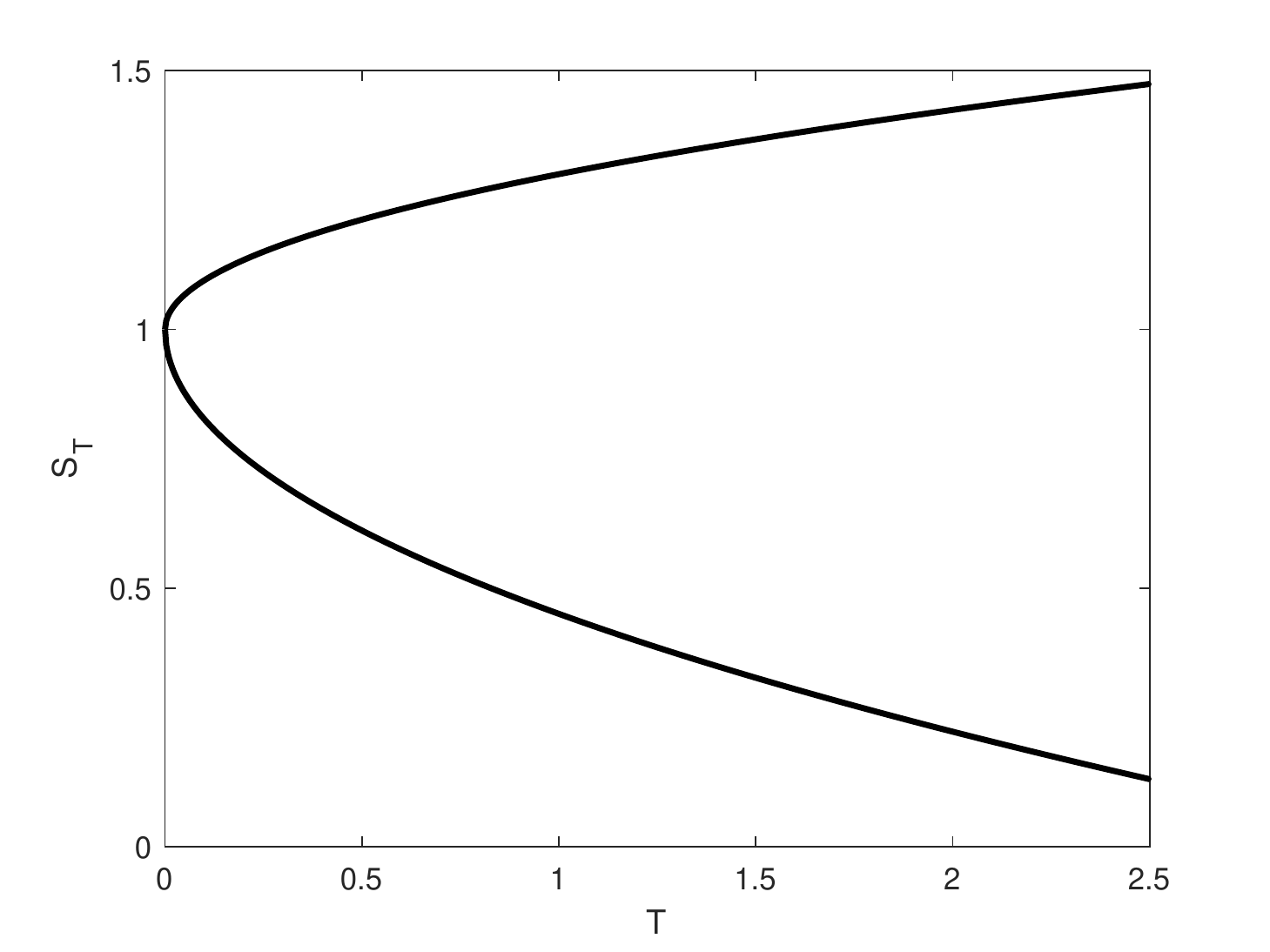}
         \caption{Span of the strike price through time.}
         \label{S_diffuse}
    \end{figure}

\noindent Given the well-known square-root scaling of price volatility, the grid specification in \cite{horvath2021deep} tends to over-sample the strike region at short times and under-sample it for long horizons. Similar considerations and practical reasons relating to Monte Carlo errors in the generation of samples prompted \cite{romer2022empirical} to an adaptive grid. Making the strike price a function of time to maturity is in fact the only way to guarantee one gathers sufficient statistics in the wings of the smile.  
In a similar spirit, we also try an adaptive grid. We choose times to maturity $\{0.01, 0.025, 0.1, 0.3, 0.6, 1.0, 1.5, 2.0\}$ years, and, for any $T_j$, we take (almost) equidistant strikes in the range $[S_0(1-l\sqrt{T_j}),S_0(1+u\sqrt{T_j})]$, with $l=0.55$ and $u=0.30$. Figure \ref{S_diffuse} provides a graphical representation of the region covered by the strike price. Please notice that the square root rule only applies until the time to maturity is not too long, at which point strikes would be prescribed negative.
Anyway, times are now much more informative and strikes more aligned with market quotes for any given maturity, especially very short ones.

\subsection{The direct approach}

\cite{hernandez2017model} tackles the calibration problem entirely differently. With some abuse of terminology, we may refer to his approach, compared to the one in \cite{horvath2021deep}, as an inverse one. He exploits machine learning to learn the map from implied volatilities to model parameters. To achieve the goal, he uses a feed-forward neural network with four layers and 64 nodes each. The sample size is larger than \cite{horvath2021deep} with almost twice as many grid surfaces in the training set. Of course the two approaches share the same training set, with the obvious difference that model parameters and implied volatilities switch their roles. Consistently, the dimension of the input layer is $n \times m$ while the output dimension is $p$.    

\subsection{The pointwise approach}

Both approaches in \cite{hernandez2017model} and \cite{horvath2021deep} (and later modifications, e.g., \cite{romer2022empirical}) require two rounds of volatility data processing. First, one needs to project the input from the market using interpolation/extrapolation routines on the same grid used for training. After calibration, one has to lift the output to the same strike-maturity pairs as the market. The natural choice to perform this step is by calling the true pricing function (by Fourier transform or Monte Carlo simulation) given the calibrated parameters. The last step may only partially preserve the quality of the fit, especially if Monte Carlo is involved,  but finer grids would help, as demonstrated in \cite{romer2022empirical}. Still, the need to run the true pricing function after calibration seems highly inefficient. One could resort to a computationally faster approach based on spline interpolators, but this would be more worrisome when preventing arbitrages and guaranteeing proper extrapolation for short maturities. \\

\noindent 
The literature supporting the grid-based approach claims that applying interpolation / extrapolation techniques to the implied volatility surface is a well-understood and easy-to-implement procedure. Nonetheless, some words of caution are necessary here: The
extrapolation of volatilities at a very short time consistently with the roughness of the market may be subtle. If the smallest expiration in the grid is shorter than the first maturity in the market, a flat extrapolation would bias the roughness parameter high, and the longer the first
maturity, the higher the bias. While a flat extrapolation is overly simplistic, the design of a polynomial extrapolator consistent with the implied volatility asymptotic behavior dictated by the model is generally not trivial. \cite{dallacqua2022rhestonlocal} and \cite{bourgey2023local} present a volatility extrapolation procedure consistent with a rough model's dynamic properties.
As it will soon become apparent, our calibration procedure proposes an alternative and entirely numerical solution.

\subsubsection{Pointwise training}

\noindent \cite{bayer2018deep} and \cite{liu2019neural} introduce the pointwise approach. This is also two-stage, but the idea is that the network learns single points in volatility rather than a grid and calibrates directly to the market. By design, the trained network can generate implied volatility without the need to interpolate over a specified grid in strike and time-to-maturity.  
Unfortunately, however, the very large shape of the network makes calibration much slower than in the grid-based approach. 
For training the network, the authors sample each combination of model parameters $\{\bm{\theta}_u\}_{u=1,\dots, N}$ concurrently with one strike-maturity pair $(K_u, T_u)$ from some contract parameter space. 
Therefore they need to run as many Monte Carlo simulations or Fourier inversions as the number of samples. For adequate training of a large network -- the number of nodes per layer ranges from a few hundred to about 4 thousands -- the number of samples must be as high as $10^6$, thus making the entire training phase quite time consuming.\\

\subsubsection{Training with random grids}

Our approach is of pointwise type too, since we are learning the underlying map from model parameters and contract specifications to (single points in) implied volatility. Figure \ref{network_our} indeed clarifies that we inherit the same network structure as in a pointwise method:

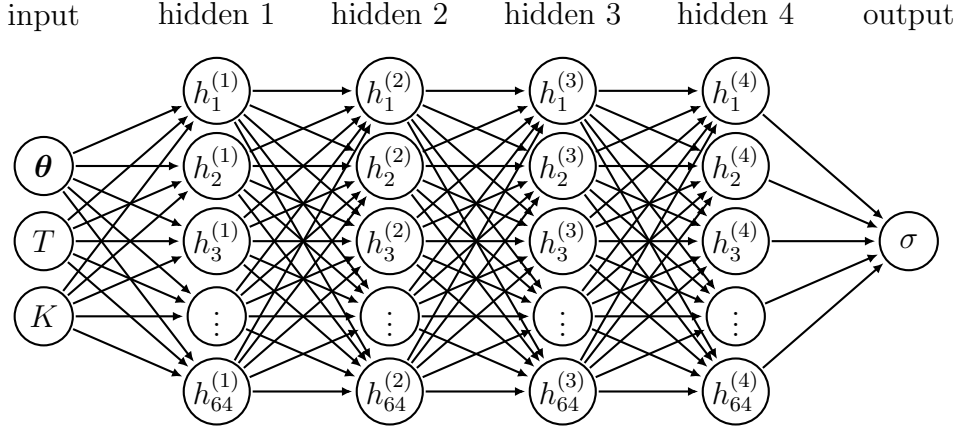
\begin{figure}

\centering

\begin{tikzpicture}[x=2.3cm,y=1cm]
  \readlist\Nnod{3,5,5,5,5,1} 
  \readlist\Nstr{3,64,64,64,64,1}
  \readlist\Cstr{\strut x,h^{(\prev)},h^{(\prev)},h^{(\prev)},h^{(\prev)},y}
  \message{^^J  Layer}
  \foreachitem \N \in \Nnod{ 
    \edef\lay{\Ncnt} 
    \message{\lay,}
    \pgfmathsetmacro\prev{int(\Ncnt-1)} 
    \foreach \i [evaluate={\y=\N/2-\i; \index=(\i<\N?int(\i):"\Nstr[\lay]"); \x=\lay; \n=\nstyle;}] in {1,...,\N}{ 
      
      \ifnum\lay=1 
        \ifnum\i=1
          \node[node \n] (N\lay-\i) at (\x,\y) {$\bm{\theta}$}; 
        \fi
        \ifnum \i=2
          \node[node \n] (N\lay-\i) at (\x,\y) {$T$}; 
        \fi
        \ifnum \i=3
          \node[node \n] (N\lay-\i) at (\x,\y) {$K$}; 
        \fi
      \else
        \ifnum\lay=6
          \node[node \n] (N\lay-\i) at (\x,\y) {$\sigma$};
        \else
          \ifnum \i=4
            \node[node \n] (N\lay-\i) at (\x,\y) {$\vdots$}; 
          \else 
            \node[node \n] (N\lay-\i) at (\x,\y) {$\Cstr[\lay]_{\index}$};
          \fi
        \fi
     \fi
      
      \ifnum\lay>1 
        \foreach \j in {1,...,\Nnod[\prev]}{ 
          \draw[connect arrow] (N\prev-\j) -- (N\lay-\i); 
        }
      \fi 
      
    }
    
  }

  \node[above=35.5,align=center] at (N1-1.90) {input};
  \node[above=8,align=center] at (N2-1.90) {hidden 1};
  \node[above=8,align=center] at (N3-1.90) {hidden 2};
  \node[above=8,align=center] at (N4-1.90) {hidden 3};
  \node[above=8,align=center] at (N5-1.90) {hidden 4};
  \node[above=64,align=center] at (N6-1.90) {output};
  
\end{tikzpicture}

\caption{Neural Network architecture as we use it in the numerical section. Boldface notation underlines that $\bm{\theta}$ stands for the entire parameter vector: the input layer is $p+2$-dimensional, where $p$ is the number of model parameters.}
\label{network_our}

\end{figure}

\noindent The NN takes model parameters $\bm{\theta}$, strike $K$ and time to maturity $T$ in input, and it outputs the implied volatility $\sigma$ of the associated option. \\

\noindent Nonetheless, we introduce an important difference with respect to \cite{bayer2018deep} and \cite{liu2019neural}. The key is that the training set that we dispense to the network results from the following procedure: We produce grid surfaces in the generation phase but pass them to the network in the form of quadruplets $(\bm{\theta}_l,K_i,T_j,\sigma_{ij})_{l,i,j=1}^{q,m,n}$ (we have $q \times m \times n$ of them). Generation is clearly reminiscent of \cite{horvath2021deep} but strikes and maturities are no longer restricted to a pre-specified fixed (or adapted, \cite{romer2022empirical}) grid. \\

\noindent Indeed, there is no particular reason why times should be confined to a few buckets of maturity and strikes predetermined given the expiration. Both of them can be taken random, and we explain how to do it in the rest of this section. \\

\noindent We sample parameters sets $\bm{\theta}_l, l=1,\dots,q$ uniformly at random from $\Theta$ and associate each particular set with $n=11$ expiries and $m=13$ strikes. In particular, maturities $T_1^l,\dots,T_n^l$ are drawn from the following set
\begin{align}\label{time sub-intervals}
[T_{\min},T_{\max}] = & [0.003,0.030) \cup [0.030,0.090) \cup [0.090,0.150) \cup [0.150,0.300) \cup [0.300,0.500) \cup \nonumber \\ & [0.500,0.750) \cup [0.750,1.000) \cup [1.000,1.250) \cup [1.250,1.500) \cup [1.500,2.000) \cup \nonumber \\ & [2.000,2.500],
\end{align}
$T_j^l$ coming from the $j$-th subinterval ($j=1,\dots,n$). This last fact ensures proper coverage of the typical market maturities in the training set. Finally, strikes $K_1(T),\dots,K_m(T)$ are sampled in range whose width is a function
\begin{align}\label{sqrt_bounds}
        [K_{\min}(T),K_{\max}(T)] = [S_0(1-l\sqrt{T}),S_0(1+u\sqrt{T})], \qquad l=0.55 \quad u=0.30.
\end{align}
of time to maturity $T$. \\
We provide a graphical description of the dataset construction procedure below:
\begin{align*}
    \begin{cases}
      \bm{\theta}_1 
      \begin{cases}
      T_1^1 \longrightarrow K_1(T_1^1) \ \dots \ K_m(T_1^1)  \\
      \vdots \\
      T_n^1 \longrightarrow K_1(T_n^1) \ \dots \ K_m(T_n^1)
      \end{cases} \\
      \vdots \\
      \bm{\theta}_q
      \begin{cases}
      T_1^q \longrightarrow K_1(T_1^q) \ \dots \ K_m(T_1^q)  \\
      \vdots \\
      T_n^q \longrightarrow K_1(T_n^q) \ \dots \ K_m(T_n^q)
      \end{cases} \\
    \end{cases}\,.
\end{align*}
Crucially, strikes are not uniform in $[K_{\min}(T),K_{\max}(T)]$ but our sampling tries to replicate similar granularity as the market. More specifically, we take 
\begin{itemize}
    \item 4 strikes in the left tail $[K_{\min}(T),S_0(1-0.20\sqrt{T})]$
    \item 7 strikes in the central region $[S_0(1-0.20\sqrt{T}),S_0(1+0.20\sqrt{T})]$
    \item 2 strikes in the rigth tail $[S_0(1+0.20\sqrt{T}),K_{\max}(T)]$
\end{itemize}

\noindent Our grids are therefore adaptive in nature but promoting them to be random allows for rearrangement in a form that is suitable for the pointwise approach, which fact is enough to calibrate on market points directly and getting rid of any interpolation/extrapolation. Notably, then, we will also see in the numerical section that such a calibration is typically very fast. \\ 

\noindent We will talk about a random-grid pointwise approach to denote our hybrid solution and distinguish it from the original proposals in \cite{bayer2018deep} and \cite{liu2019neural} - which could now be casted as truly pointwise (thus meaning that both generation of samples and submission to the network are indeed pointwise, i.e. each parameter set is associated with one strike-maturity pair only). \\ 

\noindent The benefits from our method are multiple: \begin{itemize}
    \item production of the training set is very fast compared to an approach that is truly pointwise. 
    Say one wants to produce $M = N_T \times N_K$ points overall when describing a surface ($N_T$ maturities, $N_K$ strikes per maturity) and assume - for the moment - that the CF of the asset log-price is known (e.g. the rHeston model): individual options can be priced with $N$ calls to the CF. Populating the whole surface only requires $N \times N_T$ calls to the CF with our random grid approach, as opposed to $N_T \times N_K \times N$ if one is truly pointwise. A similar reasoning applies when pricing by Monte Carlo methods. Random grids require one simulation on a time grid that contains all of the desired maturities, but a truly pointwise approach would ask for $N_T \times N_K$ simulations.

    \item proper training only needs a small number of neurons, for a network which is comparable in size with the grid-based approach from \cite{horvath2021deep}. Numerical experiments in this paper are based on a neural network with 4 hidden layers, 64 nodes each. The reason why such a thin network structure performs properly is investigated in the following subsection.
    
\end{itemize} 

\noindent Options falling outside of the contract parameters space that we used in the generation phase cannot be priced by the network. However, our square root rule for the selection of strikes is such that what is left over is typically quite illiquid and would be excluded in any case. Also, the same choice for the range of the strikes is also needed in a grid-based approach. \\

\noindent Finally, we notice that calibration with a pointwise method is very similar to what one does with the FFT (and would do with Monte Carlo). A fundamental difference remains. Every network approximation of the pricing function is much faster to evaluate than the function itself, and the calibration process faster as a consequence.\\ 

\subsubsection{Training with random smiles}

Another viable choice for pointwise training comes with the generation of random smiles. This strategy is indeed halfway between the pure pointwise approach and training with random grids. \\

 \noindent We sample parameter sets $\bm{\theta}_l \ l=1,\dots,q$ from space $\Theta$; we supplement each of these sets with random time to maturity $T$ and strikes $K_1(T),\dots,K_m(T)$; we price the corresponding options and invert for implied volatility. Expirations are such that we guarantee uniform coverage of time sub-intervals in Equation \eqref{time sub-intervals}. The usual square root bounds \eqref{sqrt_bounds} define the range of the strikes. Finally, we dispense the generated smiles to the network in a pointwise manner. \\

\begin{figure}[h!]
  \centering
  \includegraphics[width=0.75\textwidth]{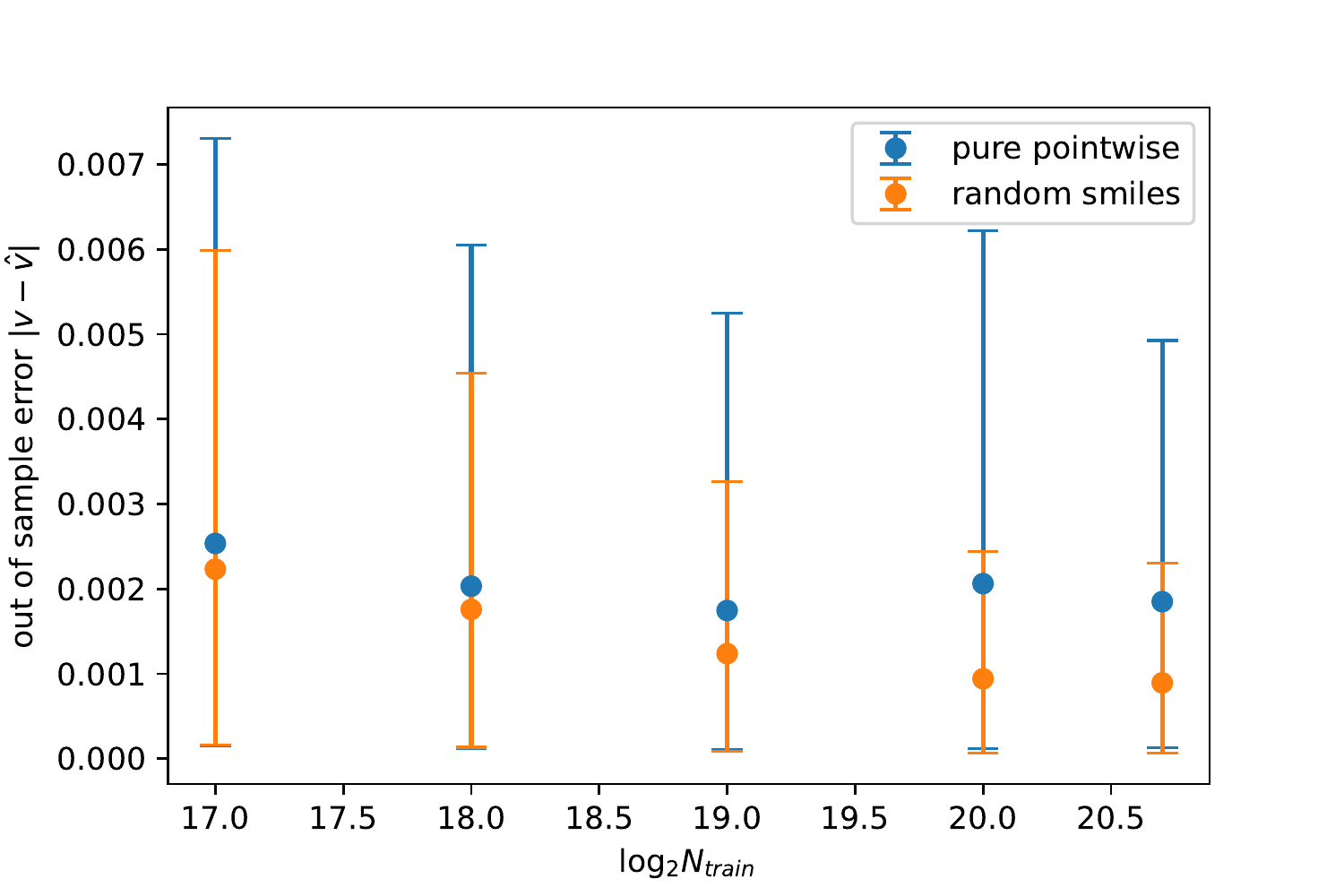}
  \caption{Evolution of the out-of-sample error as a function of the number of data points in the training set. Pure pointwise approach in blue, training with random smiles in orange. Candles cover the range between the $5^{th}$ and $95^{th}$ quantiles. Points denote the mean absolute error.}
  \label{cmp_Ntrain}
\end{figure}

\begin{figure}[h!]
  \centering
  \includegraphics[width=0.75\textwidth]{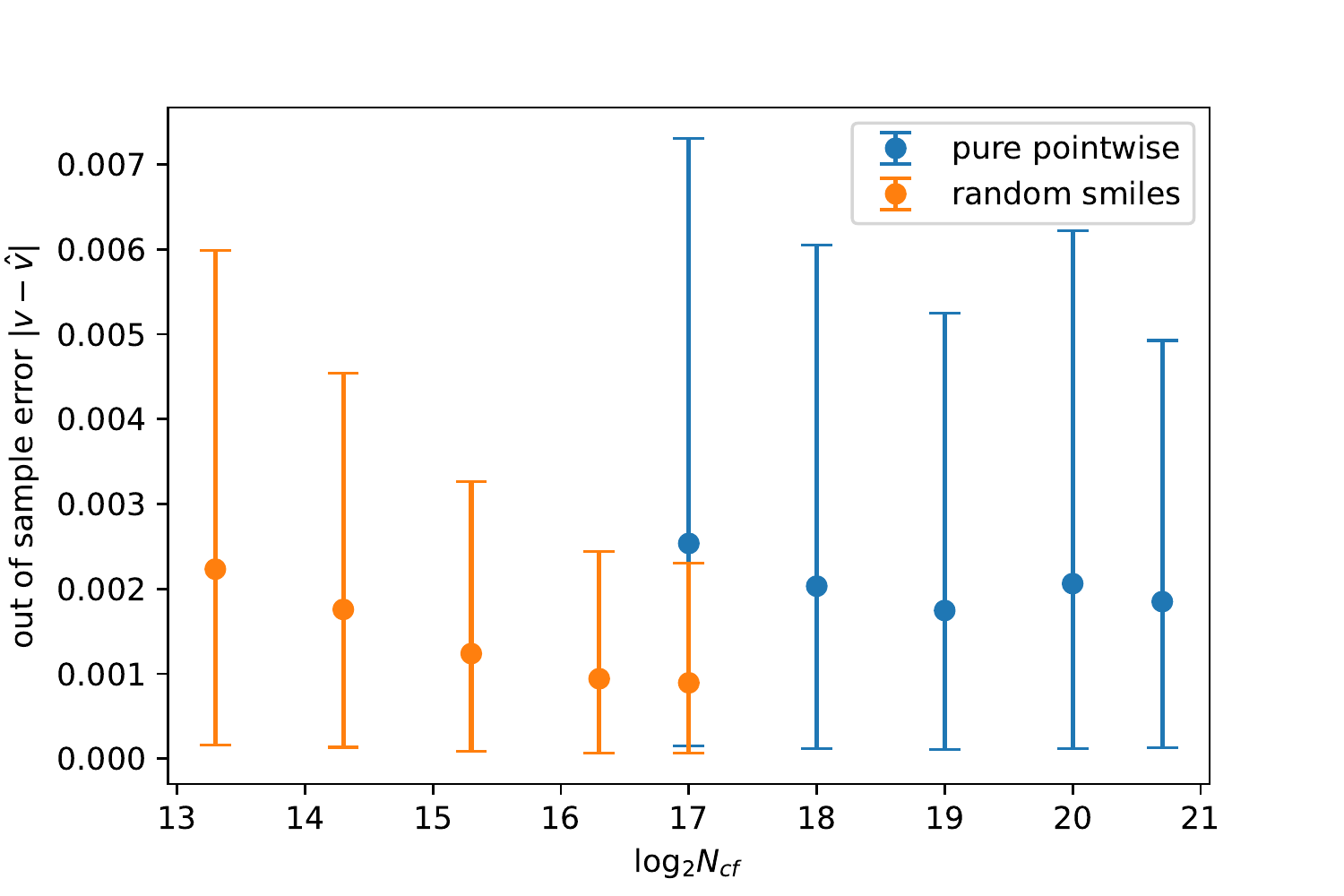}
  \caption{Evolution of the out-of-sample error as a function of the computational burden, measured in terms of calls of the CF, $N_\text{cf}$. Pure pointwise approach in blue, training with random smiles in orange. Candles cover the range between the $5^{th}$ and $95^{th}$ quantiles. Points denote the mean absolute error.}
  \label{cmp_Ntime}
\end{figure}

 \noindent A systematic comparison of the out-of-sample performance between the pure pointwise approach and training with random smiles is illustrated in Figures \ref{cmp_Ntrain} and \ref{cmp_Ntime} for the rHeston model. \\ 
 
\noindent We start with increasing the number of points in the training set and look at the evolution of the NN approximation error.    
Figure \ref{cmp_Ntrain} shows that there is a level in the size of the training set where a network which is purely pointwise stops learning: We see no decreasing trend in the blue candles after $2^{19}$ training samples have been included. Conversely, training with random smiles (in orange) exhibits monotonic decrease of the average error throughout the whole region we are considering. The two approaches behave pretty similarly for small $\log_2(N_{train})$, but the gap gets larger and larger as we proceed. \\

\noindent We therefore conclude that filling the parameter space in a purely pointwise way is not effective for proper training. Most importantly, the network needs to be exposed to multiple strikes so as to learn information about the smile, i.e. about the conditional density implied by the model. Proper coverage of typical market maturities also plays a role. \\

\noindent Figure \ref{cmp_Ntrain} also suggests that proper training can be achieved without increasing the size of the network. Layers and nodes are fixed throughout the experiment, but we are able to achieve excellent out-of-sample performance when sampling random smiles. We can work with a much thinner network than \cite{bayer2018deep} and \cite{liu2019neural} because of the information we include in the training data. Generating multiple entire volatility surfaces on random grids but training as if they were pointwise provides the network with rich information about the shape of the smiles and their evolution across different maturities. \\

\noindent The size of the training set is a crucial variable to account for, but what is more important is the computational cost of the entire generation phase (i.e. building samples for the training set). Specifically for those models where the CF is available in semi-closed form, one can perform a fair comparison between the pointwise and the random smile approach in terms of the computational burden, 
by counting the number of times the algorithm computes the CF. Plotting the NN approximation error as a function of the number of calls of the CF in Figure \ref{cmp_Ntime} makes the superior performance of the random smile approach even more apparent. For $N_\text{cf}=2^{17}$, the average error from the pure pointwise approach is larger than the $95^{th}$ quantile of the corresponding orange candle. \\

\noindent Generating random smiles rather than individual points guarantees a large gain in computational time (roughly $m$ times faster, where $m$ is the number of strikes in one smile). The CF only depends on the model parameters and time to maturity, and can be re-used for all strikes in the random smile approach, without the need to recompute it for each point as in the pointwise approach. 

\noindent A similar reasoning can be extended  to purely Monte Carlo models, such as the rBergomi, where the whole cost comes from simulation through maturity $T$ but is obviously almost insensitive to the number of options priced in a single smile. However, a proper investigation of the relative performances of the pointwise and random smile approach in a Monte Carlo setting may depend on the implementation details and model at hand. We do not consider it in the present work and leave the analysis for future research.

\subsubsection{Practitioner's corner: NN architecture and training}

The object we will be working with in the numerical section is a simple feed-forward neural network with 4 hidden layers, 64 nodes each\footnote{We did not try any optimization on the network structure and therefore simpler architectures could provide satisfactory results as well.}. Model parameters $\bm{\theta}$ and contract specifications $(T,K)$ enter as an input, and the associated implied volatility comes in the output. A graphical description of the network has been presented in Figure \ref{network_our}. \\

\noindent The Elu activation function \\
\begin{equation*}
a_{ELU} = 
\begin{cases}
    x & \text{if }  x \geq 0 \\ 
    \text{e}^x-1 & \text{if }  x < 0
\end{cases}    
\end{equation*}
is solely responsible for the non-linearity in the network. The output layer comes with the identity function as an activation: we find that there is no need to enforce positivity of the implied volatility in any way.\\

\noindent We use the RMSE as a loss and take advantage of Adam optimizer during training. We allow for 500 epochs at most, and impose a patience parameter of 50 epochs for early stopping. The validation loss is monitored. No dropout or more advanced techniques seem to be needed to prevent overfitting. Mini batches are used for performance. Other hyperparameters are unchanged with respect to the default in Keras. \\

\noindent Once the network is trained~\footnote{Proper training clearly depends on the quality of the samples one has produced during the generation phase. We do not enter the technicalities behind Monte Carlo estimation of rBergomi option prices (for which we refer to \cite{bennedsen2017hybrid} and \cite{mcCrickerd2018turbocharging}), but stress that those parameter combinations that produced prices for which Black-Scholes inversion fails need to be discarded.}, we save neural network weights to disk and use the optimal configuration for calibration. The neural approximation of the true pricing function and its gradient are hard-coded in Numpy. This allows for extra speed relative to the prediction method in Keras, as suggested by \cite{horvath2021deep}.\\

\noindent A working example for the usage of the network is available on the following GitHub page: https://github.com/fabioBaschetti/random-grid-NN-calib.

\section{Numerical experiments}

This section presents a large set of tests for the random-grid pointwise approach. We show how it naturally emerges as a solution to various problems from a grid method (be it a fixed grid or adapted), and equip it with a parametric forward variance curve before we draw our conclusions. \\

\noindent The standard with rough volatility models is that the forward variance curve is piecewise constant. Moreover, when it comes to neural network methods for pricing and calibration it is common practice to associate grid times with the discontinuities of the curve. We stick to this convention. As for the pointwise approach, we make the forward variance curve entirely equivalent to the one we use for the adaptive grid (i.e. same buckets). \\

\subsection{Wily estimate of the roughness parameter} \label{estH_rHes_fxdVSada}

\cite{horvath2021deep} apply fixed grid methods for calibration of the rBergomi model. We also try rHeston and spend a few words of caution for such an application in the present section. \\

\noindent We already noticed that extension to an adaptive grid is crucial for inclusion of very short expirations in the training set. In particular, for the rHeston model, this also results in better estimates of the Hurst exponent that are key to capture the explosive behavior of the ATM skew, as we now demonstrate. \\

\noindent For this,  we  sample a thousand parameter sets, compute implied volatilities over both a fixed grid and an adaptive grid, and calibrate using the appropriate network. The underlying forward variance curve is piecewise constant but jump times synchronized with grid maturities and therefore different. Benchmark volatilies that we calibrate on are generated with a flat forward variance curve. Having a third construction of the curve that both networks could replicate is indeed necessary not to favor any of the two and ensure that model parameters are not biased by forward variance effects - at least in principle. \\

\noindent All considerations in the present subsection find graphical support in the following plots. We refer the reader to Appendix \ref{tables} for the numbers behind them. \\

\begin{figure}[h!]
     \centering
     \begin{subfigure}[]{0.49\textwidth}
         \centering
         \includegraphics[width=\textwidth]{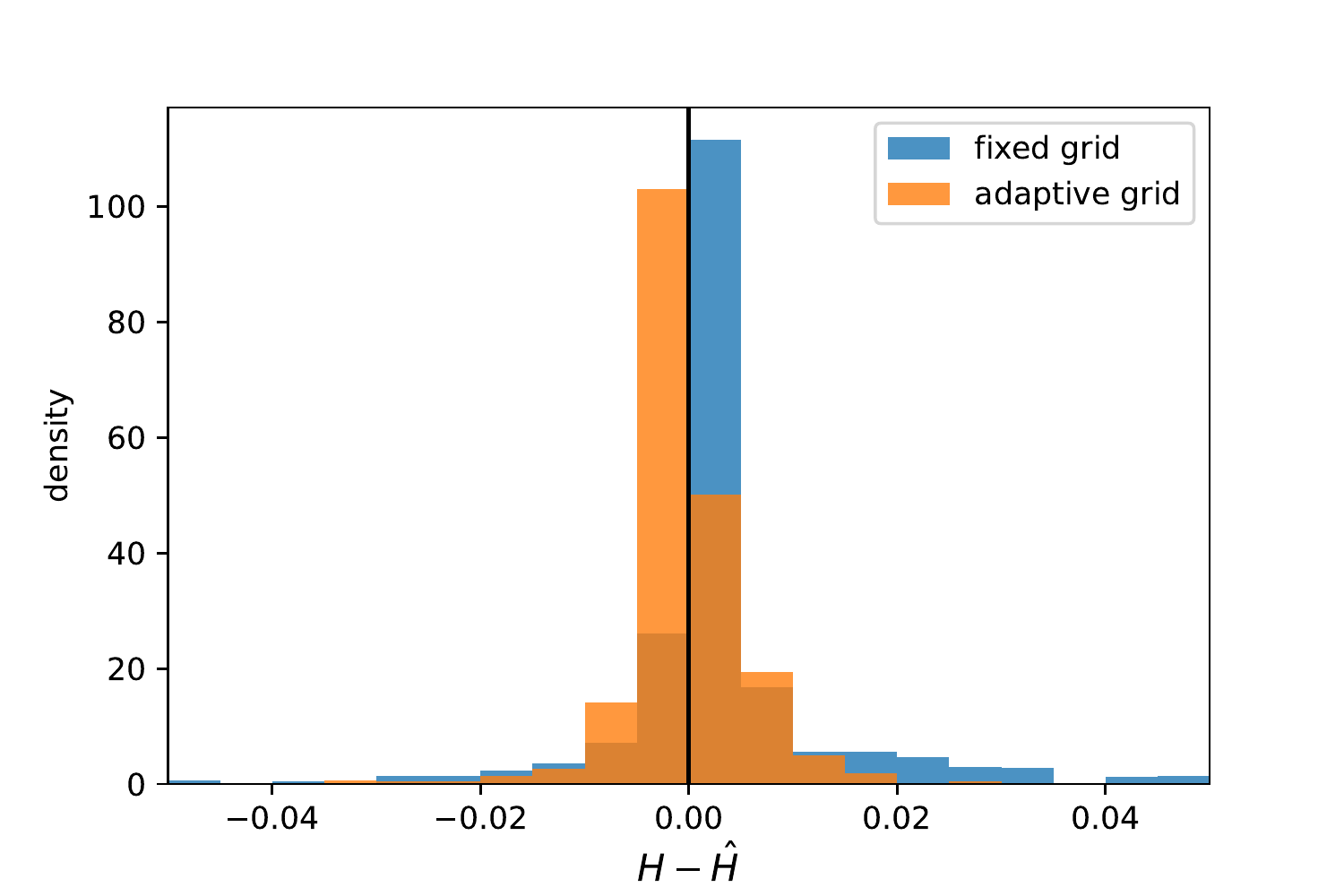}
         \caption{}
         \label{rHes_errH_fxdVSada}
     \end{subfigure}
     \hfill
     \begin{subfigure}[]{0.49\textwidth}
         \centering
         \includegraphics[width=\textwidth]{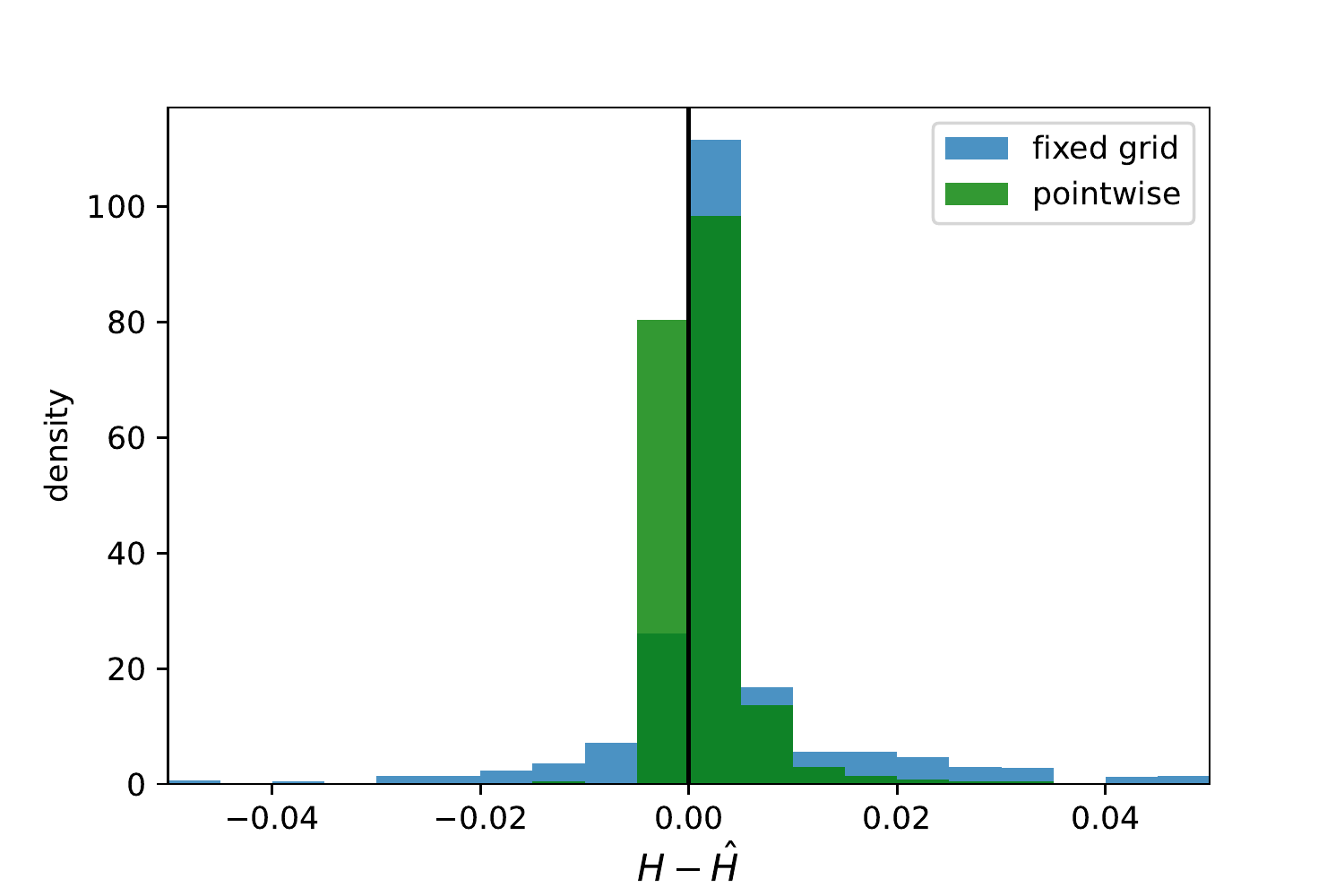}
         \caption{}
         \label{rHes_errH_fxdVSpnt}
     \end{subfigure}
     \caption{Distribution of absolute errors $e_H$ for fixed-, adaptive- and random-grid neural network estimates of the roughness parameter $H$ in the rHeston model over a sample of 1000 surfaces with flat variance curve.}
     \label{rHes_errH}
\end{figure}

\noindent While both networks seem to do a good job, a few points are worth noticing after Figure \ref{rHes_errH} and the corresponding Table \ref{eH_tab}. The range of the absolute errors $e_H = H-\hat{H}$ is much larger for the fixed grid than it is for the adapted grid and the tails of the error distribution fatter. Also, very large negative errors are observed with the fixed grid (greatly larger than the largest positive error), thus resulting in huge overestimation of the roughness parameter. Short maturities carry important information about the skew and kurtosis; ignoring them biases our estimates of the roughness parameter as a consequence. \\

\noindent The lowest variability in the estimation of $H$ is actually observed with the random-grid pointwise approach. Standard deviation of the error is slightly better than the adaptive grid, and a factor three smaller than the fixed grid (see Table \ref{eH_tab}).\\

\begin{figure}[h!]
     \centering
     \begin{subfigure}[]{0.49\textwidth}
         \centering
         \includegraphics[width=\textwidth]{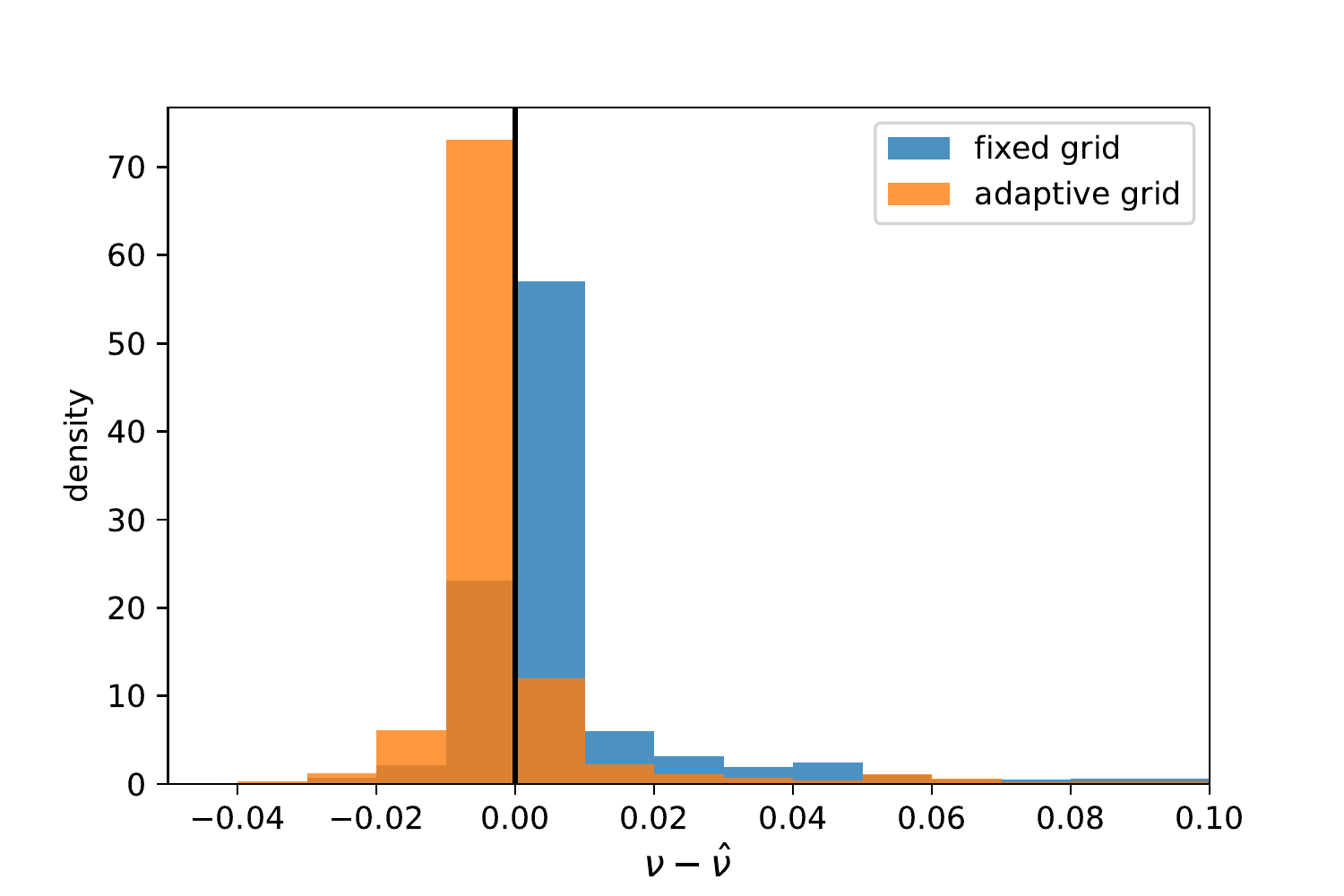}
         \caption{}
         \label{rHes_errNU_fxdVSada}
     \end{subfigure}
     \hfill
     \begin{subfigure}[]{0.49\textwidth}
         \centering
         \includegraphics[width=\textwidth]{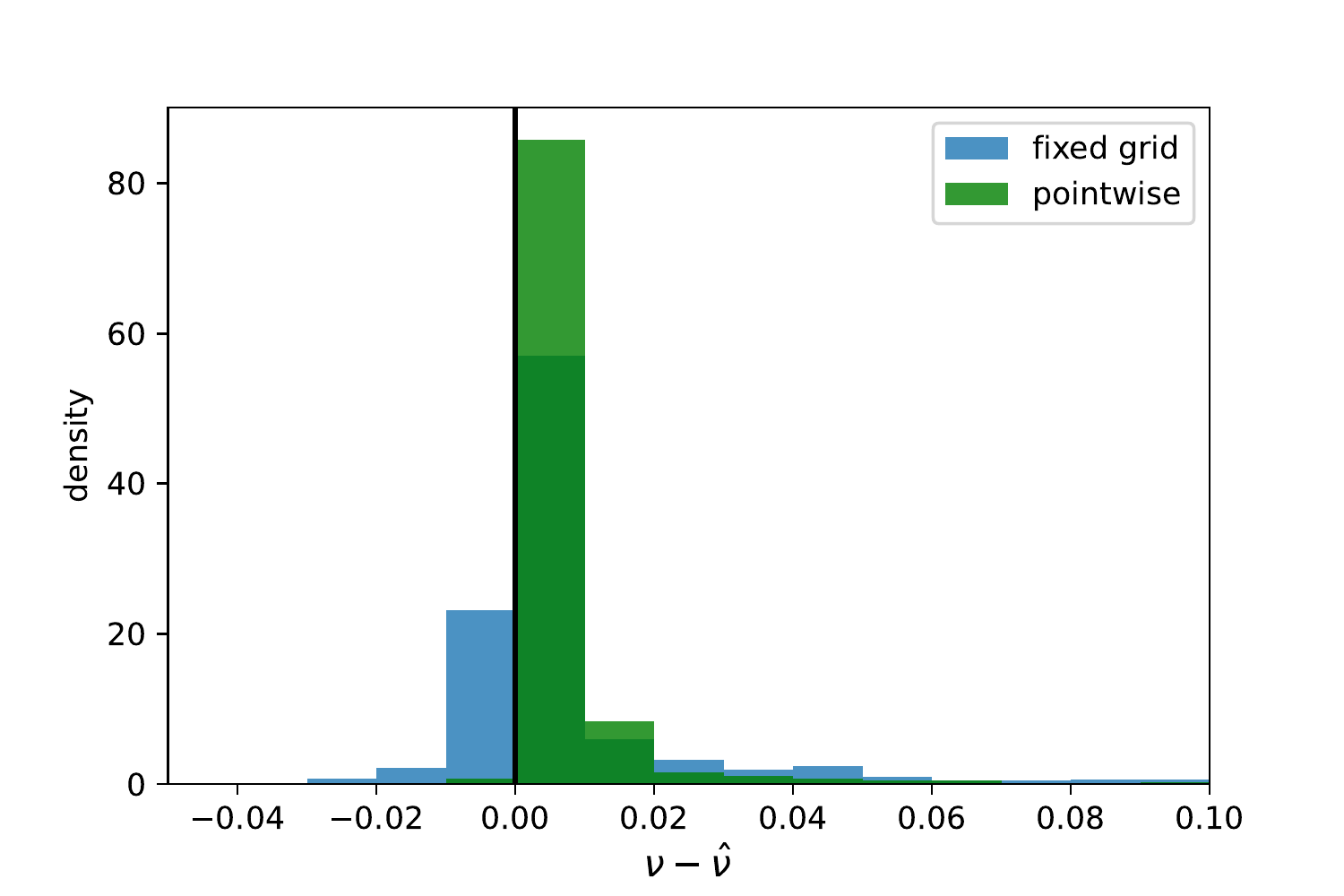}
         \caption{}
         \label{rHes_errNU_fxdVSpnt}
     \end{subfigure}
     \caption{Distribution of absolute errors $e_{\nu}$ for fixed-grid, adaptive-grid and pointwise (adaptive-grid) neural network estimates of the VOV $\nu$ in the rHeston model over a sample of 1000 surfaces with flat variance curve.}
     \label{rHes_NU_estimate}
\end{figure}

\noindent The other parameters -- volatility of volatility (VOV for short) and correlation -- require separate treatment. The average error in the estimation of $\nu$ and $\rho$ gets closer to zero when moving to the adapted grid but no reduction in the standard deviation is observed. Actually, visual inspection of Figure \ref{rHes_errNU_fxdVSada} would suggest a general overestimation (underestimation) of the VOV with the adapted (fixed) grid. It is indeed true that the estimates from the adapted grid are higher than the true value of the VOV about $80\%$ of the times (vs $26\%$ for the fixed grid) but the average error in Table \ref{eNU_tab} is very close to zero thus suggesting that such an overestimation is mild. In addition to this, the pointwise approach has (almost) always a positive bias and we should remember that it is now using the same set of points for calibration as the adapted grid. It is therefore unlikely that this erratic behavior of the overestimation ratio when moving from one network to another is explained by the particular set of expirations and strikes that we have selected. Most importantly, then, the average error is always small and the pointwise approach is the one solution than makes the standard deviation the smallest - as it is now evident in Figure \ref{rHes_errNU_fxdVSpnt} and confirmed in Table \ref{eNU_tab}.  \\

\begin{figure}[h!]
     \centering
     \begin{subfigure}[]{0.49\textwidth}
         \centering
         \includegraphics[width=\textwidth]{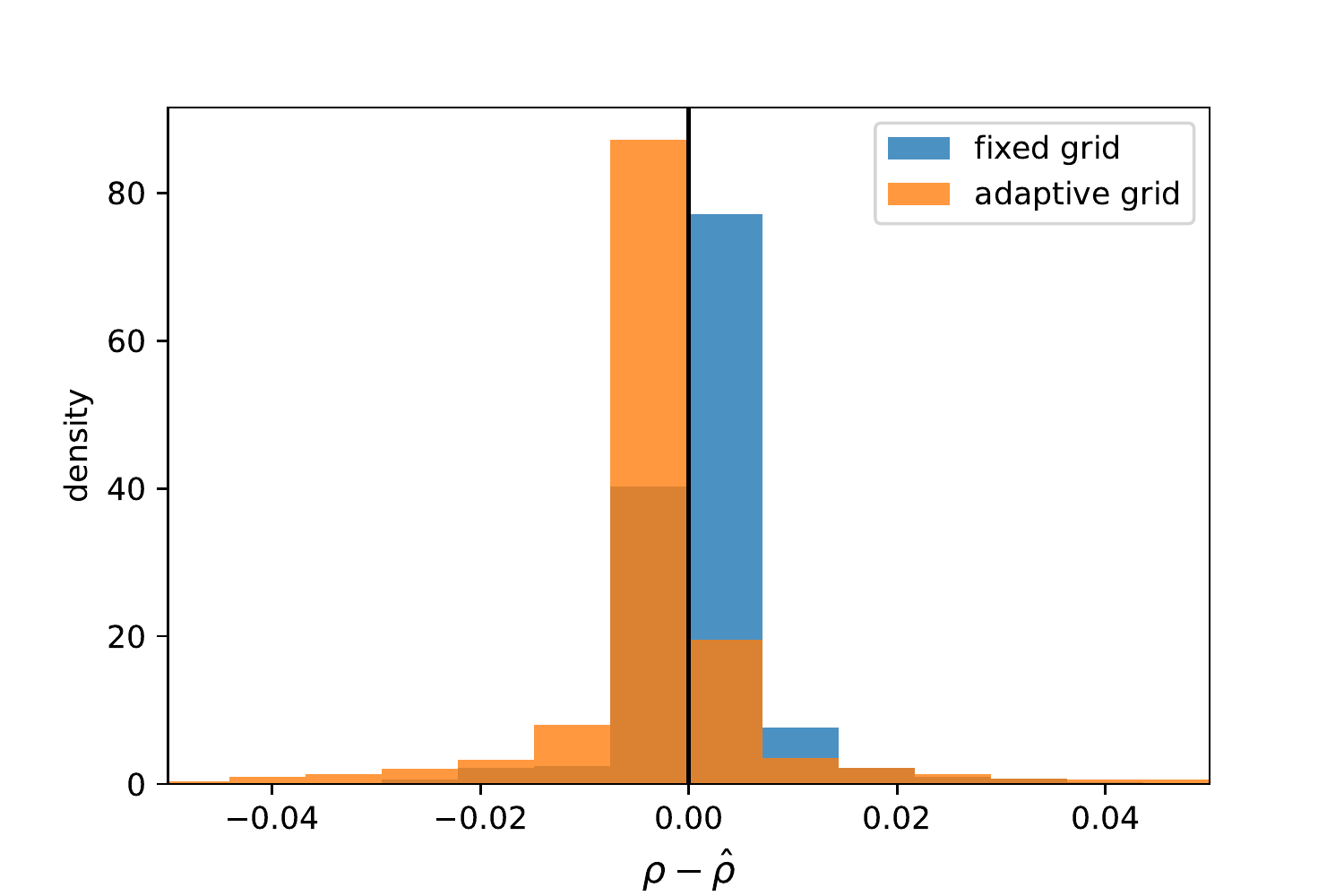}
         \caption{}
         \label{rHes_errRHO_fxdVSada}
     \end{subfigure}
     \hfill
     \begin{subfigure}[]{0.49\textwidth}
         \centering
         \includegraphics[width=\textwidth]{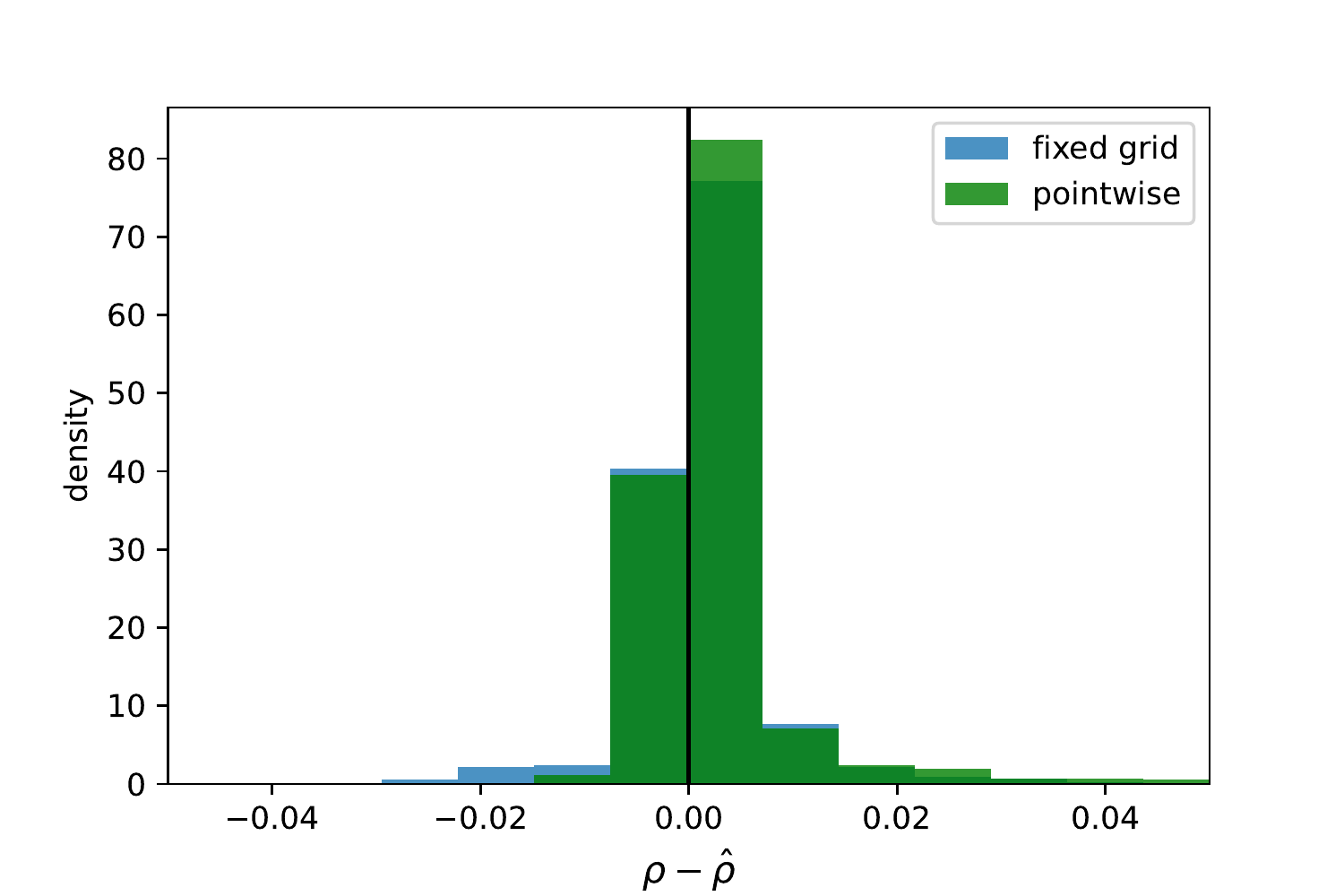}
         \caption{}
         \label{rHes_errRHO_fxdVSpnt}
     \end{subfigure}
     \caption{Distribution of absolute errors $e_{\rho}$ for fixed-grid, adaptive-grid and pointwise (adaptive-grid) neural network estimates of the correlation $\rho$ in the rHeston model over a sample of 1000 surfaces with flat variance curve.}
     \label{rHes_RHO_estimate}
\end{figure}

\noindent As for the correlation parameter $\rho$, the adaptive grid exhibits more variability than the fixed grid (Figure \ref{rHes_errRHO_fxdVSada}) but the pointwise approach fixes this (Figure \ref{rHes_errRHO_fxdVSpnt}). Table \ref{eRHO_tab} reveals that the average error with the pointwise approach is very similar to the fixed grid but the range much smaller and the standard deviation halved. \\

\noindent Hence, the pointwise approach emerges as the best compromise from this comparison, but its true potentialities will be further explored in the subsequent sections. For the moment, we only conclude that making the range of the strikes depend on the maturity of the option is not only more practical under simulation, but more financially sound and particularly important to get an accurate estimate of the roughness of the volatility. We believe this is an important piece of information as an entire stream of literature is especially interested in the subject: from possible estimation methods to the description of the empirical behavior of the ATM skew.

\subsection{Pricing off the grid}

Whenever a grid-based method is used for calibration, separate pricing of all market points follows. Points off the grid may be achieved either via interpolation techniques or with the true pricing function. Either ways, the estimation issues we have outlined above impact on the quality of the fit. \\

\noindent Because the pointwise approach calibrates to all available data (at least those in the domain where the network has been trained) we expect it to perform better than the combination of a grid-based method plus extension to the market points. Then, it is natural to take the pointwise approach as a benchmark. Evidence in the previous section suggests that the general tendency from the fixed grid is an underestimation of the roughness parameter being accompanied with lower volatility of volatility and more negative spot-vol correlation. Market data confirms this tendency. March 16, 2021 is an example, with the following calibration parameters in Table \ref{rHes_params_fxdVSadaVSpnt} for the rHeston model: \\

\begin{table}[h!]
    \centering
    \begin{tabular}{|c|c|c|c|}
        \hline
        model parameters & fixed grid & adaptive grid & pointwise \\
        \hline
        $H$    &  0.0100 &  0.0933 &  0.0808 \\
        $\nu$  &  0.3563 &  0.5065 &  0.5355 \\
        $\rho$ & -0.8085 & -0.6575 & -0.6678 \\
        \hline
    \end{tabular}
    \caption{Estimated rHeston parameters as of March 16, 2021 using both types of grid and the pointwise approach.}
    \label{rHes_params_fxdVSadaVSpnt}
\end{table}

\begin{figure}[h!]
     \centering
     \begin{subfigure}[]{0.49\textwidth}
         \centering
         \includegraphics[width=\textwidth]{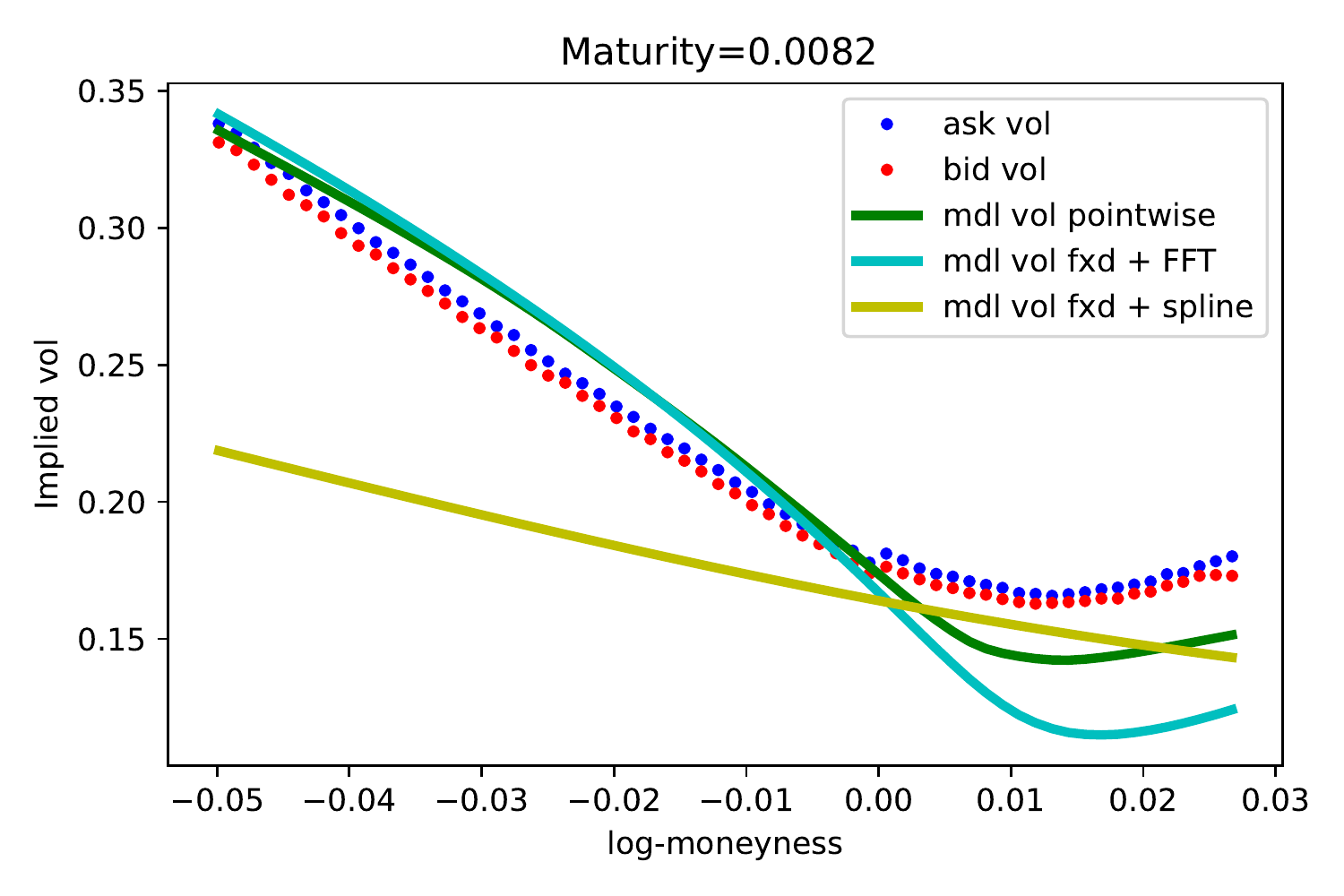}
     \end{subfigure}
     \hfill
     \begin{subfigure}[]{0.49\textwidth}
         \centering
         \includegraphics[width=\textwidth]{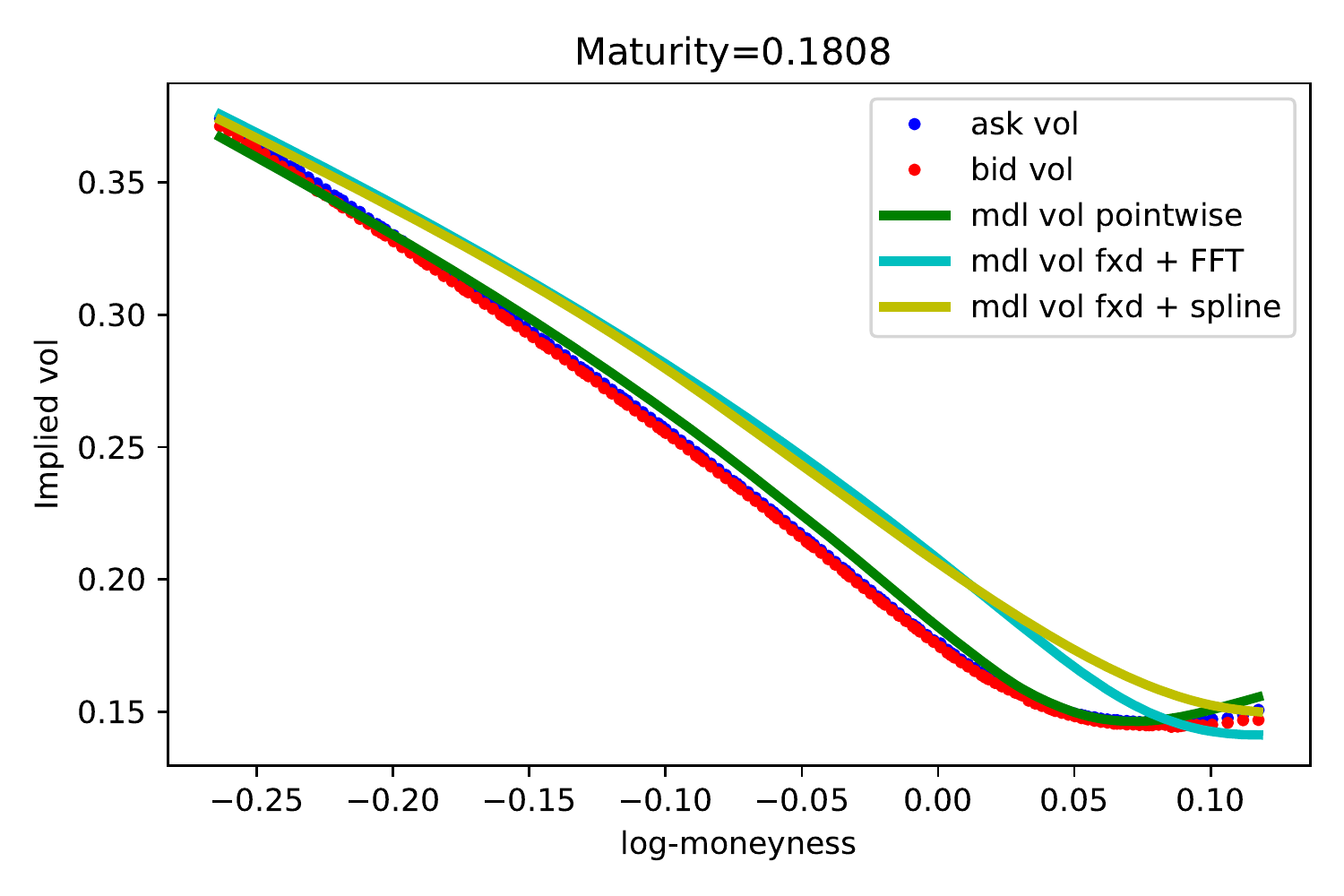}
     \end{subfigure}
     \begin{subfigure}[]{0.49\textwidth}
         \centering
         \includegraphics[width=\textwidth]{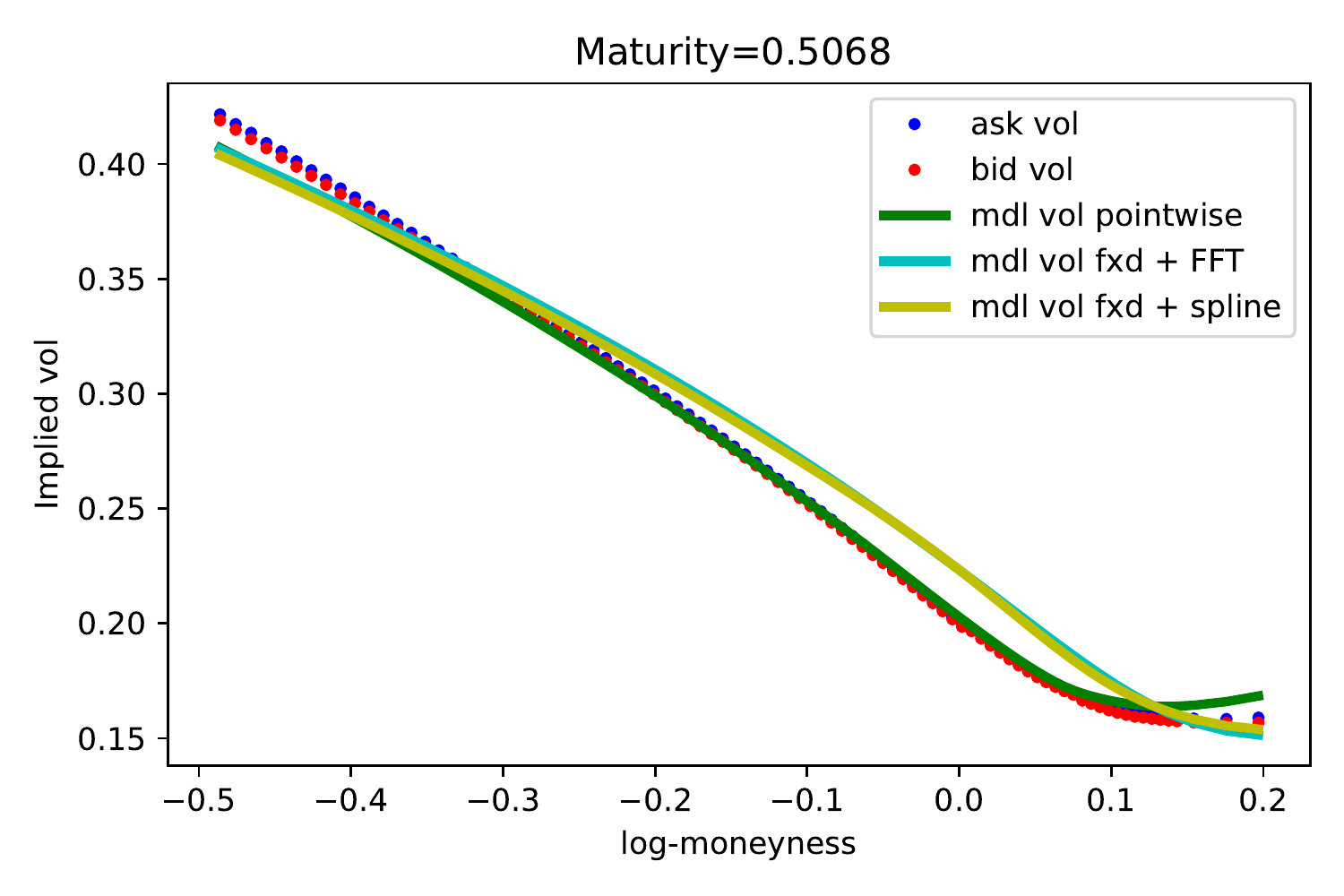}
     \end{subfigure}
     \hfill
     \begin{subfigure}[]{0.49\textwidth}
         \centering
         \includegraphics[width=\textwidth]{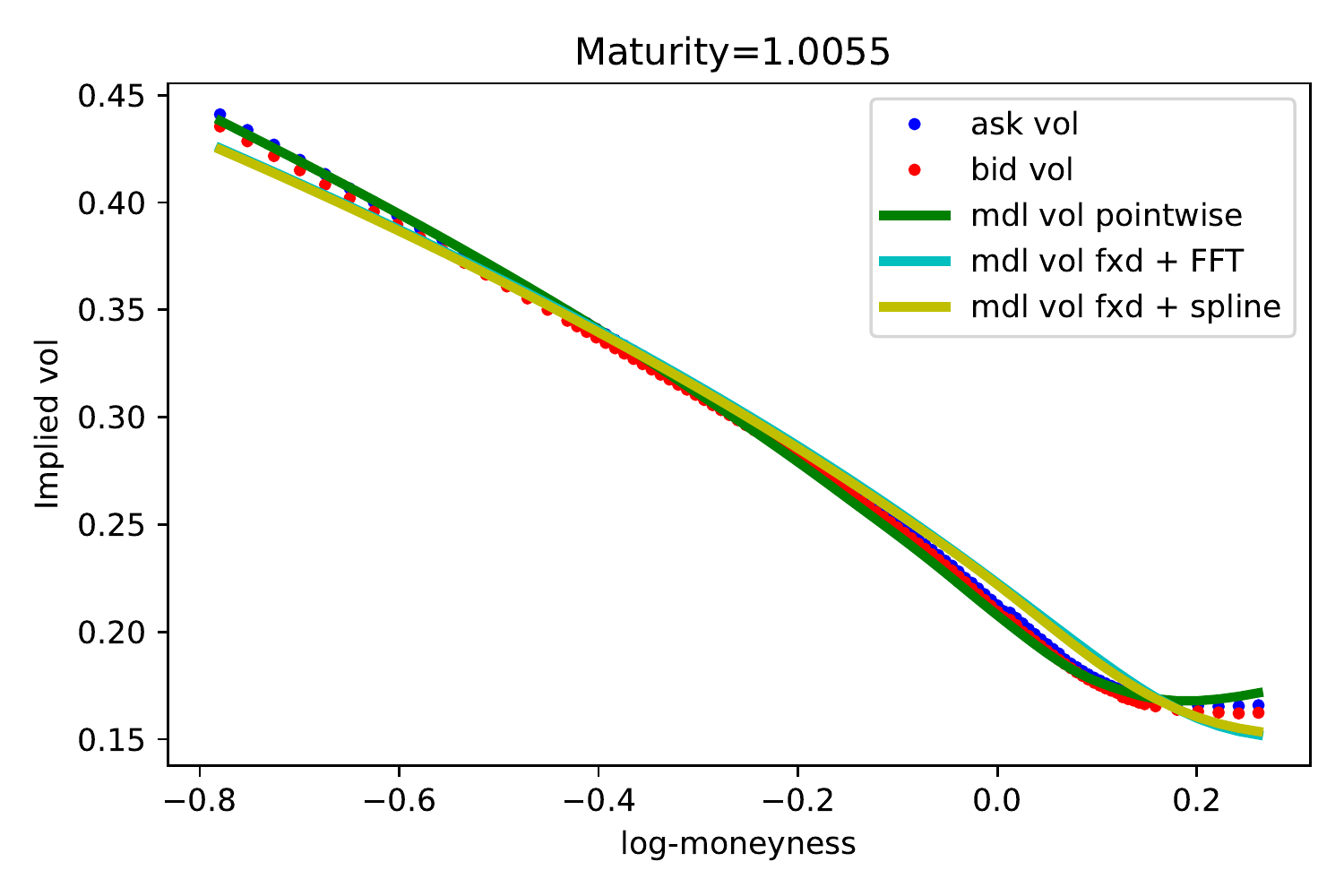}
     \end{subfigure}
     \caption{Calibration to the market vol. surface as of March 16, 2021. Green is pointwise (random grid), cyan is fixed grid (fxd) plus interpolation/extrapolation by the true pricing function (FFT for rHeston), and yellow is fixed grid plus splines.}
     \label{interp_fxd}
\end{figure}

\noindent Extension to all market points with the true pricing function is biased as a consequence of the calibrated parameters from the grid being biased. Specifically, the roughness parameter $H$ being lower than it should results in a steeper smile at very short times. Not only, problems may also arise with the level of the smile. The optimal level of the forward variance curve for the first grid maturity is not necessarily optimal for shorter market times thus shifting the smile vertically. Both effects are visible in the upper left corner of Figure \ref{interp_fxd} (green vs cyan lines). \\

\begin{figure}[h!]
\centering
\includegraphics[width=\textwidth, trim=0 400pt 0 0, clip]{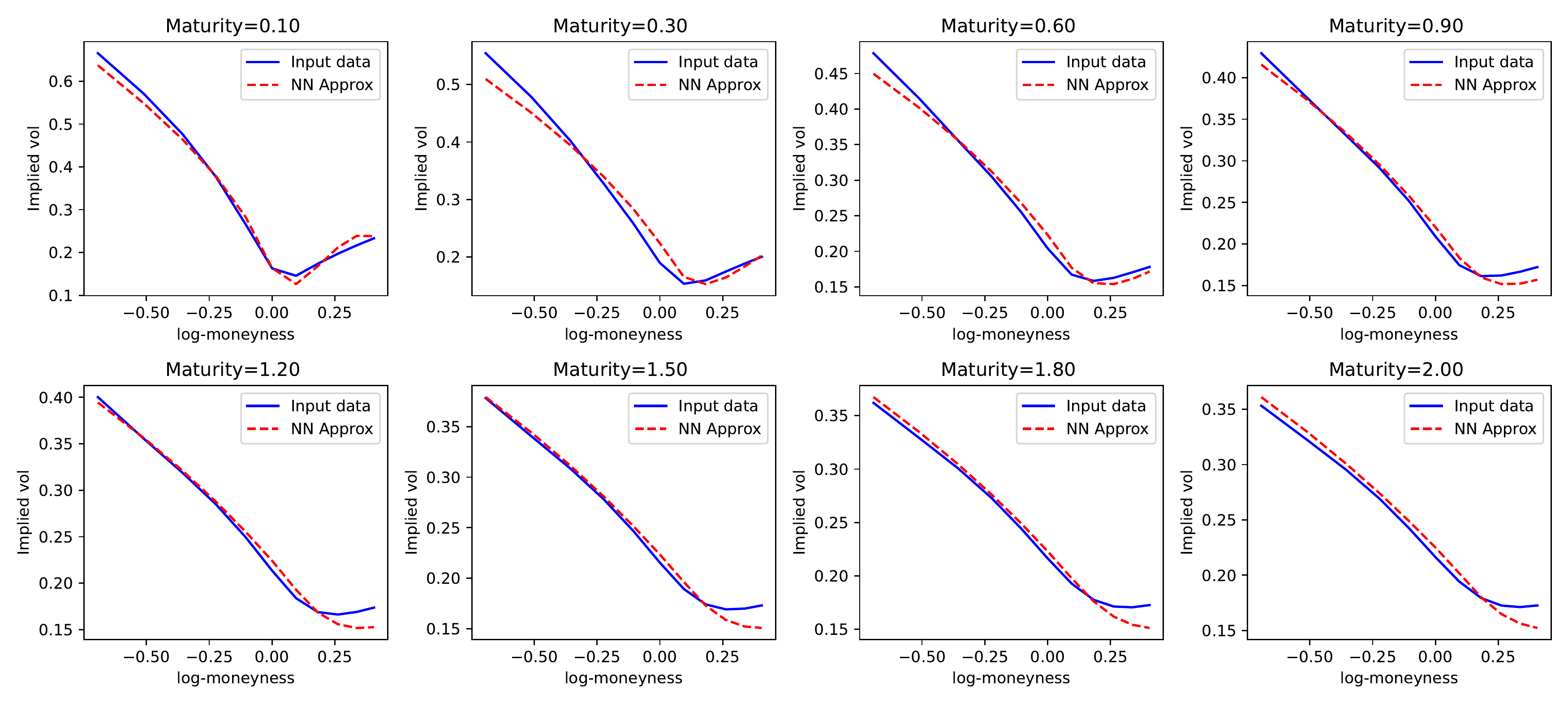}
\caption{Neural network calibration of the market projection on a fixed grid as of March 16, 2021.}
\label{fxdGrid2mkt}
\end{figure}

\noindent Poor quality of the fit also extends to longer maturities. It is important to notice, though, that the real matter with such later times is not extension itself but the quality of the calibration to the grid. If the fit to the grid is not good enough, extension to the whole market cannot certainly do better. Figure \ref{fxdGrid2mkt} in fact reveals substantial difference in the shape of the smile when comparing grid volatilities with the fit from the network, especially around the ATM. Needless to say, problems on the grid propagate to the whole market irrespectively of the method one uses for extension. The yellow curves in Figure \ref{interp_fxd} that we obtain via interpolation with spline techniques are also very far from the market. \\

\noindent Extrapolation via spline techniques needs a special mention. Standard practice in the literature is to extrapolate the smiles flat before the first maturity, but doing so from 0.1 years may result in short-dated smiles to loose all of their curvature. This effect is clear in the top left corner of Figure \ref{interp_fxd} (yellow curve). \\

\noindent All in all, then, a fixed grid must be recognized to be possibly inappropriate for calibration to the market for at least two reasons. Firstly, it does not allow for the inclusion of very short expiries that are crucial for proper estimation of the roughness and extrapolation via the true pricing function. Secondly, minimizing a distance over implied volatilities and making the range of the strikes so large for short maturities as a couple of months or so places much of the weight on the wings of the smile at the cost of the ATM region. The price for this is poor interpolation as we have described above. \\

\noindent The case of an adaptive grid is very much different. Model parameters are now closer to those we obtain with the pointwise approach -- if we come back to Table \ref{rHes_params_fxdVSadaVSpnt}. Also, the forward variance curve being more versatile at short times~\footnote{Recall that we have included very short maturities in the adaptive grid and ensured the forward variance curve changes level over the same buckets.} partially fixes possible issues with the level of the first smiles. \\

\noindent The fact that including short-dated options provides us with improved estimates of the roughness parameter is no surprise after the discussion in the previous section. VOV and correlation are also more sound and the overall fit to the market is definitely superior with respect to the one we are able to achieve with a fixed grid. \\

\begin{figure}[h!]
\centering
\includegraphics[width=\textwidth, trim=0 400pt 0 0, clip]{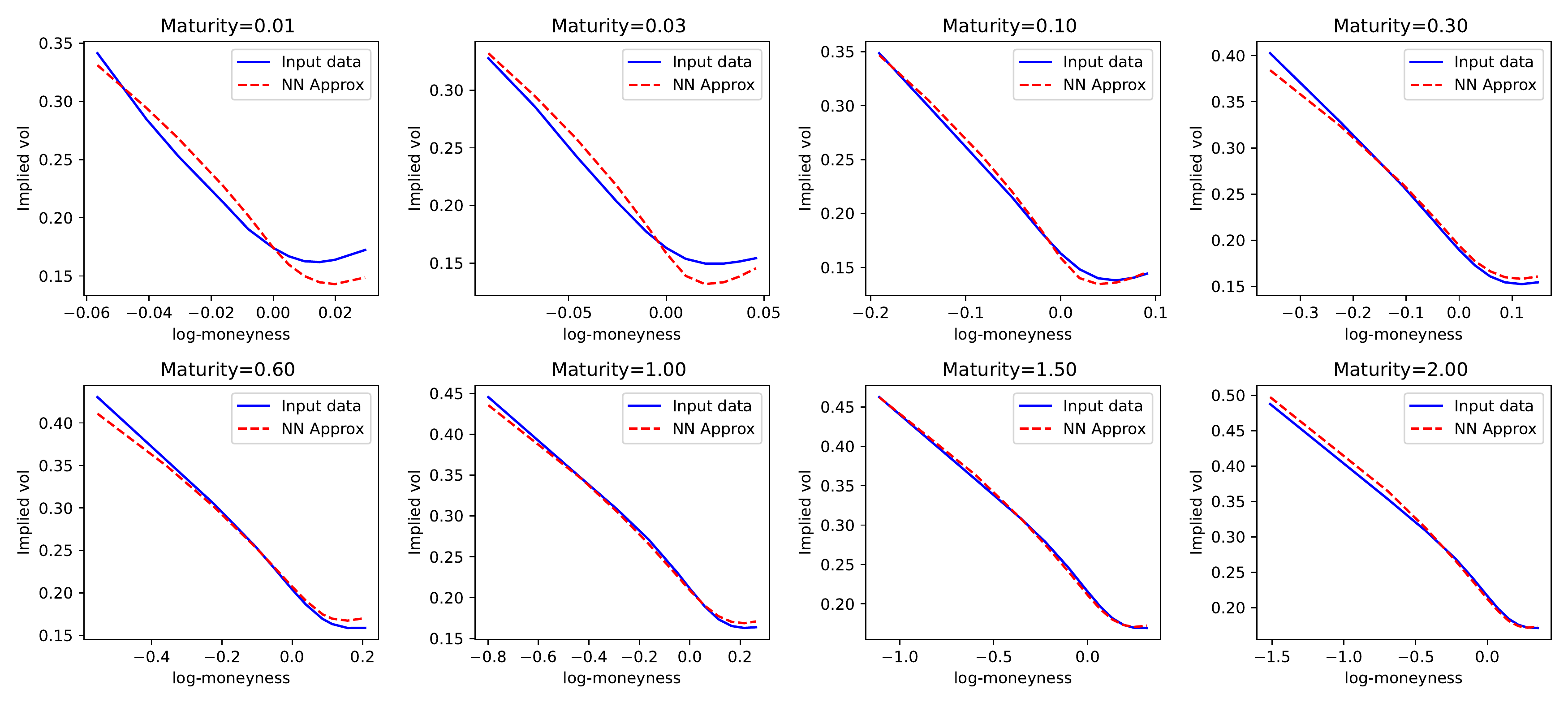}
\caption{Neural network calibration of the market projection on an adapted grid as of March 16, 2021.}
\label{adaGrid2mkt}
\end{figure}

\begin{figure}[h!]
     \centering
     \begin{subfigure}[]{0.49\textwidth}
         \centering
         \includegraphics[width=\textwidth]{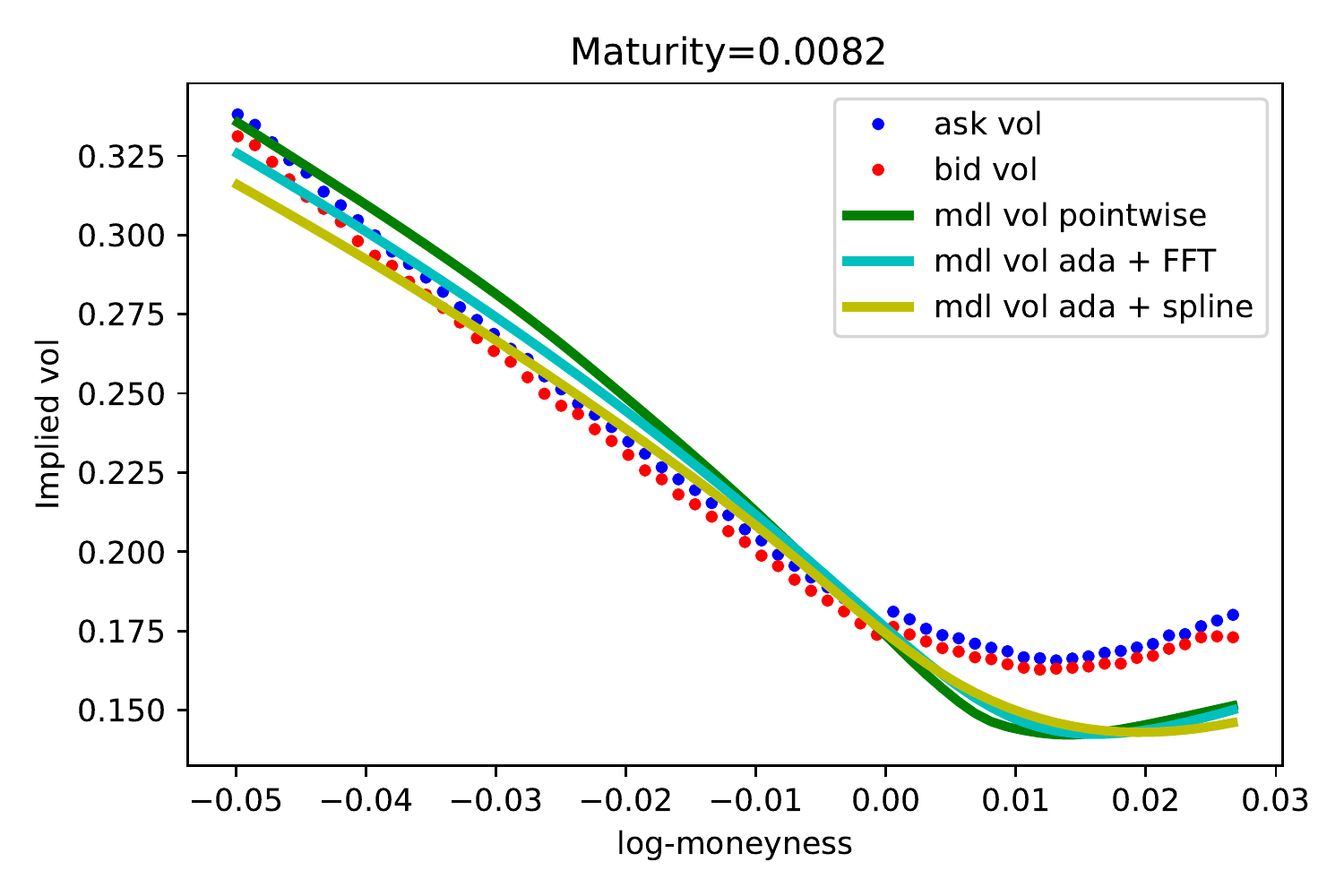}
     \end{subfigure}
     \hfill
     \begin{subfigure}[]{0.49\textwidth}
         \centering
         \includegraphics[width=\textwidth]{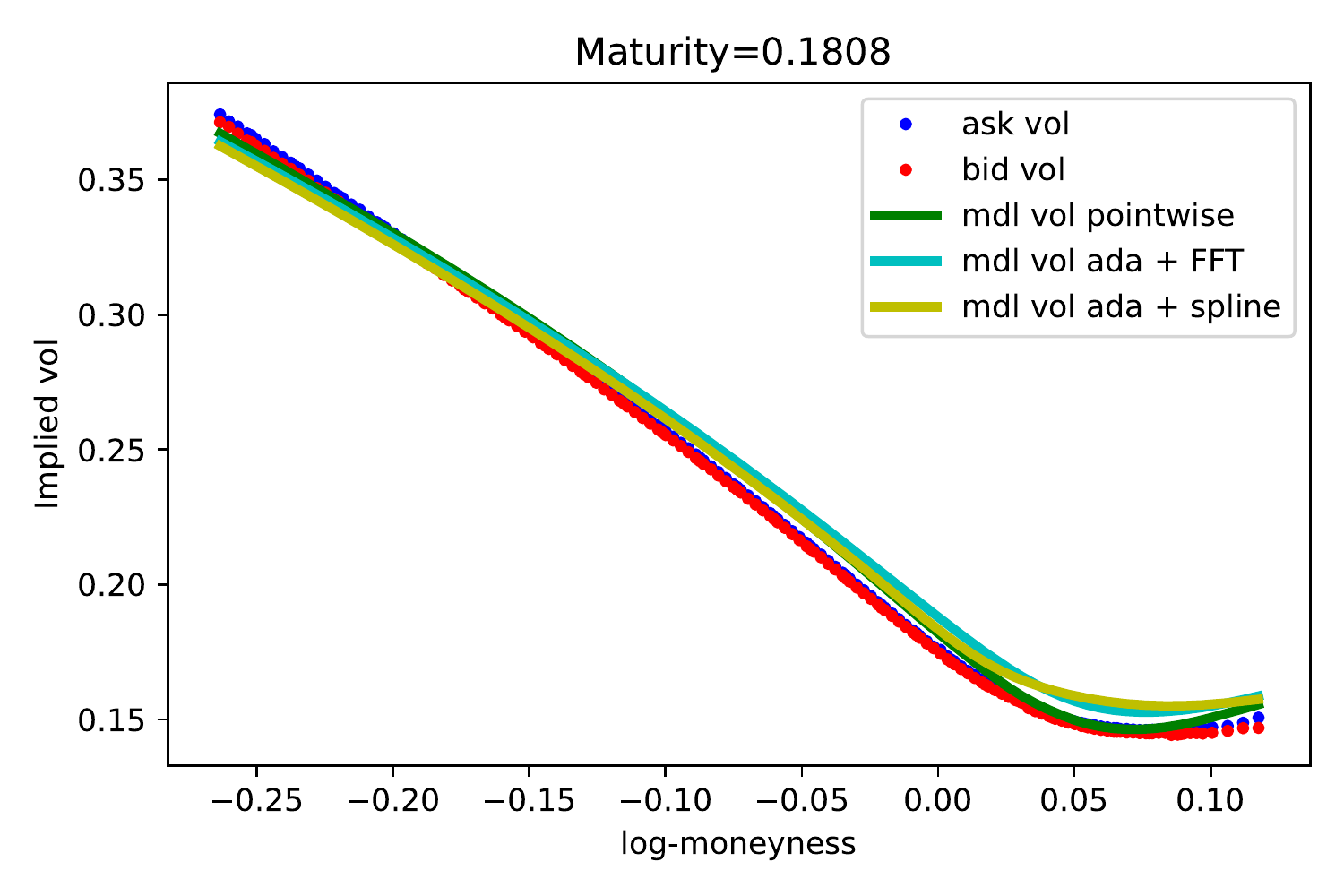}
     \end{subfigure}
     \begin{subfigure}[]{0.49\textwidth}
         \centering
         \includegraphics[width=\textwidth]{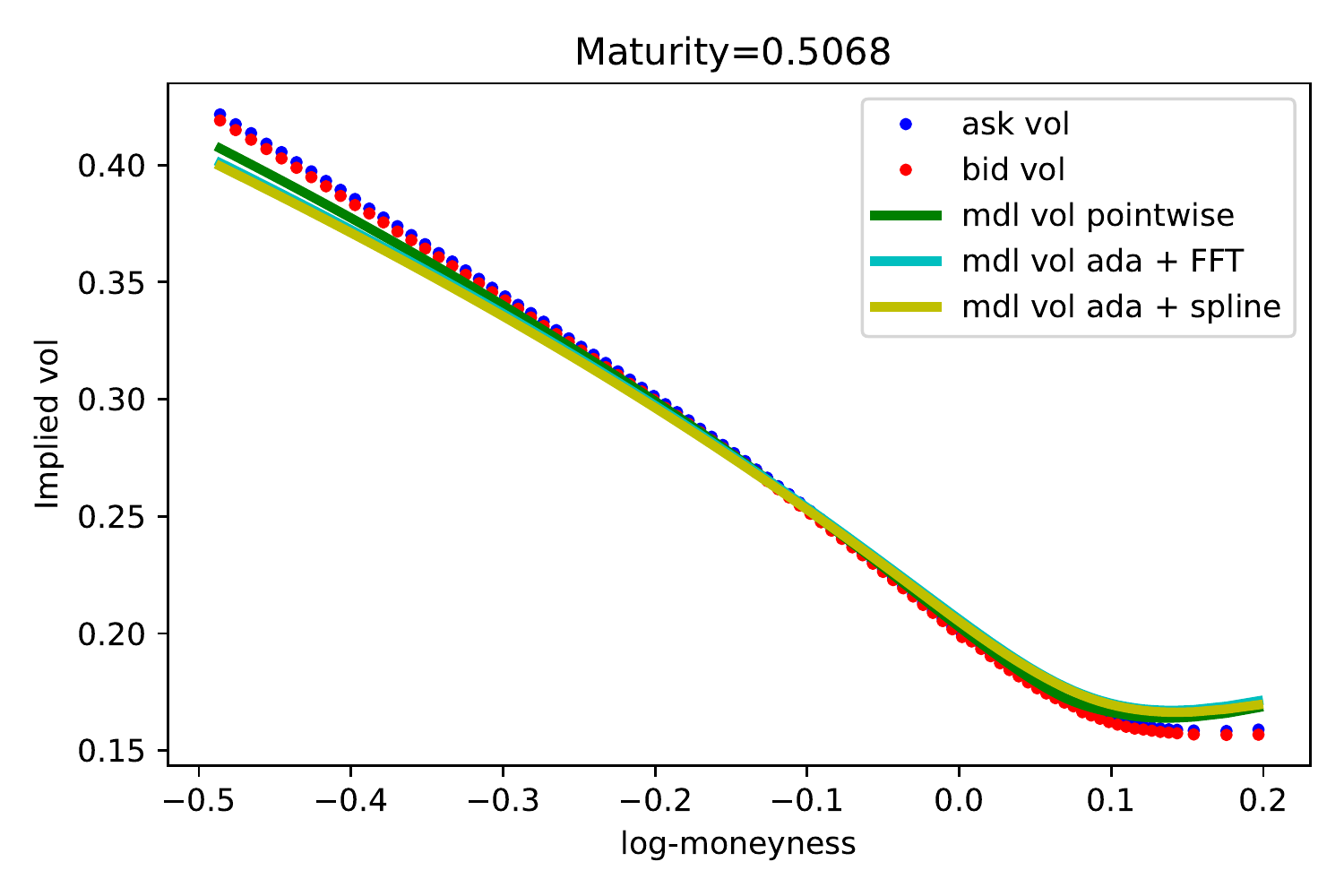}
     \end{subfigure}
     \hfill
     \begin{subfigure}[]{0.49\textwidth}
         \centering
         \includegraphics[width=\textwidth]{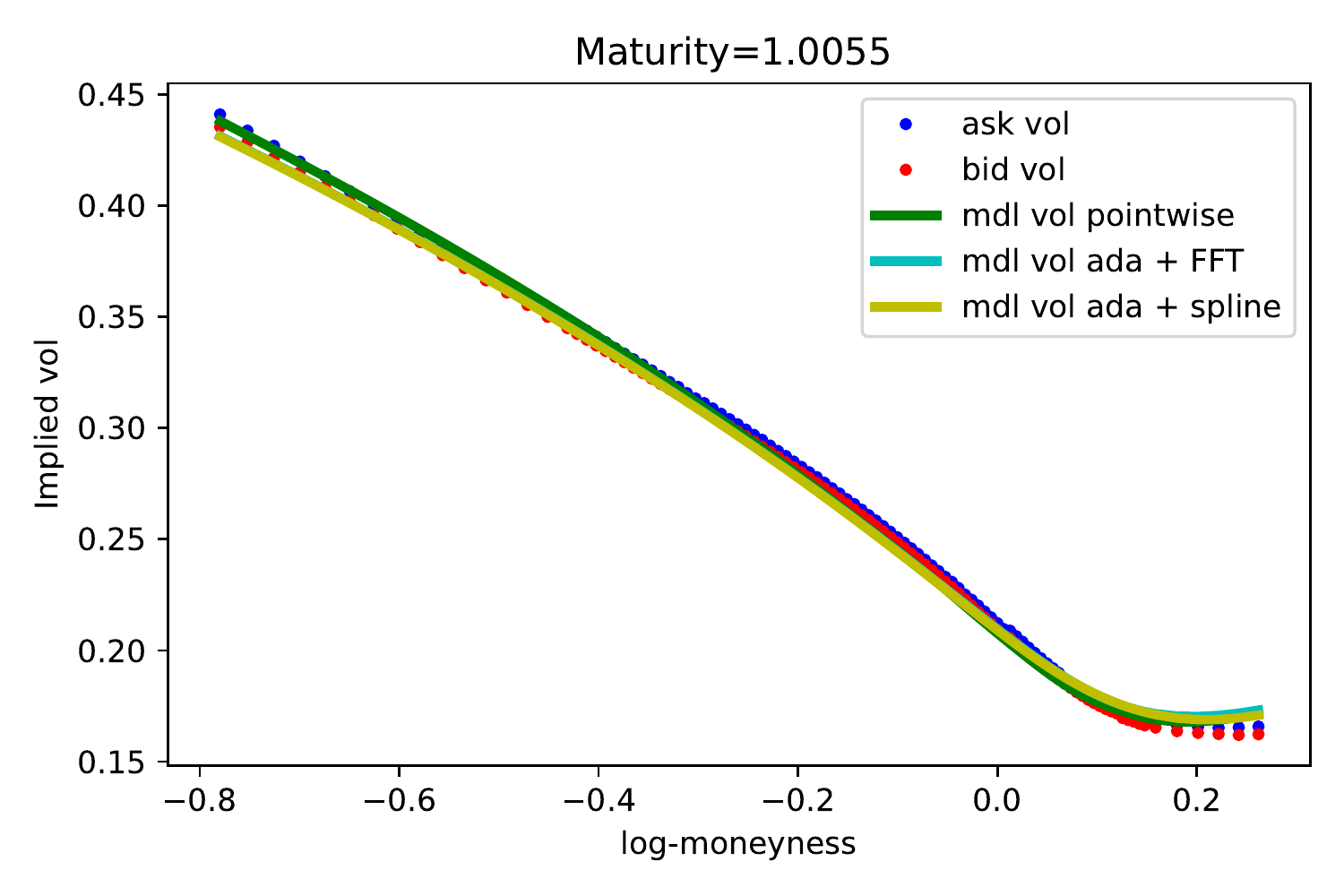}
     \end{subfigure}
     \caption{Calibration to the market vol. surface as of March 16, 2021. Green is pointwise (random grid), cyan is adaptive grid (ada) plus interp./extrap. by the true pricing function (FFT for rHeston) and yellow is adapted grid plus splines.}
     \label{interp_ada}
\end{figure}

\noindent Figure \ref{adaGrid2mkt} shows those problems around the ATM that we reported with a fixed grid have now disappeared for monthly maturities and longer. We therefore expect interpolation to be grounded on a solid base and do its job when moving to the market volatility surface. This is confirmed in Figure \ref{interp_ada} smiles are not as good as with the pointwise approach (at least not always) but far better than they used to be with a fixed grid. \\

\noindent Extrapolation now needs to bridge a much shorter interval and it therefore performs better than before. Actually, it happens, in this particular case, that the local performance of a mixture method grid plus interpolation/extrapolation is superior to the pointwise approach: The yellow curve in the top left corner of Figure \ref{interp_ada} fits the market slightly better than the green one. \\

\noindent This is clearly a random effect which cannot be made systematic and cannot become a global phenomenon. 
Conversely, theoretical reasons for the pointwise approach to perform better are obviously found in the fact that it sees all of the dataset, by its nature. \\

\begin{figure}[h!]
     \centering
     \begin{subfigure}[]{0.49\textwidth}
         \centering
         \includegraphics[width=\textwidth]{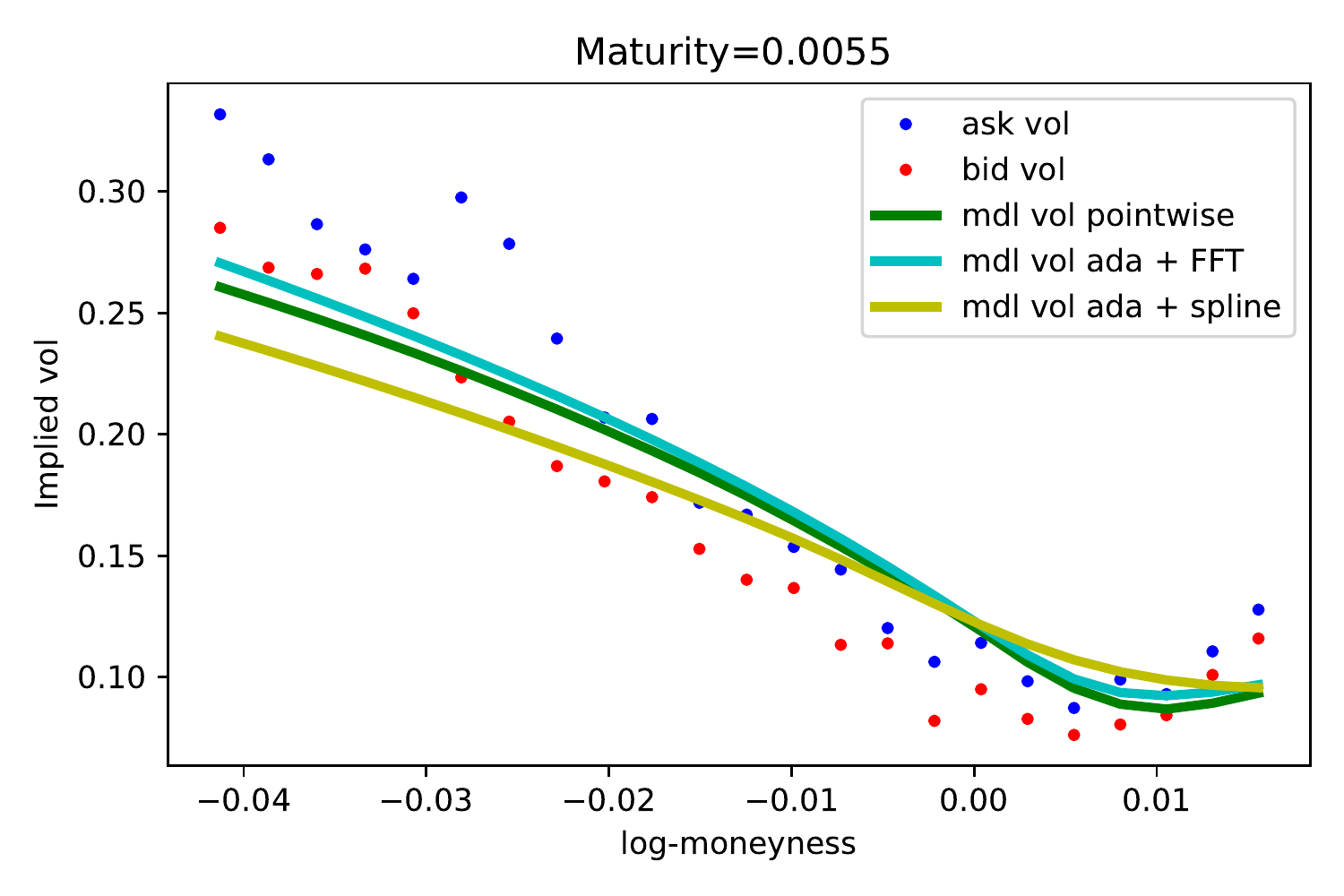}
     \end{subfigure}
     \hfill
     \begin{subfigure}[]{0.49\textwidth}
         \centering
         \includegraphics[width=\textwidth]{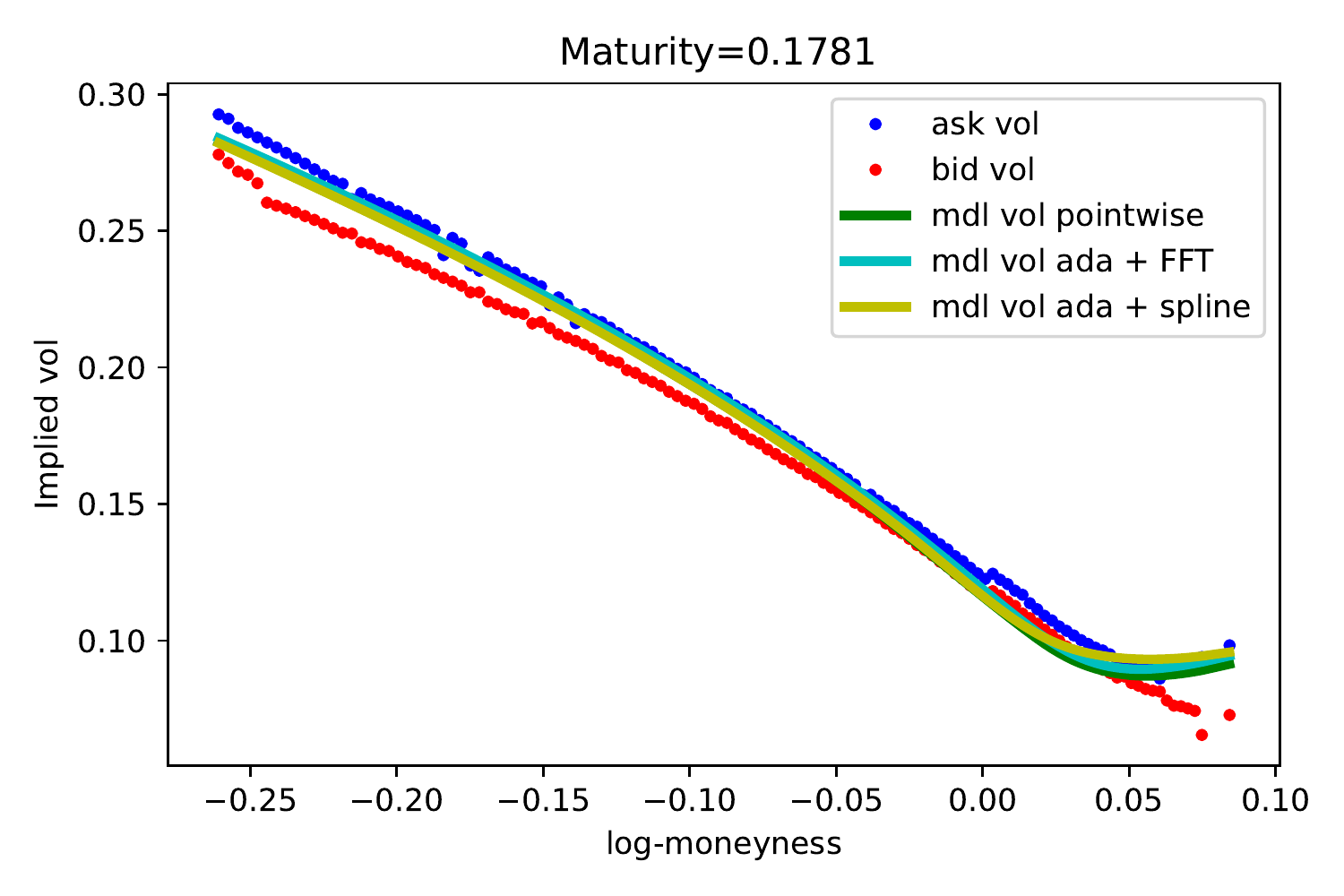}
     \end{subfigure}
     \begin{subfigure}[]{0.49\textwidth}
         \centering
         \includegraphics[width=\textwidth]{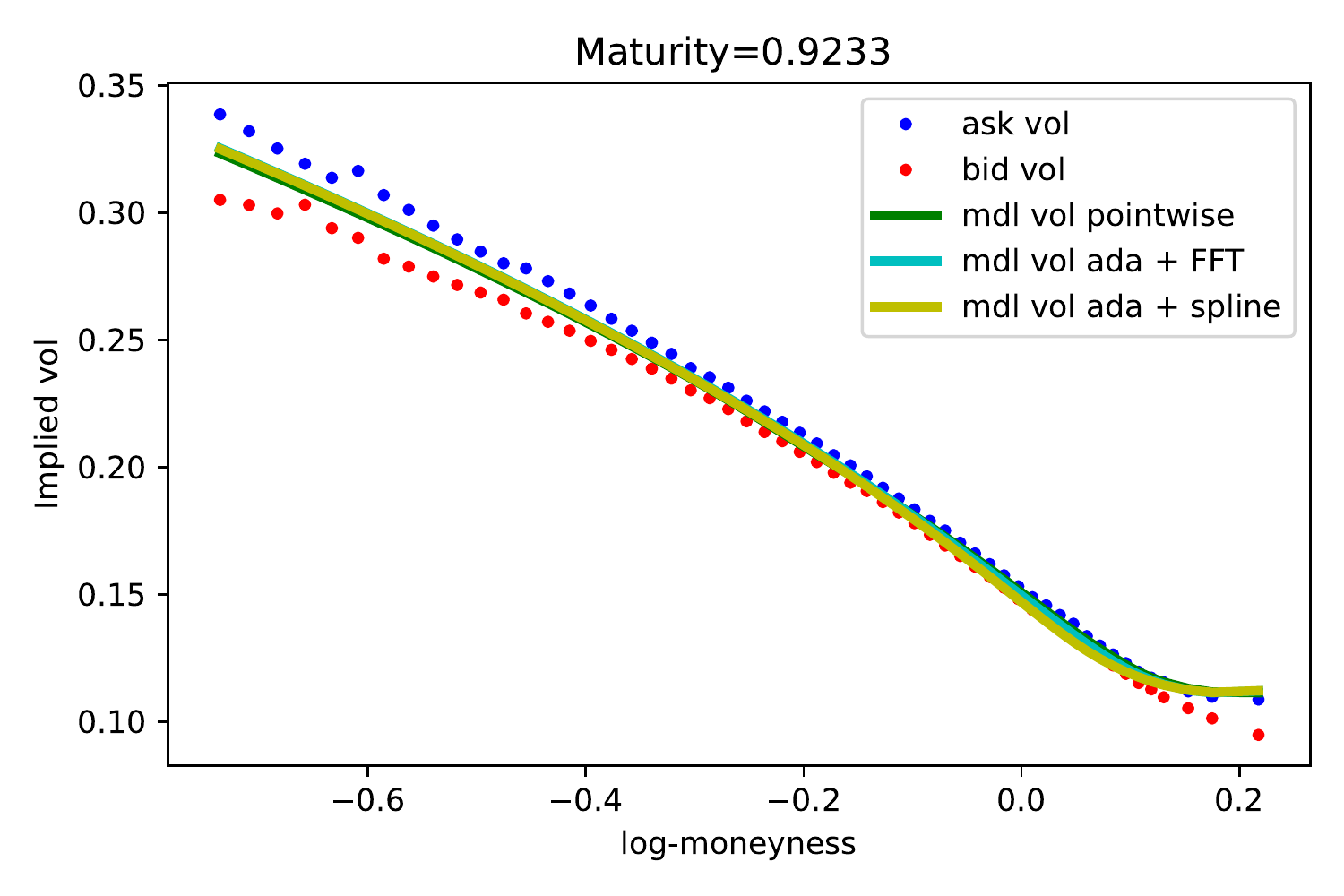}
     \end{subfigure}
     \hfill
     \begin{subfigure}[]{0.49\textwidth}
         \centering
         \includegraphics[width=\textwidth]{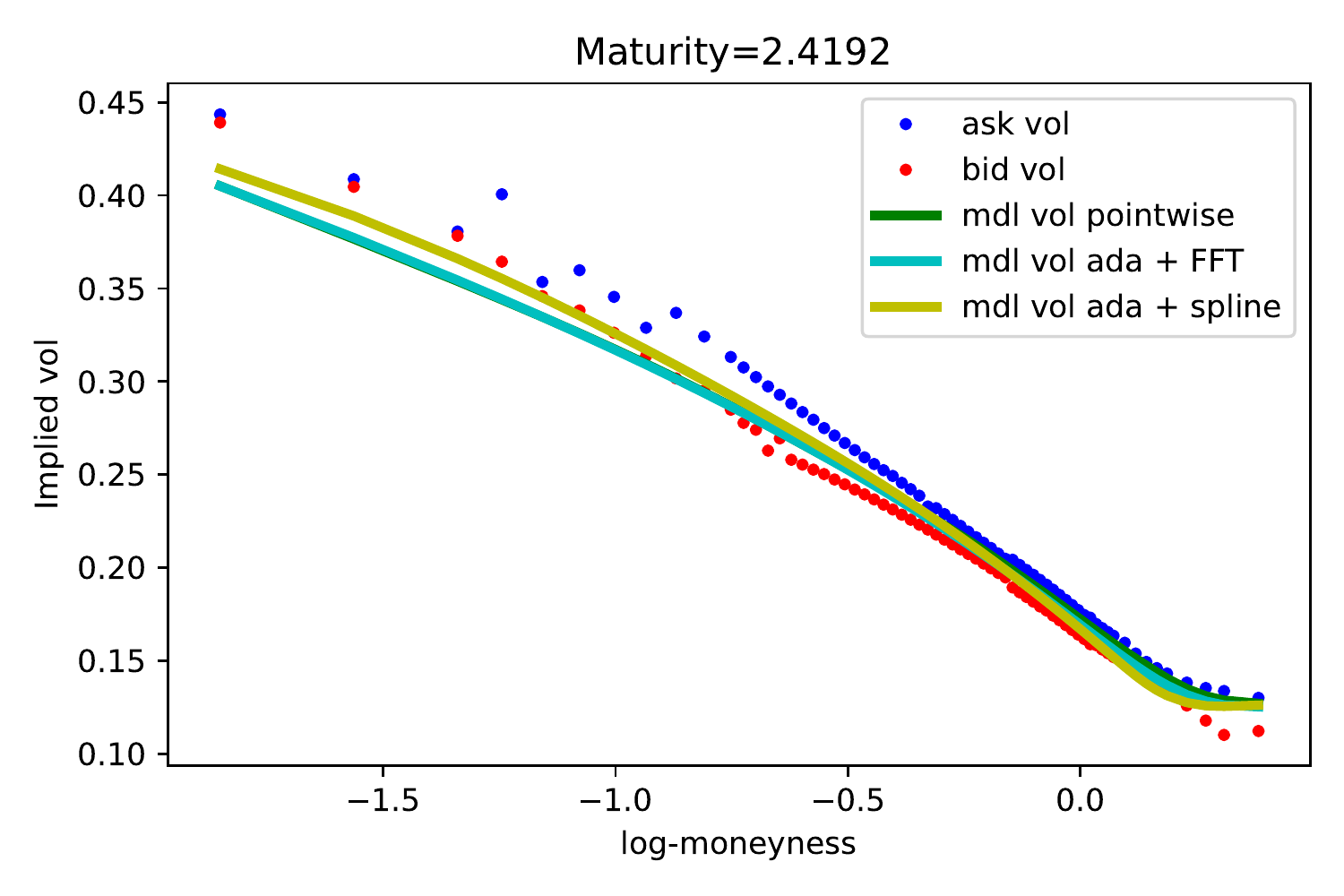}
     \end{subfigure}
     \caption{Calibration to the market vol. surface as of September 17, 2014. Green is pointwise (random grid), cyan is adaptive grid (ada) plus interpolation/extrapolation by the true pricing function (FFT for rHeston) and yellow is adapted grid plus splines.}
     \label{interp_ada_bis}
\end{figure}

\noindent In addition to this, different extrapolation methods suffer from different problems. The true pricing function is too slow to be competitive and spline techniques could make the smile  too flat at very short times. The adaptive grid alleviates this problem but does not solve it entirely. September 17, 2014 is an example of this as seen in Figure \ref{interp_ada_bis}. \\

\noindent We therefore conclude that the pointwise approach is more grounded and robust than a grid approach and, most importantly, it is self-contained, in the sense that it does not require any additional tools but it deals with both pricing and calibration alone (and in very decent time). 

\subsection{Forward variance specifications}

Option pricing and calibration under stochastic volatility models like rHeston and rBergomi - that we consider here - require precise choices for the treatment of the forward variance curve. \\

\noindent One can build it from variance swaps\footnote{Variance swaps can be priced in terms of an infinite log-strip of out of the money European options as explained in \cite{gatheral2011volatility}. The idea is that one comes up with a parametrization of the volatility surface and uses it to extend the market to a continuum of option prices to integrate over. Once variance swaps are priced on the market maturities, the forward variance curve can be readily computed by differentiation.} and make it a state variable following the example of \cite{el2019roughening} or - and this is more standard nowadays - one just fixes $c$ points in time and calibrates this many levels of a step-function together with model parameters. 
Such a number $c$ may be significantly reduced with respect to the market maturities with almost no effect on the quality of the fit. Also, making the forward variance curve to jump at specified times opens to neural network pricing of rough volatility models.
The dimensionality of the calibration problem is clearly increased but this is largely justified by the boost one obtains from the network - as compared to FFT inversion or Monte Carlo. \\

\noindent In the same spirit of the forward variance curve as a calibration object, we propose a simple parametrization that also comes with very good fits to the market. \\ 

\noindent The logic behind this is that we calibrate the rHeston model to a number of volatility surfaces for different days in history between 2010 and 2022 and plot the estimated piecewise constant forward variance curve to get a sense of the `typical' shape. \\

\noindent We report a few such curves as an example in Figure 
\begin{figure}[h!]
     \centering
     \begin{subfigure}[]{0.49\textwidth}
         \centering
         \includegraphics[width=\textwidth]{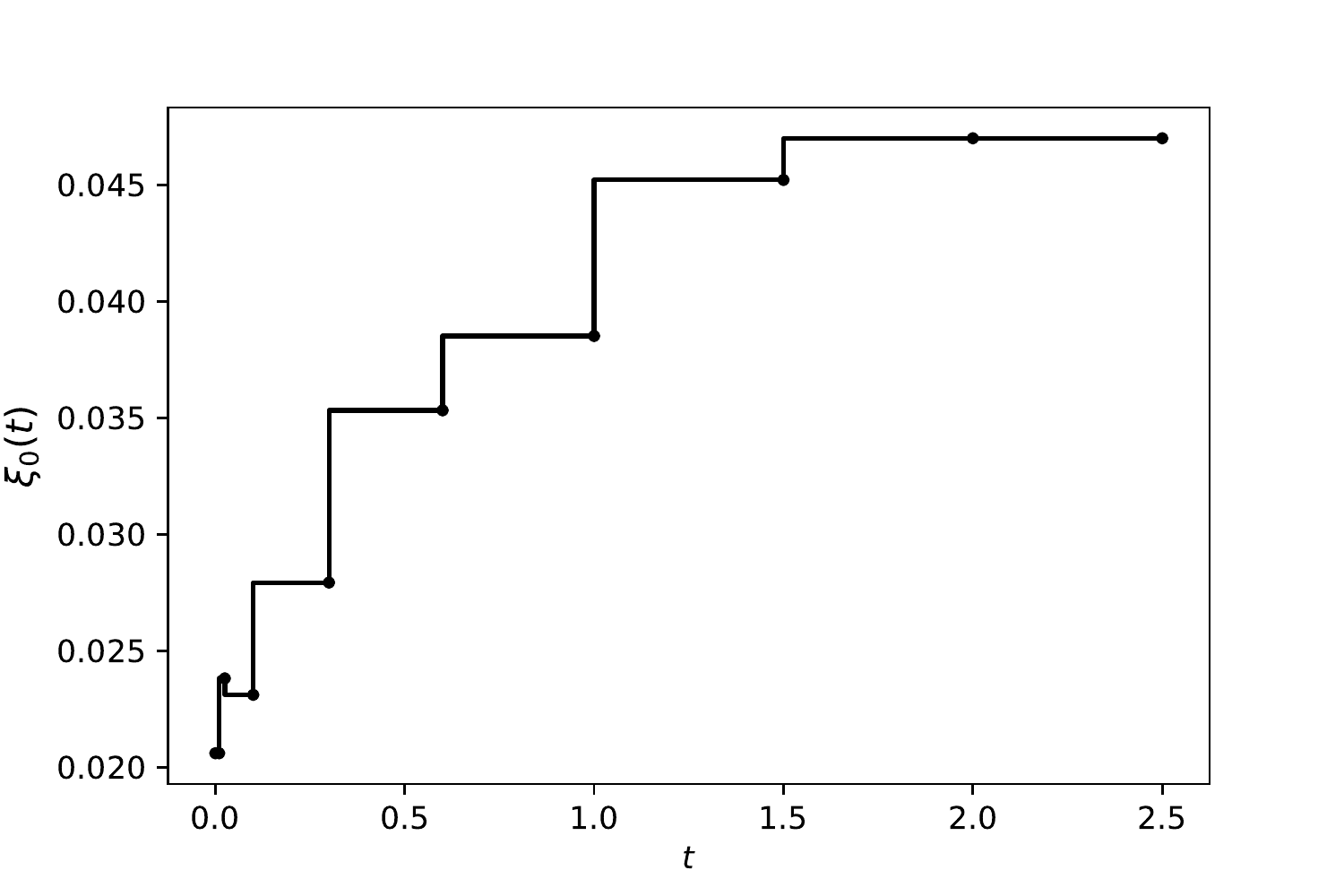}
         \caption{2014/03/19}
         \label{xi_pwc_a}
     \end{subfigure}
     \hfill
     \begin{subfigure}[]{0.49\textwidth}
         \centering
         \includegraphics[width=\textwidth]{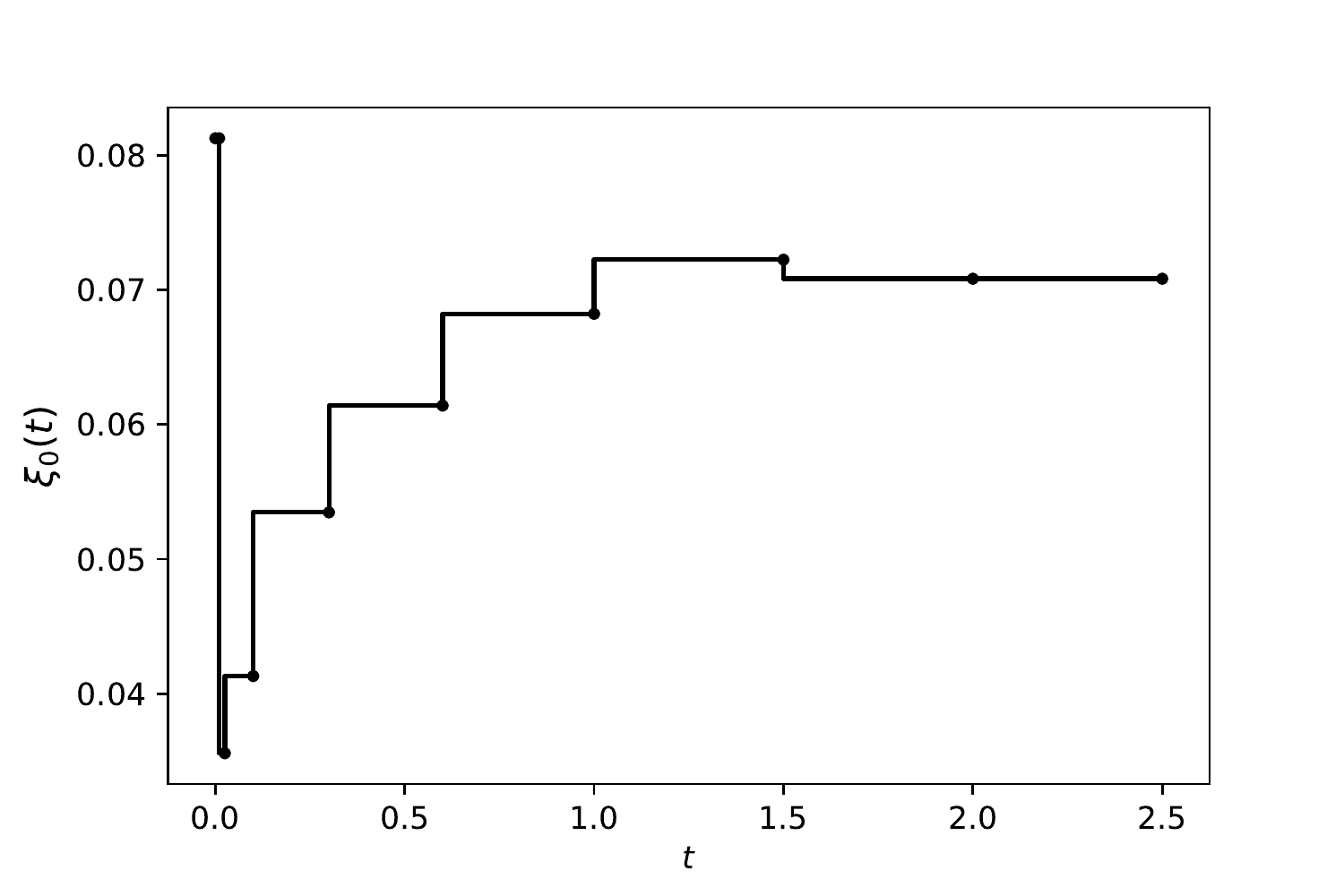}
         \caption{2011/02/23}
         \label{xi_pwc_b}
     \end{subfigure}
     \caption{Estimated piecewise constant forward variance curve as a result of calibration with the pointwise (random grid) approach under the rHeston model.}
     \label{rHes_xi_pwc}
\end{figure}
\ref{rHes_xi_pwc} and recognize that the vast majority of the surfaces only exhibits a little bit of a term structure at short times - if any. Then, monotonic growth as in Figure \ref{xi_pwc_a} may be handled with a straightforward exponential form but the presence of humps and/or bumps in Figure \ref{xi_pwc_b} requires functional forms that are somehow more involved, for example:
\begin{align}\label{fvc_par}
    \xi_0(t) = \beta_0 + \beta_1 \exp \bigg( -\frac{t}{\tau_1} \bigg) + \beta_2 \bigg(\frac{t}{\tau_2}\bigg) \exp \bigg( -\frac{t}{\tau_2} \bigg)
\end{align}
so that
\begin{align*}
& \lim_{t \to 0} \xi_0(t) = \beta_0 + \beta_1 > 0\\       
& \lim_{t \to +\infty} \xi_0(t) = \beta_0 \hspace{5.7mm} > 0.
\end{align*}
The reader would recognize in our Equation \eqref{fvc_par} a slight generalization of the \cite{nelson1987parsimonious} model, where the long-term component $e^{-x}$ and the mixed term $x e^{-x}$ are now allowed to evolve at a different rate. It is precisely the interplay of those two terms that is responsible for the humps and/or bumps that we were trying to replicate. For this, we observe that two stationary points are allowed under Equation \eqref{fvc_par} when $\beta_1 \beta_2 < 0$ and $\tau_2$ prescribes sufficiently fast increase/decrease at very short times. \\

\noindent We identify the mixed component to be a short-term one because we confine $\tau_2$ to a region such that the contribution from this is exhausted in about one year. In particular, we take forward variance parameters uniformly at random from the hypercube in Table \ref{fvc_par_range}
\begin{table}[h!]
    \centering
    \begin{tabular}{|c|c|}
        \hline
        forward variance parameters & \\
        \hline
        $\beta_0$ &  $\mathcal{U}([0.025,0.160])$ \\
        $\beta_0 + \beta_1$ & $\mathcal{U}([0.005,0.130])$ \\
        $\beta_2$ & $\mathcal{U}([-.150,0.250])$\\
        $\tau_1$ & $\mathcal{U}([0.001,1.350])$\\
        $\tau_2$ & $\mathcal{U}([0.001,0.125])$ \\
        \hline
    \end{tabular}
    \caption{Range of the forward variance parameters}
    \label{fvc_par_range}
\end{table}
for generation and training. We therefore calibrate over this same region during step 2. We do not waste computer time dealing with parameter sets where (at least one of) the $\beta$ coefficients are approximately zero, because the associated curves in those regions are practically indistinguishable from one another. Yet, the network will most likely be able to bridge this gap and let us calibrate over the whole hypercube. \\

\noindent Because nothing prevents our specification of the forward variance curve to become negative for particular choices of the parameters, we simply discard those combinations while producing the samples and only train the neural network on well-behaved parameters sets. The true market curve being strictly positive, this restriction should be no worry\footnote{We indeed observe the calibrated forward variance curve to be negative at very short times (only) if no such short maturities as a week or so are included in the volatility surface, in which case the surface would be of reduced practical interest (no price for typical hedging options, no clue as to the exploding behavior of the ATM skew and so on). In any case, a non-linearly constrained optimization may be set up to avoid problems of this sort.}. \\

\begin{figure}[h!]
     \centering
     \begin{subfigure}[]{0.49\textwidth}
         \centering
         \includegraphics[width=\textwidth]{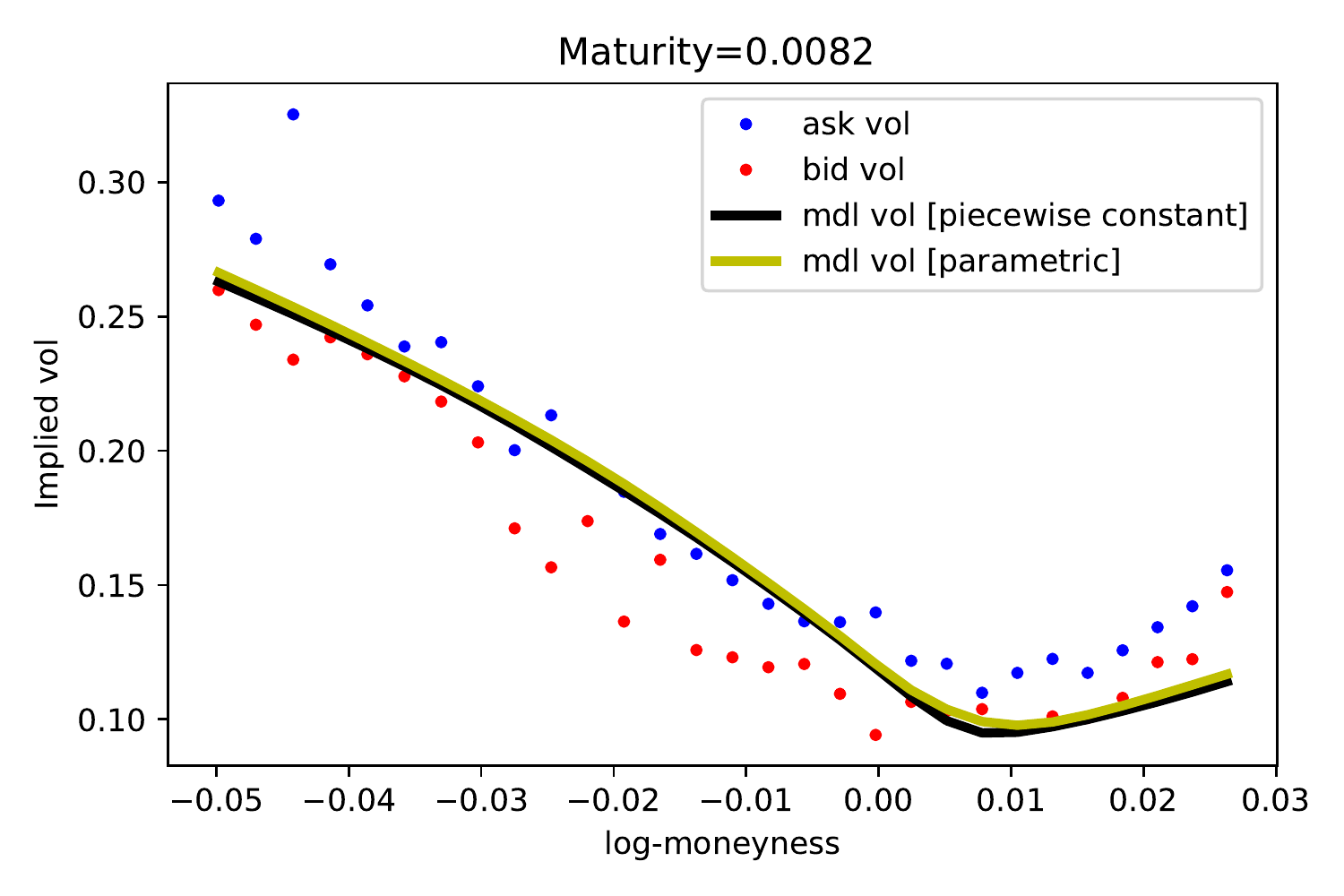}
     \end{subfigure}
     \hfill
     \begin{subfigure}[]{0.49\textwidth}
         \centering
         \includegraphics[width=\textwidth]{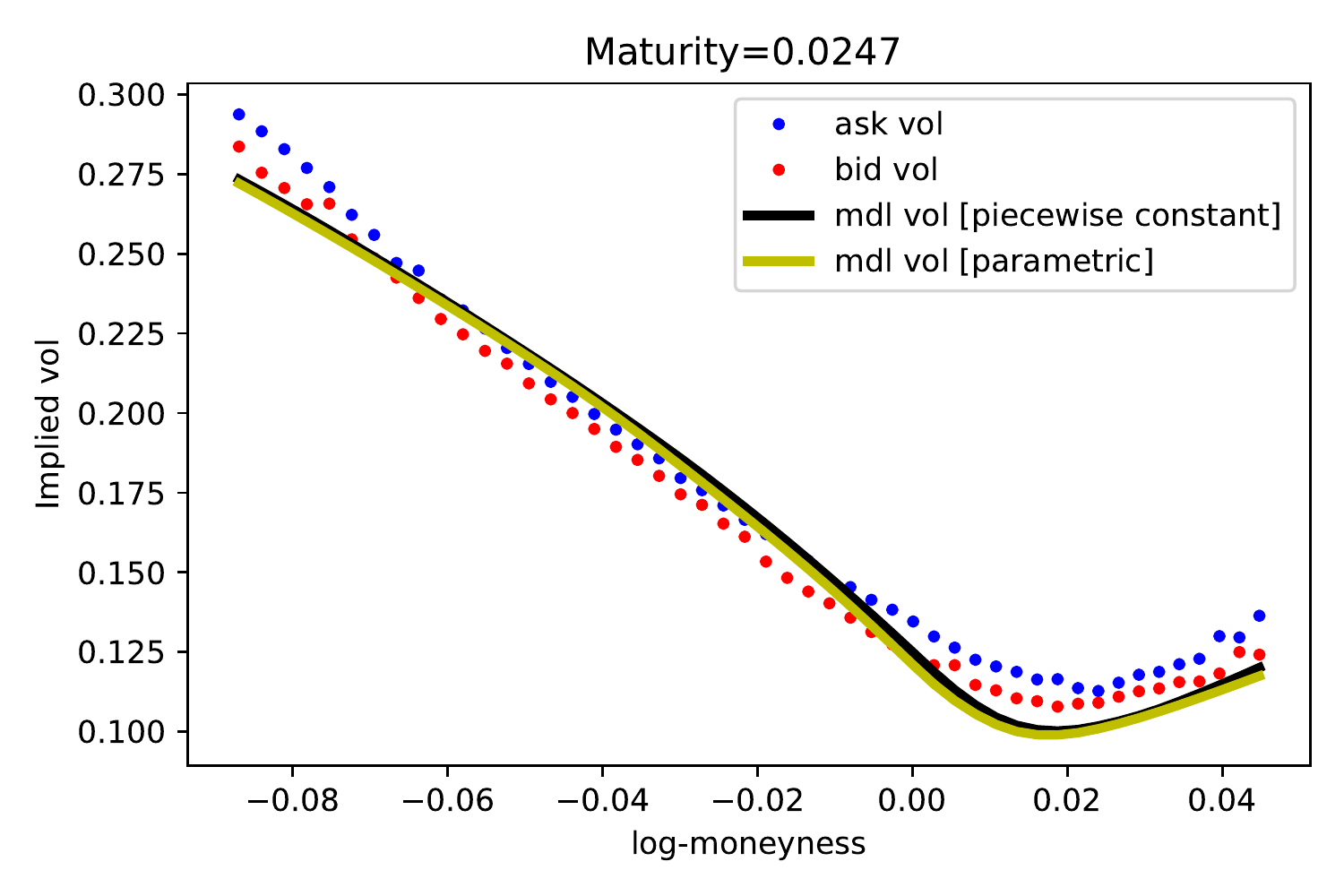}
     \end{subfigure}
     \begin{subfigure}[]{0.49\textwidth}
         \centering
         \includegraphics[width=\textwidth]{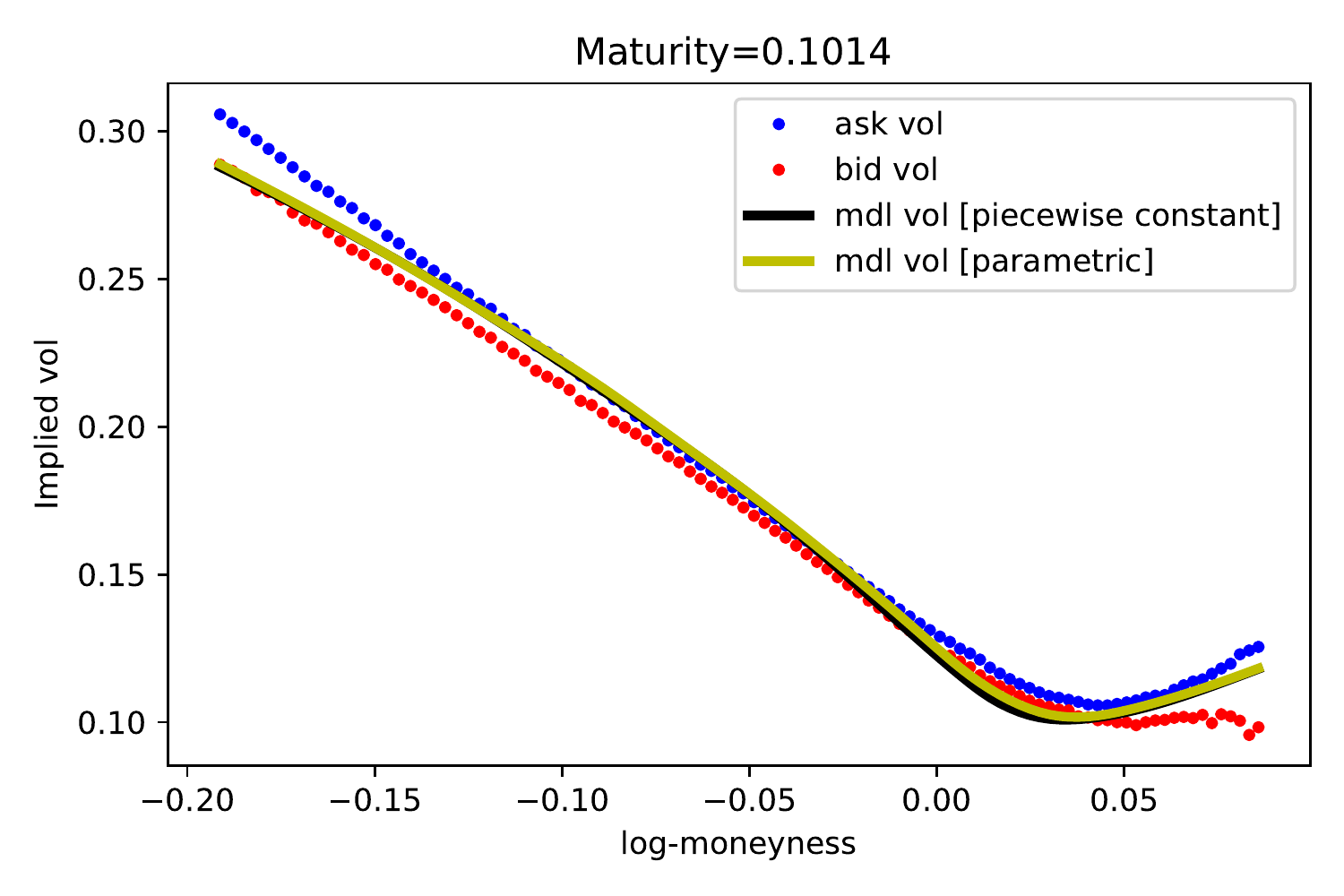}
     \end{subfigure}
     \hfill
     \begin{subfigure}[]{0.49\textwidth}
         \centering
         \includegraphics[width=\textwidth]{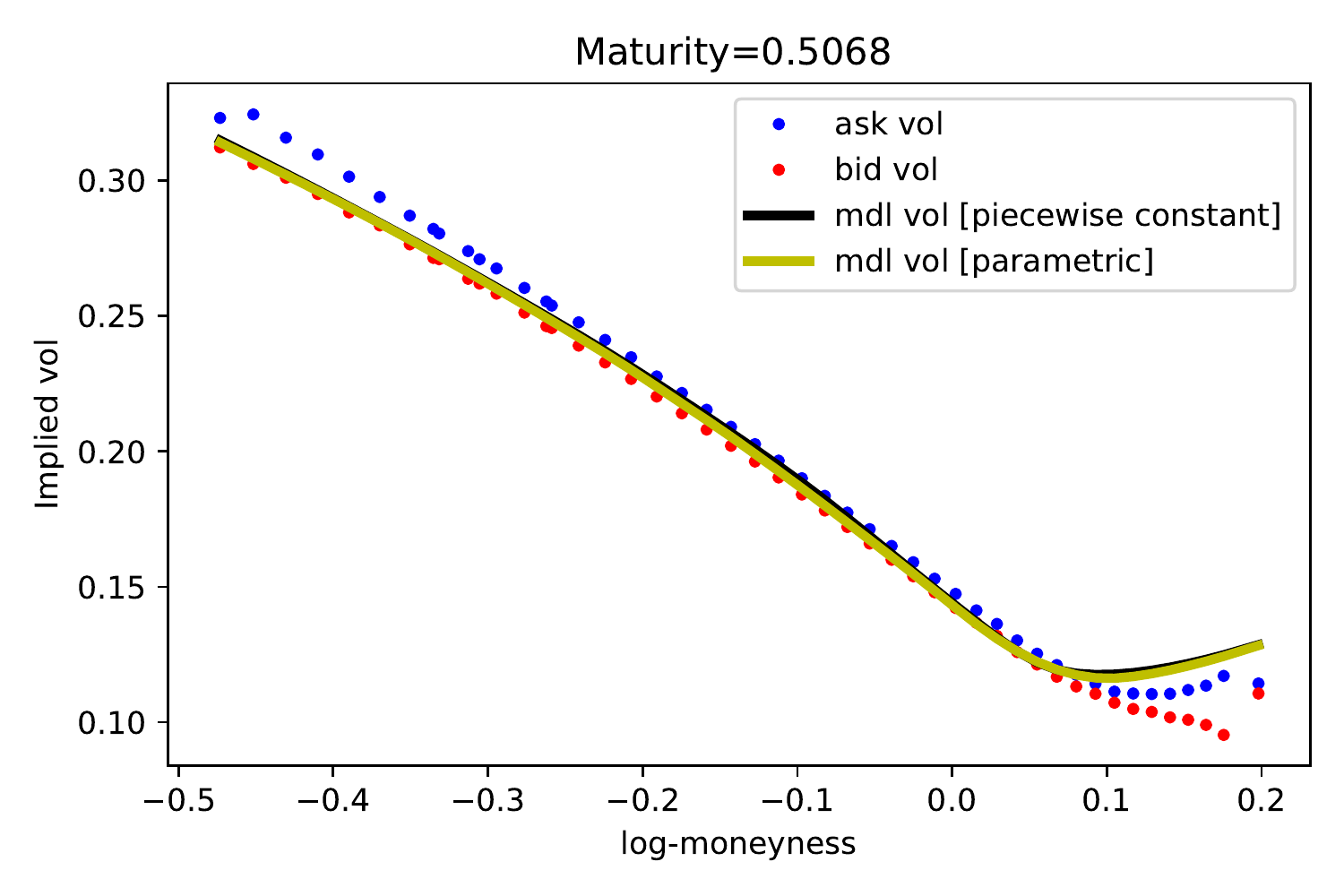}
     \end{subfigure}
     \caption{Comparing optimal rHeston fits (selected smiles) under a piecewise constant specification of the forward variance curve (black) vs the parametric model of Equation \eqref{fvc_par} (yellow) as of 2014/03/19.}
     \label{rHes_pwc_vs_par_20140319}
\end{figure}

\begin{figure}[h!]
     \centering
     \begin{subfigure}[]{0.49\textwidth}
         \centering
         \includegraphics[width=\textwidth]{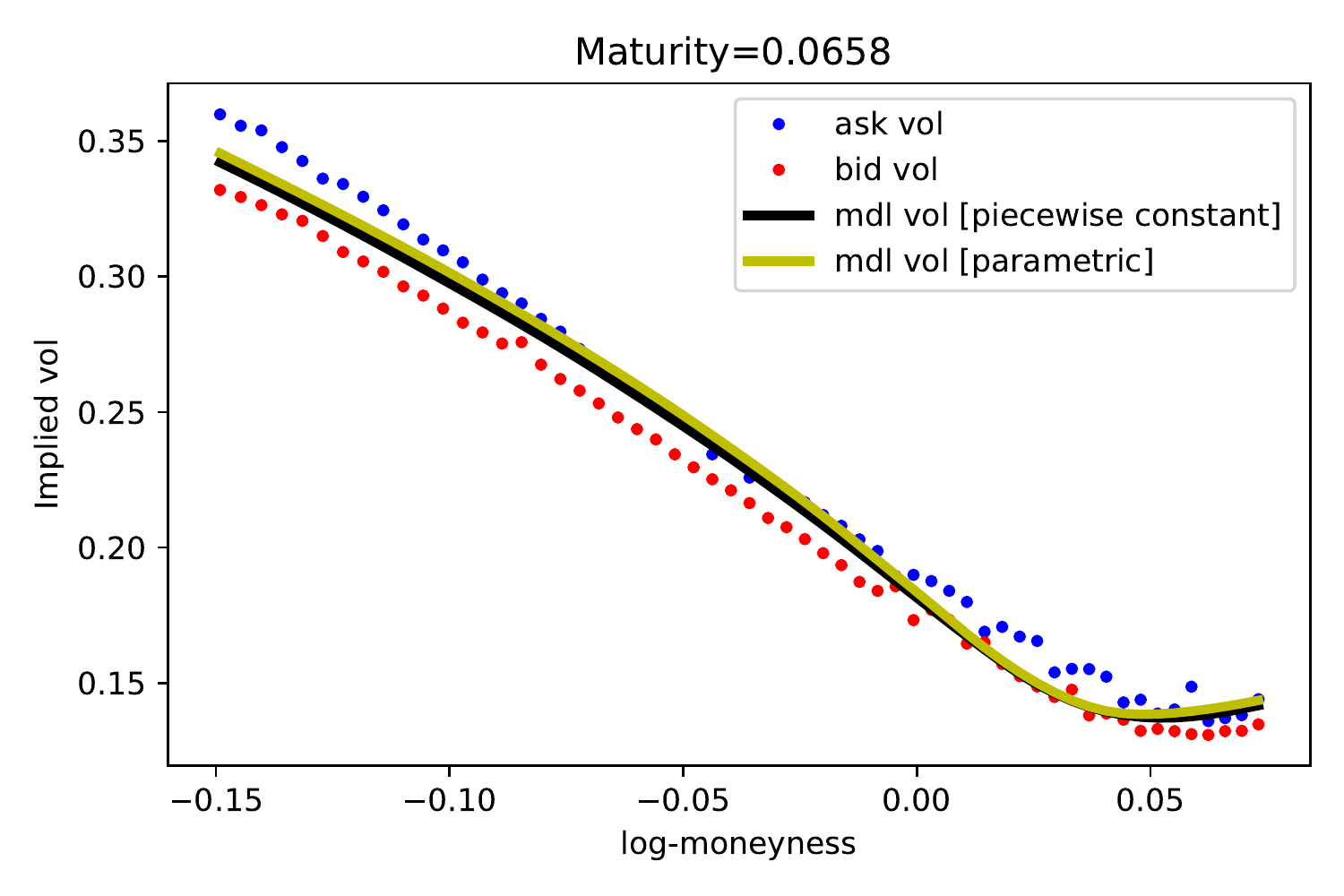}
     \end{subfigure}
     \hfill
     \begin{subfigure}[]{0.49\textwidth}
         \centering
         \includegraphics[width=\textwidth]{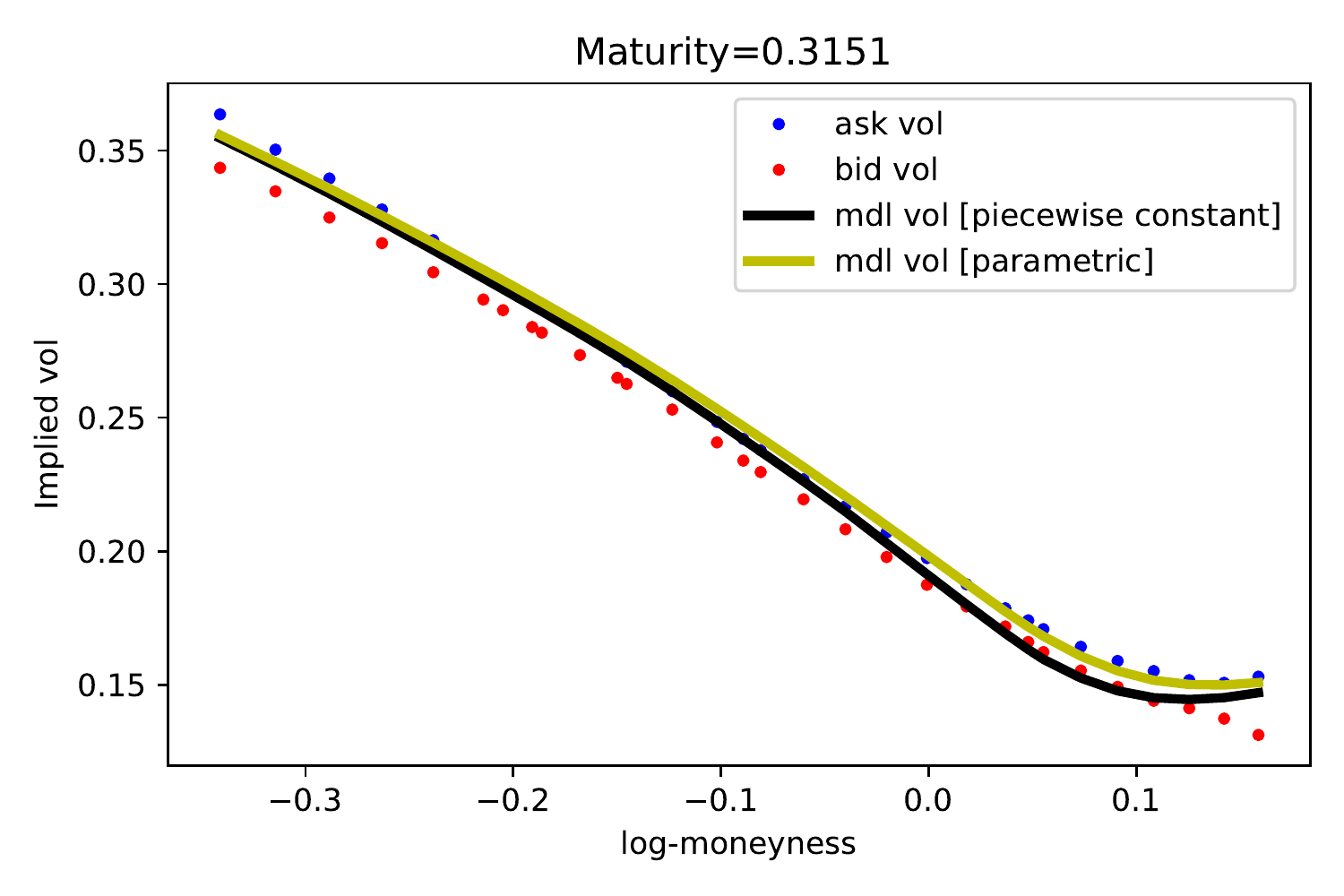}
     \end{subfigure}
     \begin{subfigure}[]{0.49\textwidth}
         \centering
         \includegraphics[width=\textwidth]{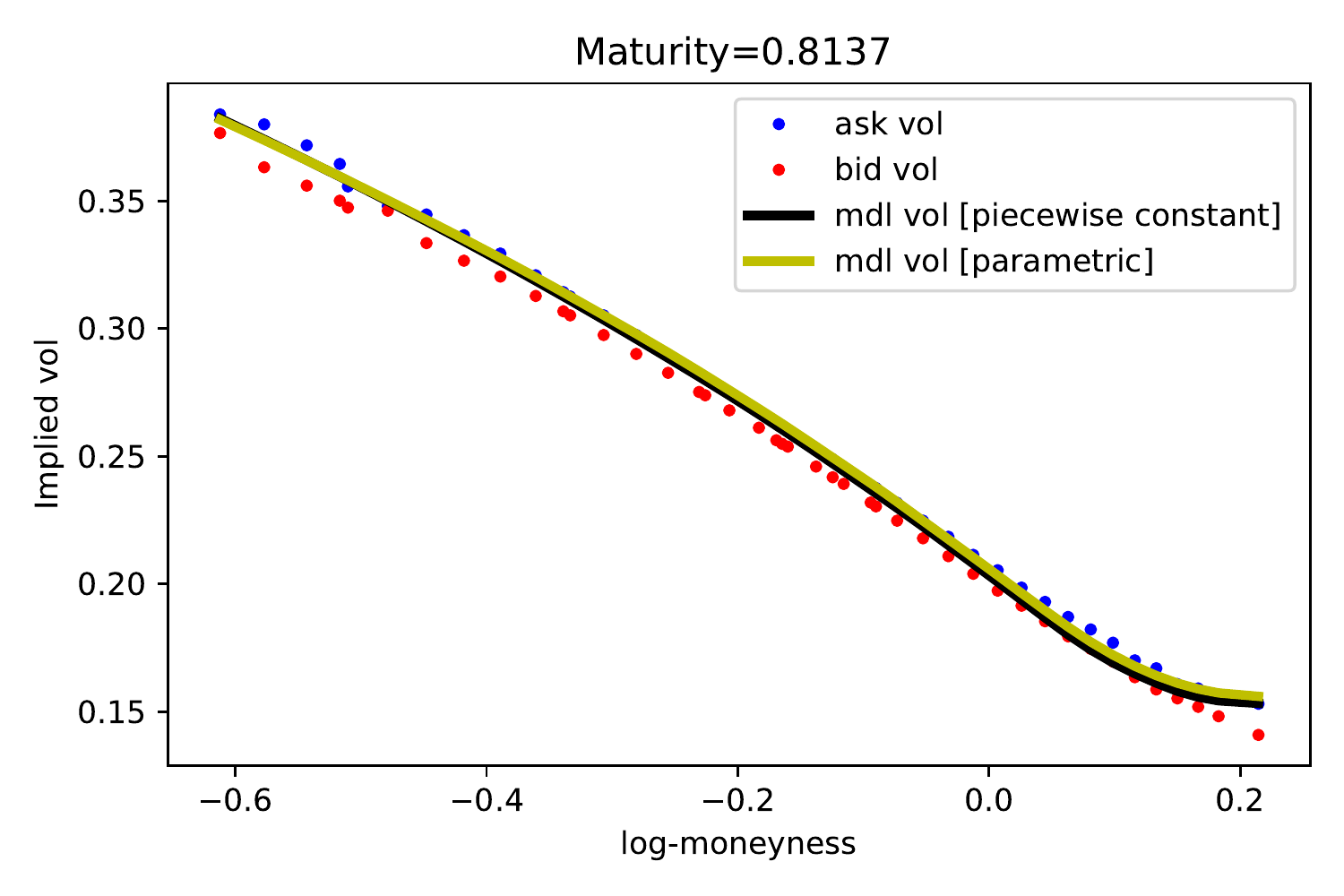}
     \end{subfigure}
     \hfill
     \begin{subfigure}[]{0.49\textwidth}
         \centering
         \includegraphics[width=\textwidth]{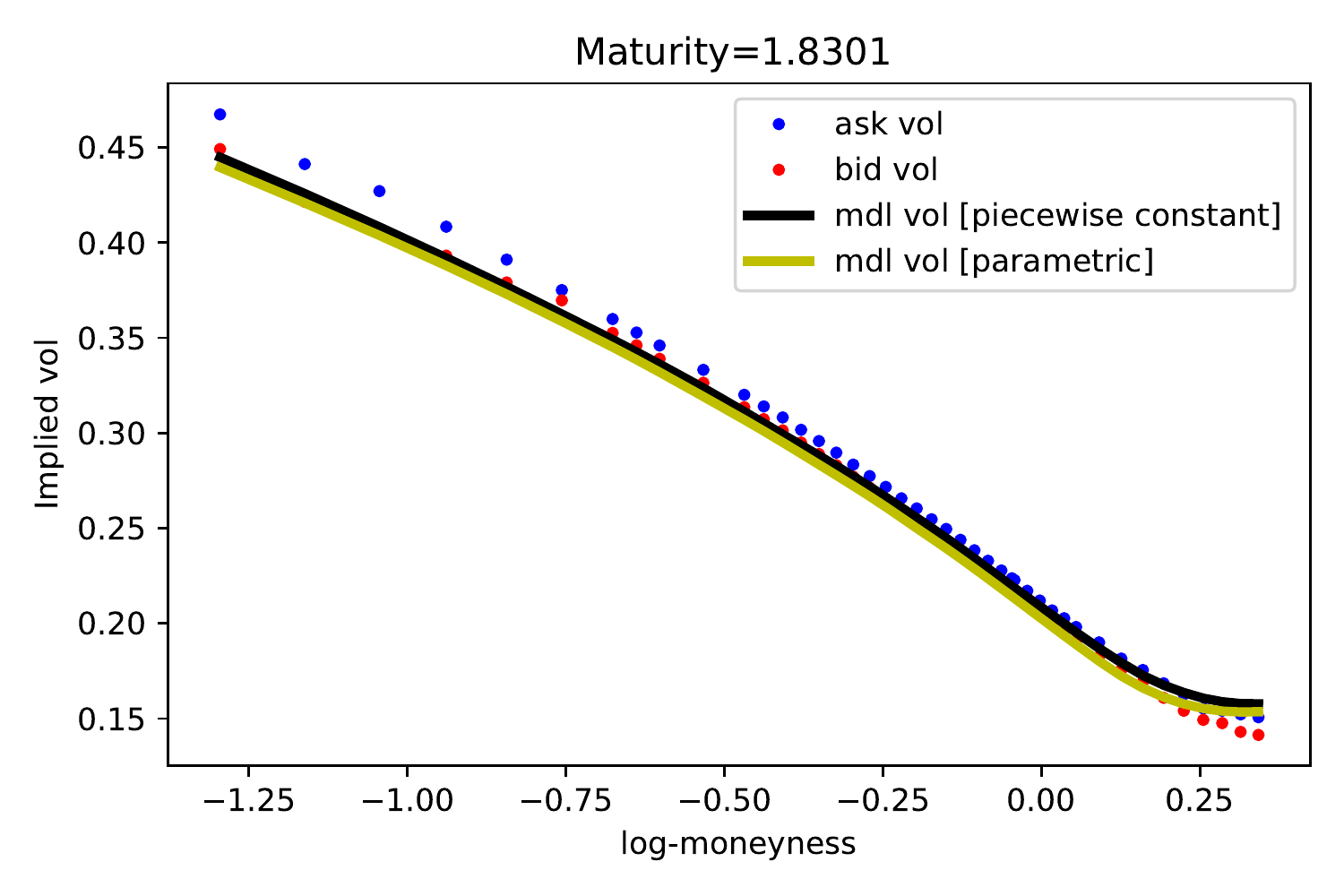}
     \end{subfigure}
     \caption{Comparing optimal rHeston fits (selected smiles) under a piecewise constant specification of the forward variance curve (black) vs the parametric model of Equation \eqref{fvc_par} (yellow) as of 2011/02/23.}
     \label{rHes_pwc_vs_par_20110223}
\end{figure}

\noindent We test the adequacy of this parametrization by checking the optimal fit with the one obtained with a piecewise constant forward variance curve (8 levels and the synchronized with grid times in our adaptive grid). Not only Figures \ref{rHes_pwc_vs_par_20140319} and \ref{rHes_pwc_vs_par_20110223} show that the fits from the two approaches are typically very similar but the model parameters always consistent, as one can see from Tables \ref{rHes_mdl_params_20140319} and \ref{rHes_mdl_params_20110223}. \\

\begin{table}[h!]
    \centering
    \begin{tabular}{|c|c|c|c|}
        \hline
        & $H$ & $\nu$ & $\rho$ \\
        \hline
        pwc & 0.0454 & 0.3054 & -0.6434 \\
        \hline
        par & 0.0496 & 0.3076 & -0.6514 \\
        \hline
    \end{tabular}
    \caption{Optimal rHeston parameters under a piecewise constant (pwc) and parametric (par) forward variance curve as of 2014/03/19.}
    \label{rHes_mdl_params_20140319}
\end{table}

\begin{table}[h!]
    \centering
    \begin{tabular}{|c|c|c|c|}
        \hline
        & $H$ & $\nu$ & $\rho$ \\
        \hline
        pwc & 0.0803 & 0.3777 & -0.7209 \\
        \hline
        par & 0.0780 & 0.3701 & -0.7233 \\
        \hline
    \end{tabular}
    \caption{Optimal rHeston parameters under a piecewise constant (pwc) and parametric (par) forward variance curve as of 2011/02/23.}
    \label{rHes_mdl_params_20110223}
\end{table} 

\begin{figure}[h!]
     \centering
     \begin{subfigure}[]{0.49\textwidth}
         \centering
         \includegraphics[width=\textwidth]{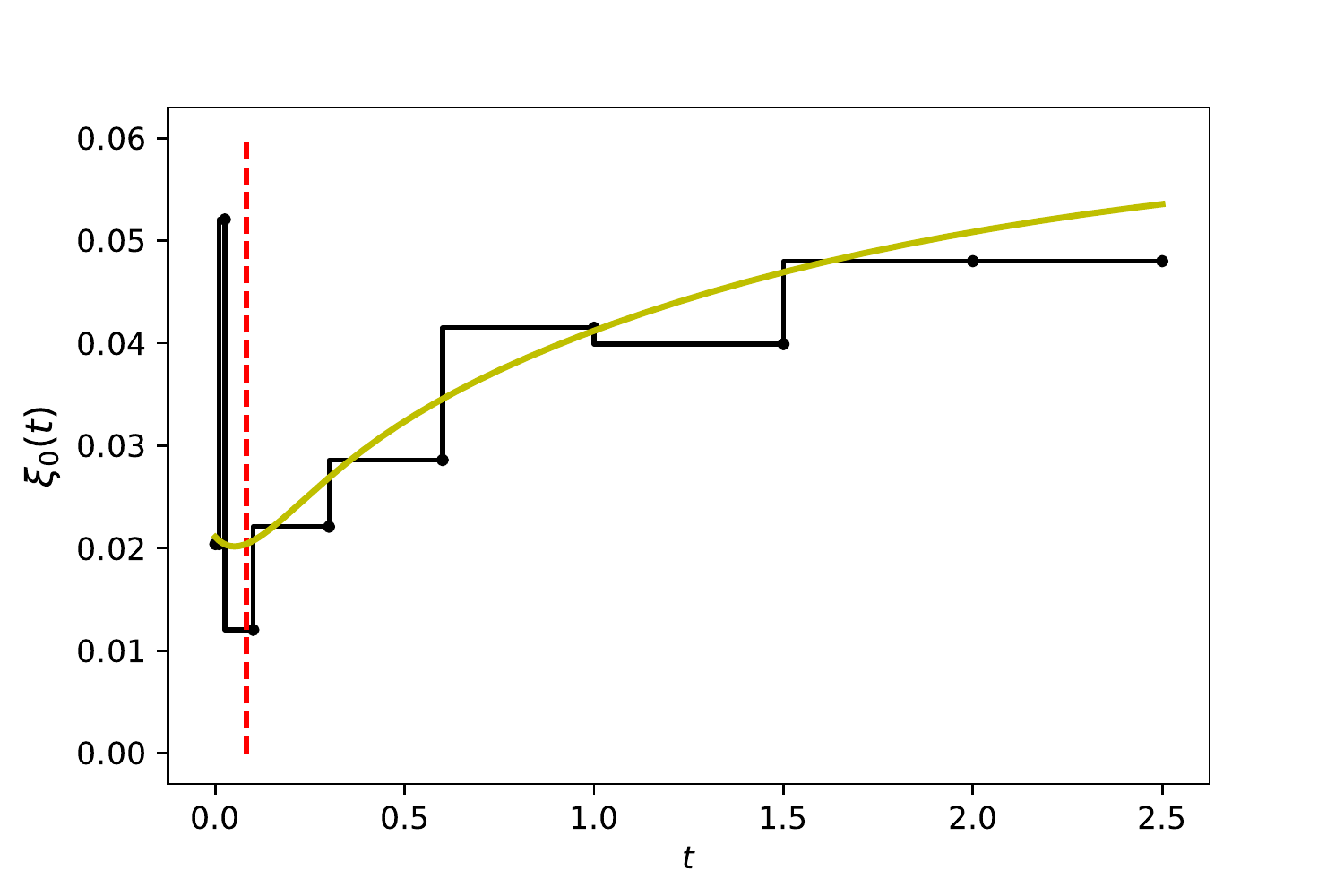}
         \caption{calibrated forward variance curve}
         \label{xi_pwc_20140717}
     \end{subfigure}
     \hfill
     \begin{subfigure}[]{0.49\textwidth}
         \centering
         \includegraphics[width=\textwidth]{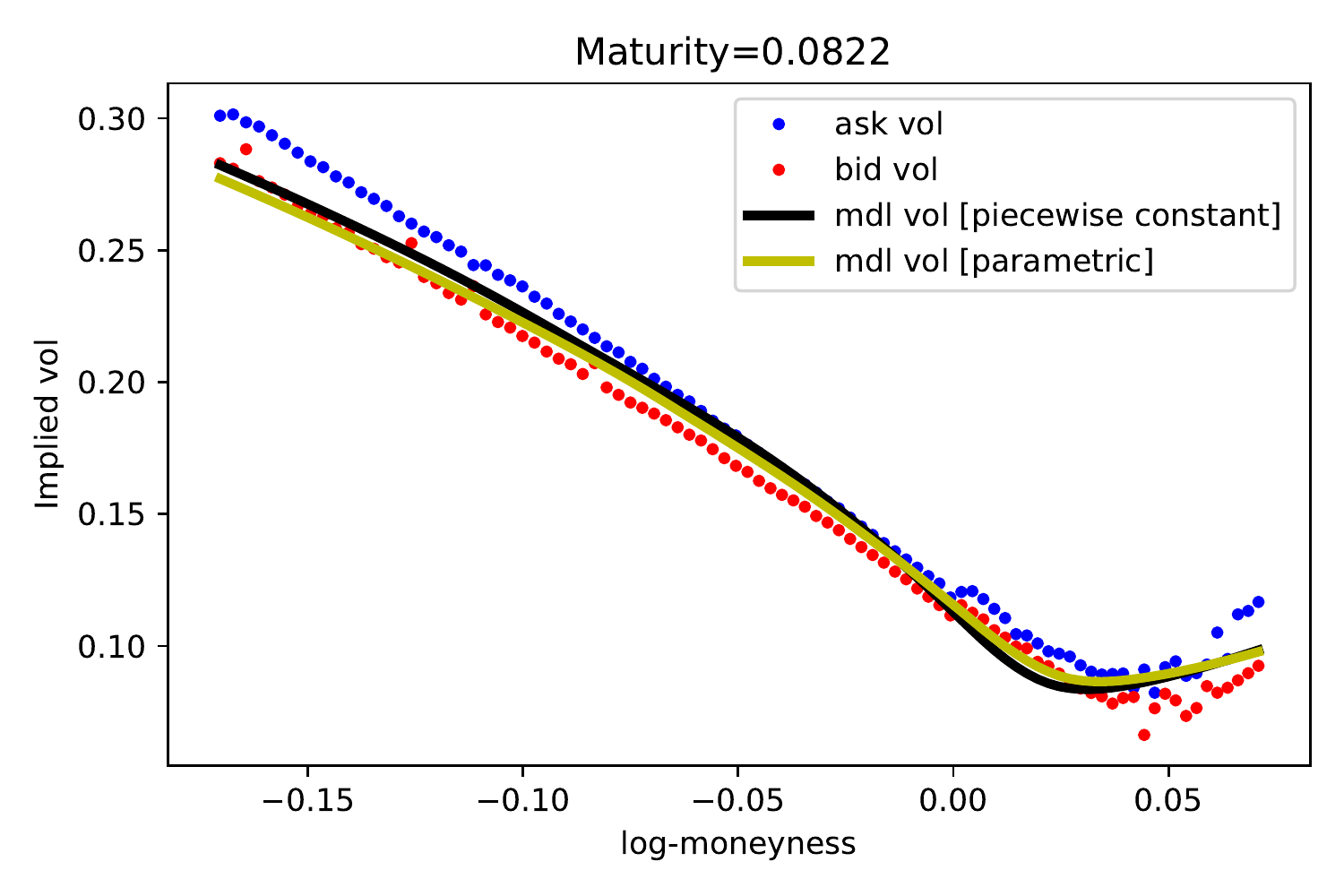}
         \caption{smile at maturity $T=0.0822$}
         \label{t001_20140717}
     \end{subfigure}
     \caption{Calibrated piecewise constant (black) and parametric (yellow) forward variance curve as of 2014/07/17 and associated fit to the market at maturity $T=0.0822$ under the rHeston model. The red dashed line indicates the position of the selected maturity on the forward variance curve.}
     \label{curve_20140717}
\end{figure}

\noindent If this seems to be convincing about the potentiality of Equation \eqref{fvc_par} as a description of the forward variance curve - in the rHeston model, for now - we should maybe spend a few words commenting the practical need for a parametric form to slavishly follow the calibrated piecewise constant curve. The reader may indeed have encountered forward variance curves that exhibit much more of a jumpy behavior at short times compared to what we have plotted in Figure \ref{rHes_xi_pwc} and may be worried about the impossibility for our modeling framework to accommodate for this. While reducing the number of levels introduces some regularization where market maturities are denser, we still sometimes observe term structures which are particularly rich during the first month or so. Figure \ref{curve_20140717} is a clear example for this, but it also demonstrates that all those ups and downs in the forward variance curve are not essential for proper fit to the market. We place ourselves right after the double pick in the piecewise constant curve at time $T=0.0822$ and observe that the generated smile is incredibly similar to the one we get with a parametric forward variance curve which is much more flat over the relevant period $(0,T]$.

\noindent Because a `flat' initial term structure could have been achieved by a piecewise constant forward variance curve as well, the insensibility of the smile to the shape of the curve is to be read as a sign of possible local minima in the objective function. We indeed recall that the log of the characteristic function in the rHeston model is a convolution between a fractional Riccati equation and the forward variance curve (up to the maturity of the option), which is therefore acting as a set of weights. Not surprisingly, then, different weights may combine to produce similar option prices whenever more than one forward variance level concur to the price itself. The other face of this phenomenon with our parametric form in Equation \eqref{fvc_par} is that different parameters may produce similar curves. Anyway, consistently with market standards and the view of the forward variance curve as a state variable, we actually only care about the shape of the curve and do not worry about the parameters that have produced it. \\

\noindent Also because the log of the characteristic function is a convolution, the value of the forward variance curve at large times gets less and less important relative to the whole past of the curve. It may consequently happen - in the insufficiency of long maturities from the market - that the calibrated levels at large times seem to be inconsistent with the previous history of the curve (because of a sudden drop/increase, for example). Our parametric form will not be fooled by this inconvenience and still produces very good fits. \\

\noindent There is therefore no need for us to conceive more complicated parametric forms in the attempt to follow the entire zoo of possible piecewise constant forward variance curves. \\

\noindent The reason why we started with the rHeston model is that the forward variance curve seems to play much more an important role here than it does in rBergomi. In fact, the calibrated parametric forward variance curve is typically very different from its piecewise constant counterpart but such a difference is not visible in terms of the fits to the market implied volatilities. We omit any visual representation of this fact - in the interest of space - but
recognize that it will certainly result in increased freedom for the modeler to come up with a valuable shape of the curve. \\ 

\noindent On account of this, we propose the same parametrization in Equation \eqref{fvc_par} for the rBergomi model as well. \\

\noindent Notably, however, we will be able to provide evidence for the validity of our functional form in a way which is completely model-independent in a few pages.

\section{Our recipe at work}

The need for an interpolation/extrapolation-free neural network-based pricer -- together with the evidence from the previous sections -- suggests using a random-grid pointwise approach. Making the forward variance curve a parametric function may also be useful when dealing with rough volatility models. We will always have this construction in mind when talking about a pointwise approach in this section. \\

\noindent We therefore devise a last round of tests for the specific framework described above. The idea is that we place ourselves under the rHeston model and check consistency of our solution with the same procedure as described in \cite{el2019roughening}. We derive a piecewise constant forward variance curve from quoted implied volatilities, pass it to an FFT pricer and optimize over the model parameters alone. The FFT method is taken as a benchmark, and the performance of our NN pointwise approach tested against it. We also superimpose optimal rBergomi fits for visual comparison, but a systematic description of the models' ability to calibrate to the volatility surface under different market scenarios is beyond the scope of this paper. \\

\noindent We restrict ourselves to the following set of model parameters in Table \ref{mdlpar_range} for the experiments that follow. \\

\begin{table}[h!]
    \centering
    \begin{tabular}{|c|c|c|c|}
        \hline
        rHeston parameters & & rBergomi parameters & \\
        \hline
        $H$ &  $\mathcal{U}([0.01,0.25])$ &
        $H$ &  $\mathcal{U}([0.025,0.50])$ \\
        $\nu$ & $\mathcal{U}([0.15,0.65])$ &
        $\eta$ & $\mathcal{U}([0.50,4.00])$\\
        $\rho$ & $\mathcal{U}([-.95,-.50])$ & 
        $\rho$ & $\mathcal{U}([-.95,-.10])$\\
        \hline
    \end{tabular}
    \caption{Domain of definition of the models parameters (rHeston and rBergomi) as seen from the neural network.}
    \label{mdlpar_range}
\end{table}

\noindent If the quality of the calibration is obviously important, the time associated with it is also a variable that we need to keep an eye on. For this, we believe comparison with the large adapted grid of \cite{romer2022empirical} and a version of the pointwise method where the forward variance curve is piecewise constant are relevant for the discussion. \\

\begin{figure}[h!]
     \centering
     \begin{subfigure}[]{0.49\textwidth}
         \centering
         \includegraphics[width=\textwidth]{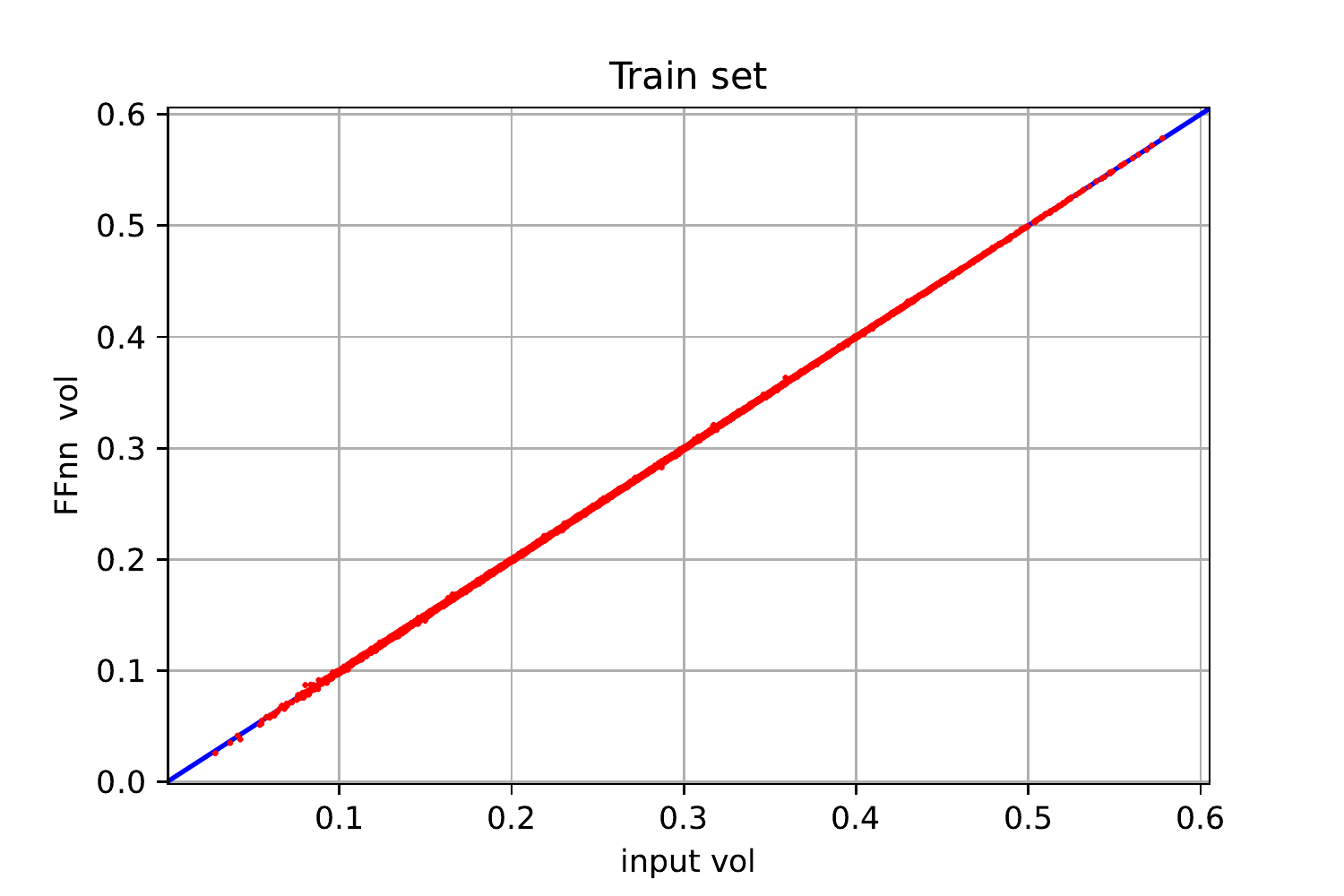}
         \end{subfigure}
     \hfill
     \begin{subfigure}[]{0.49\textwidth}
         \centering
         \includegraphics[width=\textwidth]{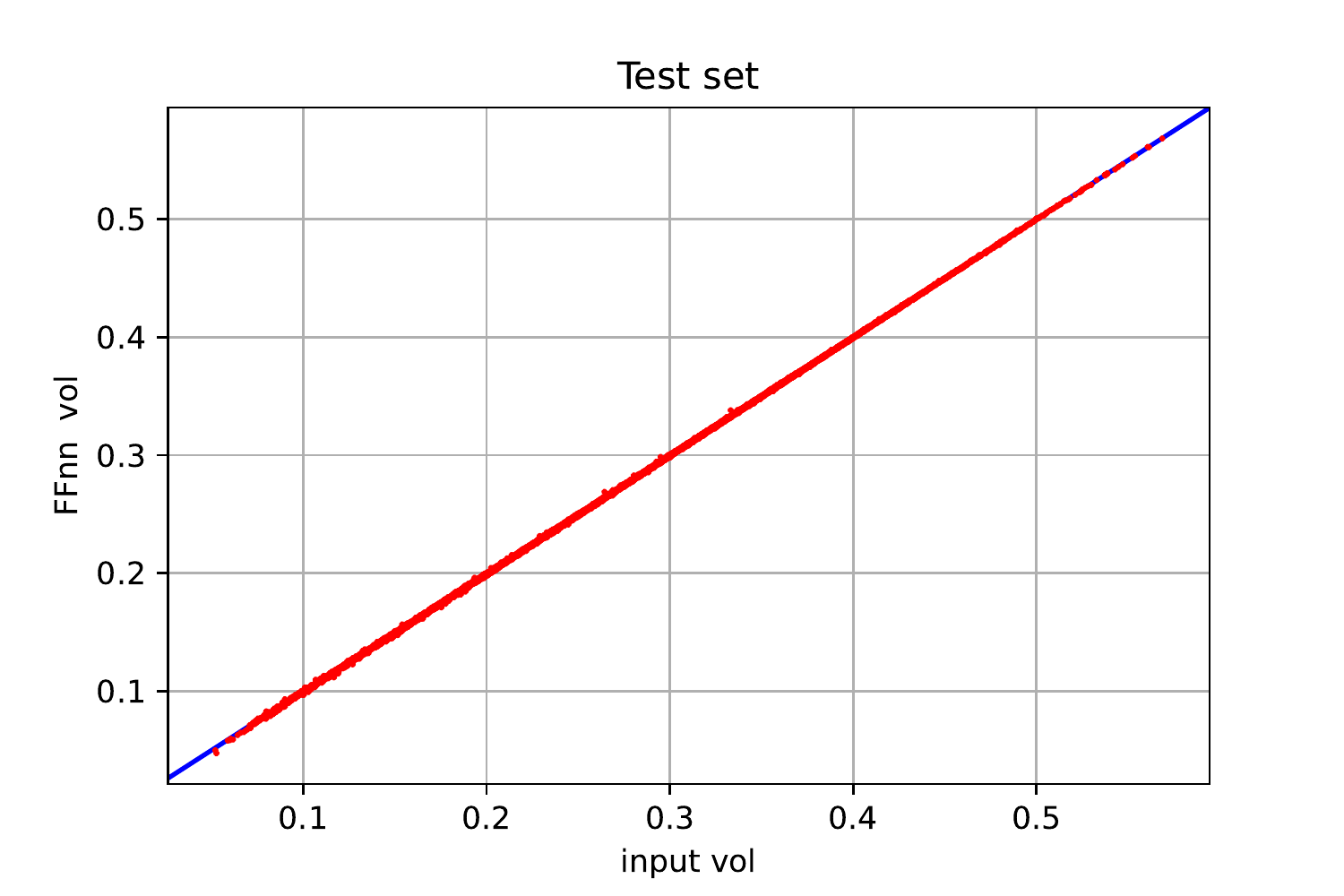}
         \end{subfigure}
     \caption{(model,network) point pairs vs 45-degree line for the rHeston model with parametric forward variance curve. In-sample performance on the left, out-of-sample on the right.}
     \label{rHes_45}
\end{figure}

\begin{figure}[h!]
     \centering
     \begin{subfigure}[]{0.49\textwidth}
         \centering
         \includegraphics[width=\textwidth]{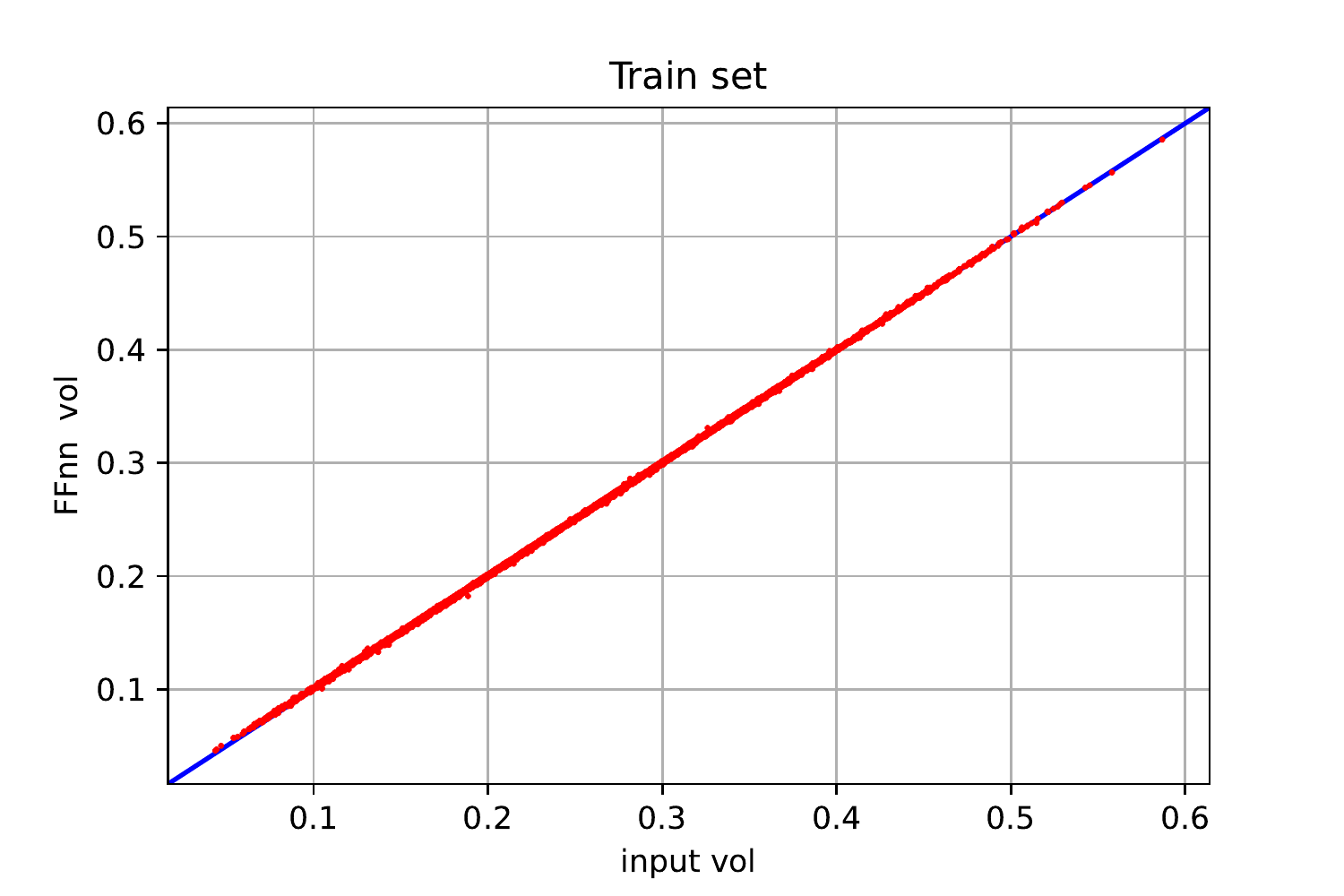}
         \end{subfigure}
     \hfill
     \begin{subfigure}[]{0.49\textwidth}
         \centering
         \includegraphics[width=\textwidth]{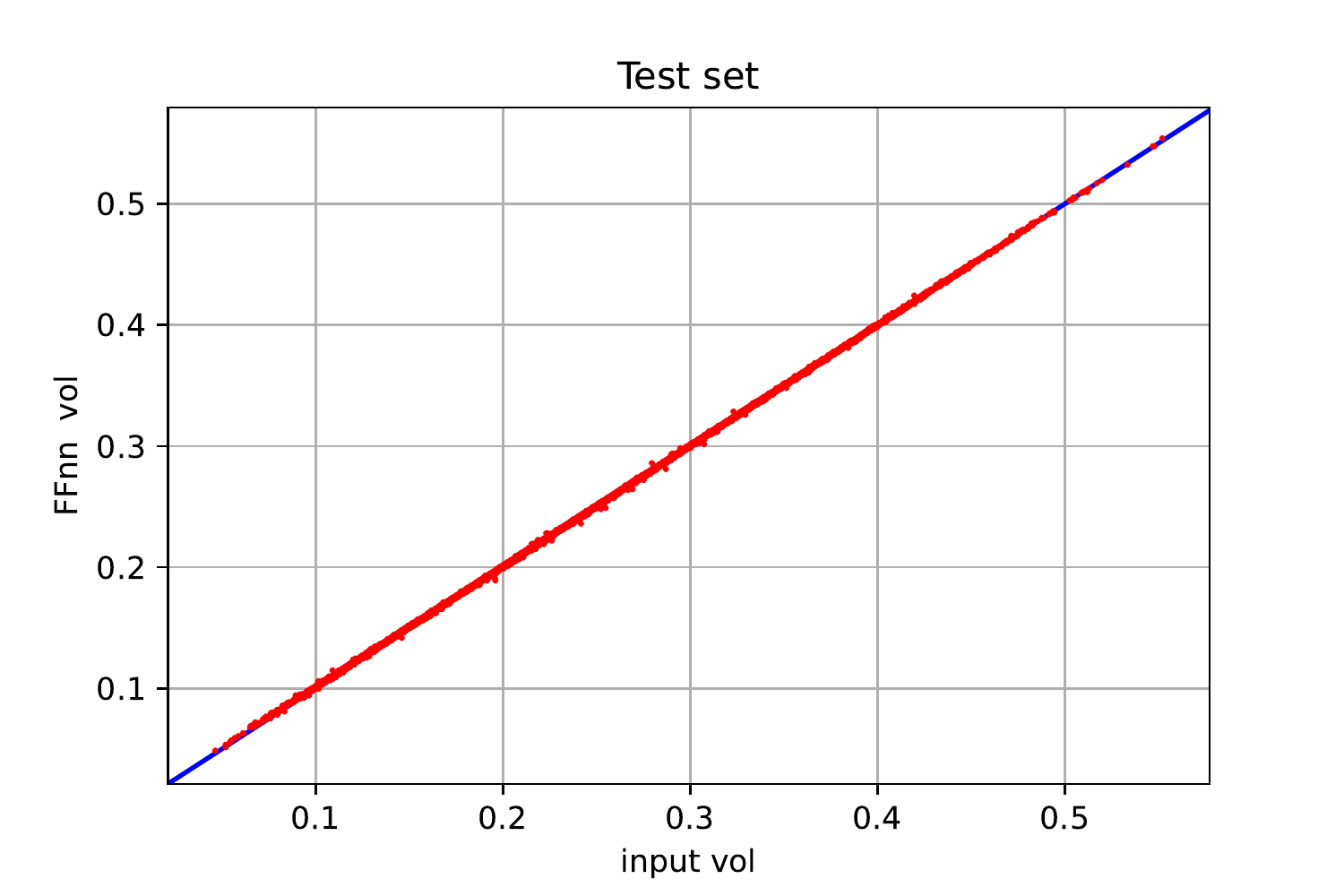}
        \end{subfigure}
     \caption{(model,network) point pairs vs 45-degree line for the rBergomi model with parametric forward variance curve. In-sample performance on the left, out-of-sample on the right.}
     \label{rBer_45}
\end{figure}

\noindent Before we start, however, it is maybe the case to understand how good the network is in replicating the underlying pricing function that we are passing it and generalizing beyond the training set. Figures \ref{rHes_45} and \ref{rBer_45} show that the network does its job for the models we are considering, with (model,network) point pairs almost perfectly lying on the 45-degree line both in-sample and out-of-sample.   

\noindent Because the pricing functions associated with the rHeston and rBergomi are free of arbitrage by construction and we have reached such an accurate approximation of them, we are confident that the well-trained NN is not prone to any arbitrage opportunity. We report in Appendix \ref{Absence of arbitrage} an additional robustness check supporting our conclusion and the reliability of the entire procedure.

\subsection{Calibration of rough volatility models}

Once we have checked that the neural network learnt the true pricing function and using a parametric forward variance curve is consistent with standard piecewise constant specifications, the last step to the validation of our pricing tool is to confront it with a completely different approach which comes after \cite{el2019roughening}. As we already mentioned, the authors suggest that the forward variance curve is estimated from option prices and used as a state variable in a subsequent calibration procedure which only involves model parameters. No network can be advocated with this approach and availability of a closed-form solution for the characteristic function is therefore essential to prevent calibration times to explode. If such an alternative is available, however, neural network approaches should clearly confront with it, which fact is often times forgotten in the literature. \\

\begin{figure}[h!]
     \centering
     \begin{subfigure}[]{0.49\textwidth}
         \centering
         \includegraphics[width=\textwidth]{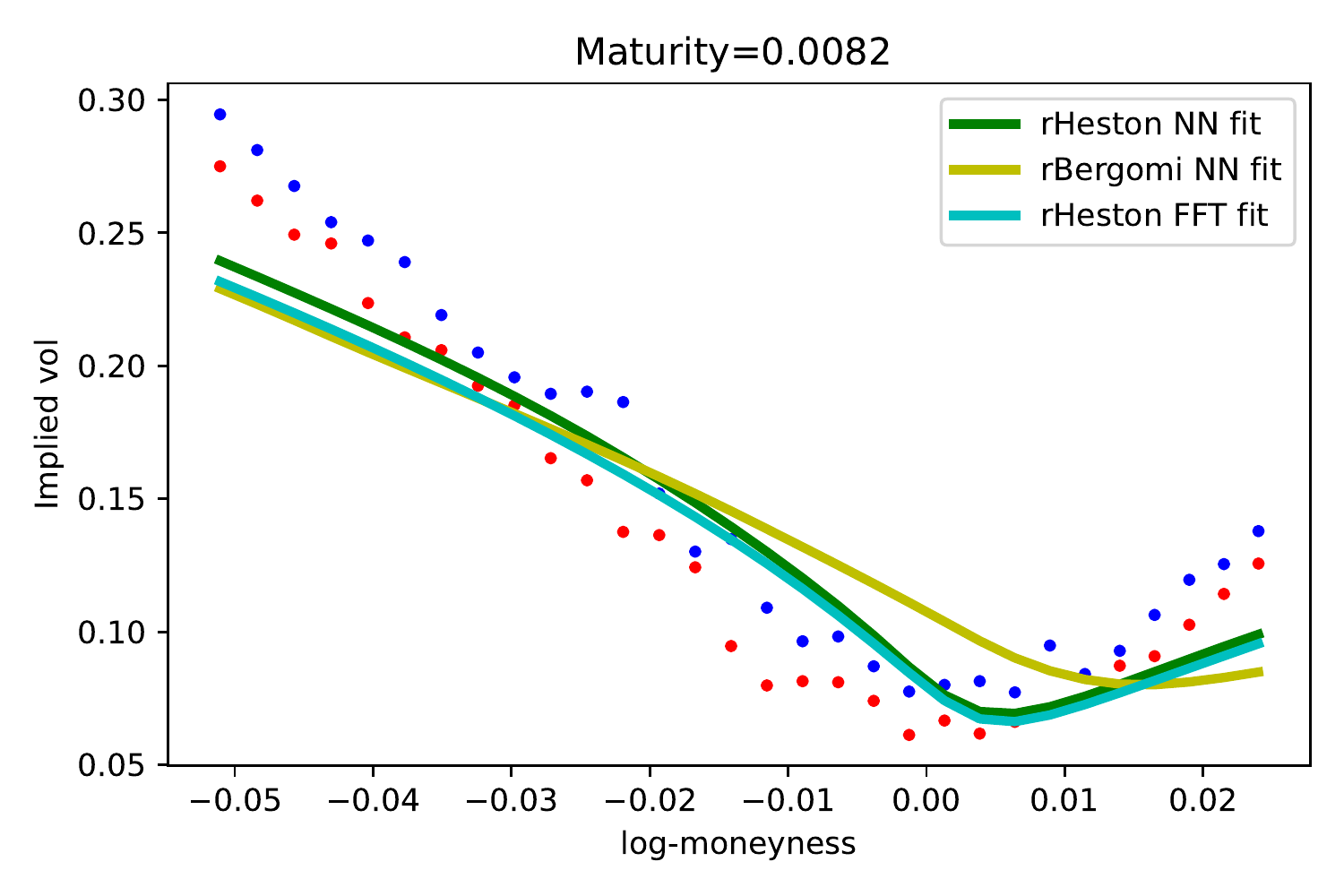}
     \end{subfigure}
     \hfill
     \begin{subfigure}[]{0.49\textwidth}
         \centering
         \includegraphics[width=\textwidth]{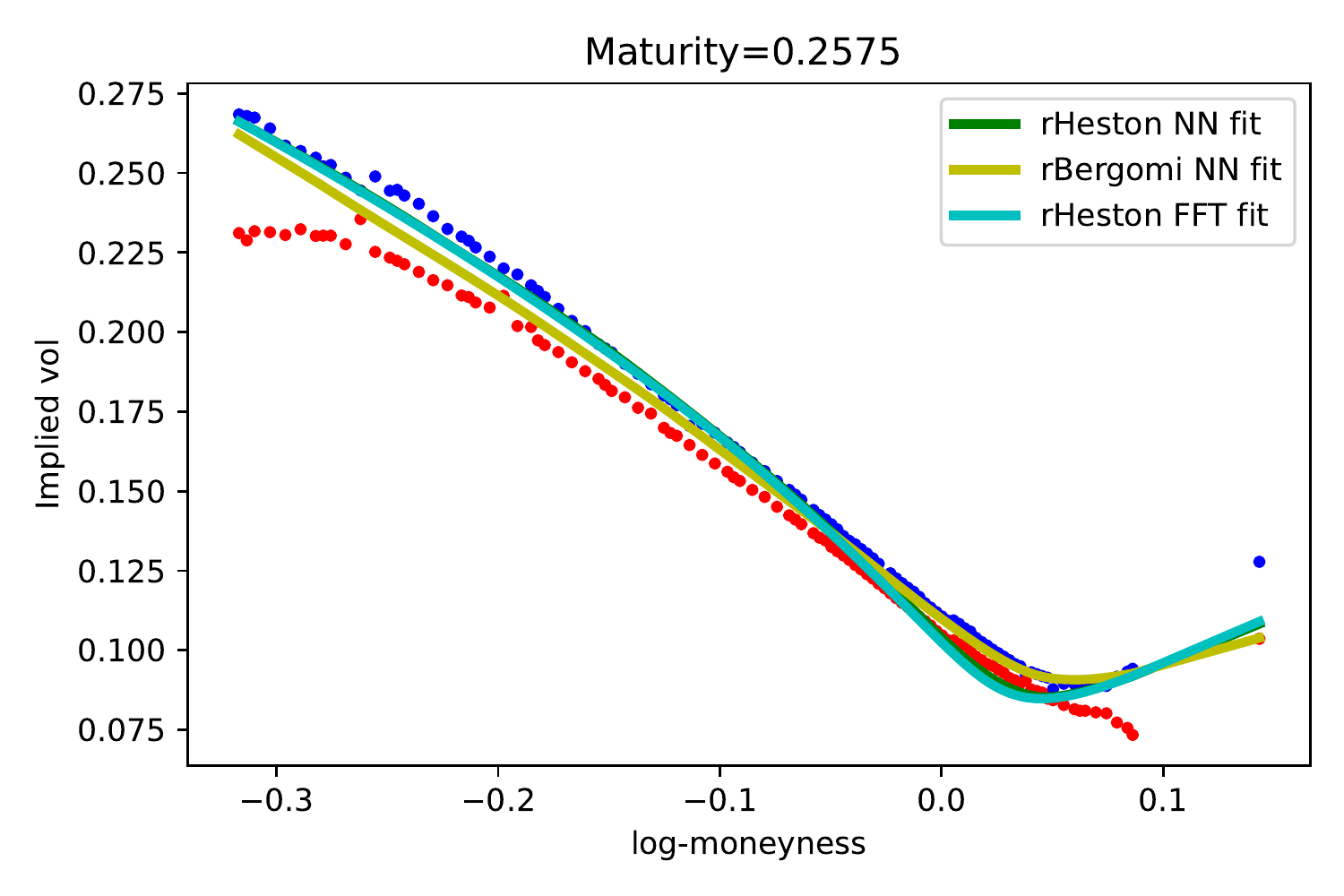}
     \end{subfigure}
     \begin{subfigure}[]{0.49\textwidth}
         \centering
         \includegraphics[width=\textwidth]{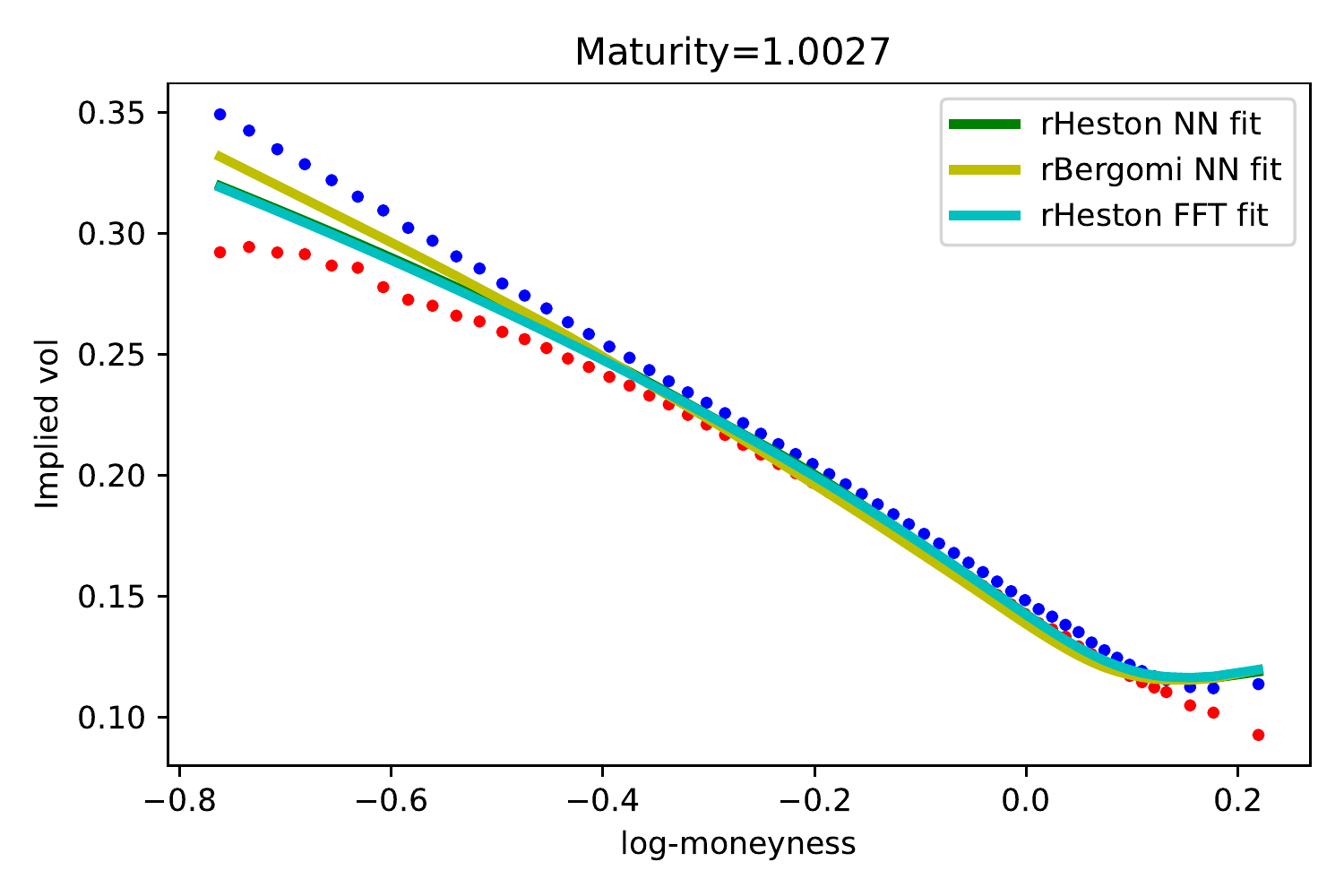}
     \end{subfigure}
     \hfill
     \begin{subfigure}[]{0.49\textwidth}
         \centering
         \includegraphics[width=\textwidth]{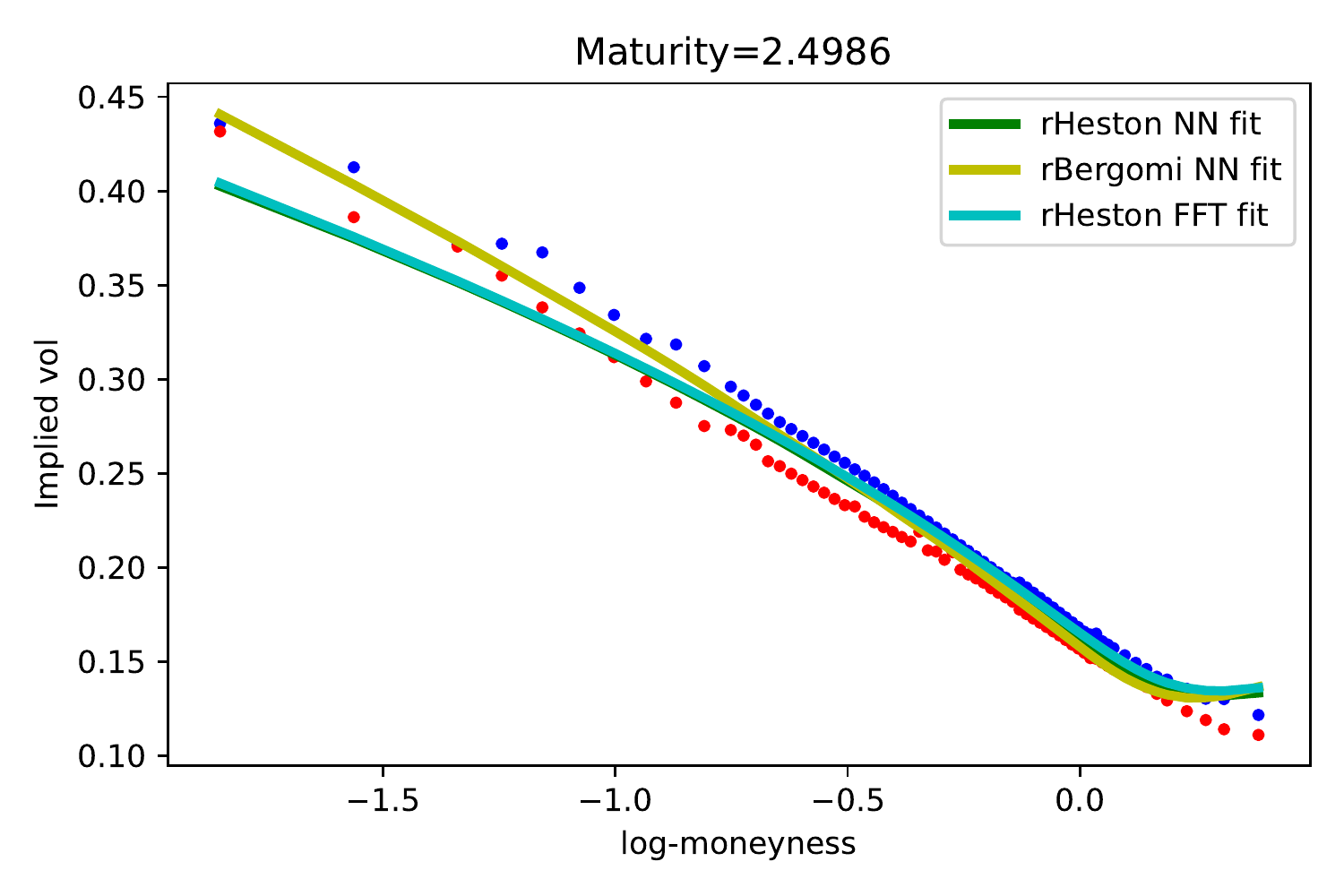}
     \end{subfigure}
     \caption{rBergomi (yellow) and rHeston fits as of June 18, 2014. Calibration via neural networks prescribes a parametric forward variance curve as in Equation \eqref{fvc_par} to be calibrated together with model parameters. FFT pricing pre-computes a piecewise constant forward variance curve and makes it a state variable. Those correspond to the green curve and the cyan curve respectively, for the rHeston model.} 
     \label{COvsSV_20140618}
\end{figure}

\noindent Preliminary construction of the forward variance curve requires evaluation of variance swaps via integration of option prices over a continuum of strikes\footnote{This is made possible by the tools from Vola Dynamics.}. Such an integration is usually performed on the market maturities only, and a piecewise constant forward variance curve obtained by differentiation. Fixing the resulting curve allows for separate calibration of the model parameters via the FFT. We do that in Figure \ref{COvsSV_20140618} and prove that our pointwise approach yields very similar results. \\

\begin{figure}[h!]
     \centering
     \begin{subfigure}[]{0.49\textwidth}
         \centering
         \includegraphics[width=\textwidth]{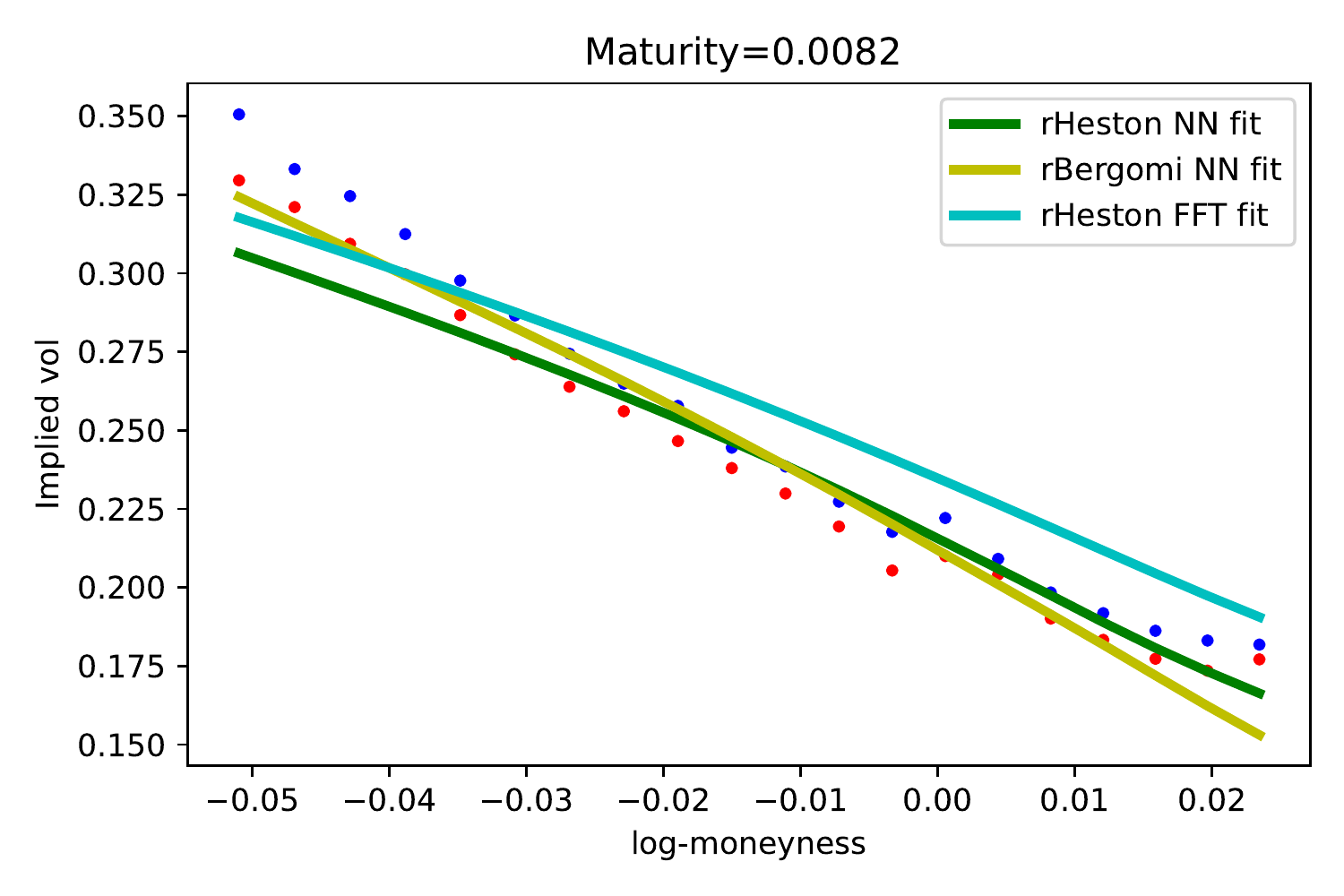}
     \end{subfigure}
     \hfill
     \begin{subfigure}[]{0.49\textwidth}
         \centering
         \includegraphics[width=\textwidth]{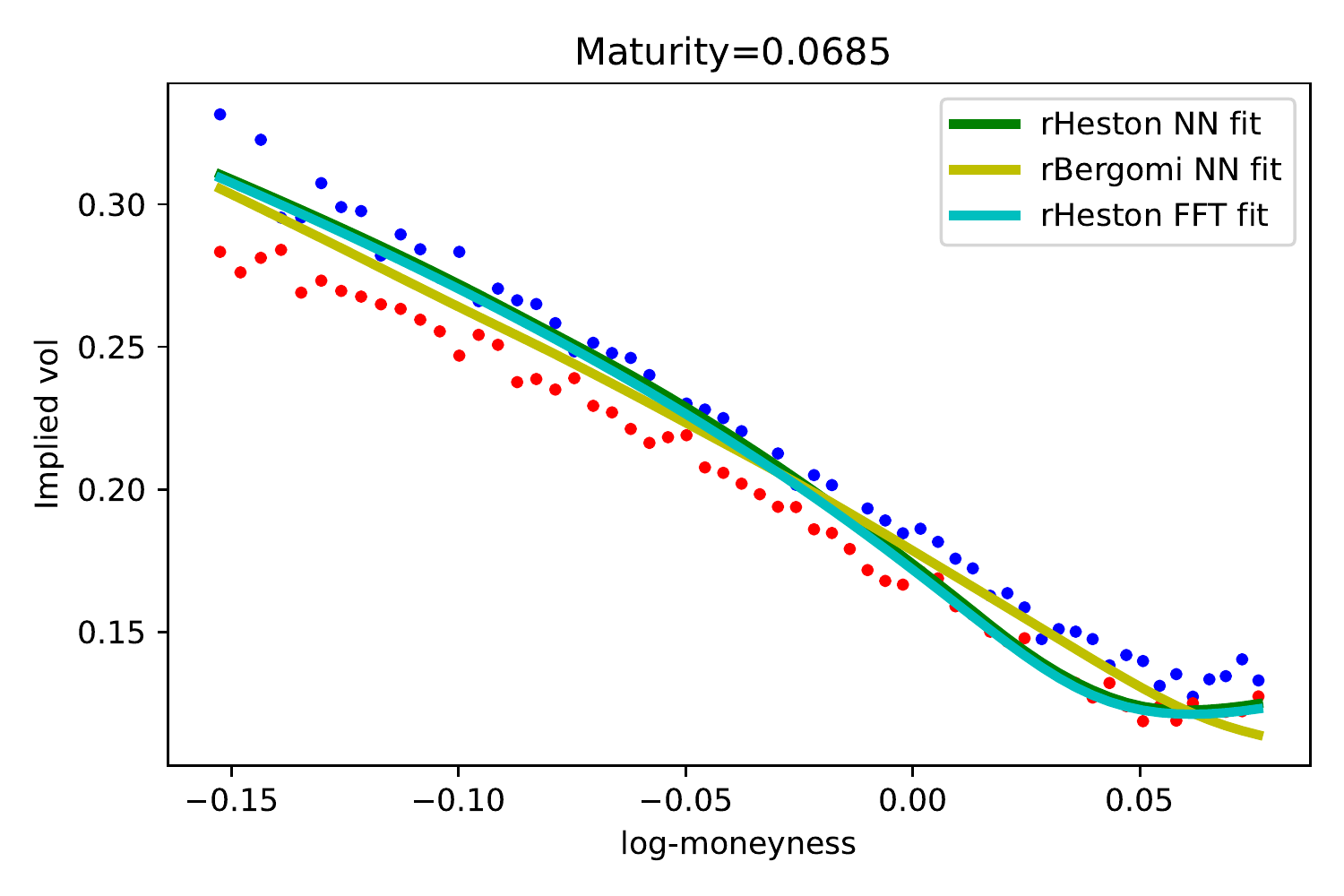}
     \end{subfigure}
     \begin{subfigure}[]{0.49\textwidth}
         \centering
         \includegraphics[width=\textwidth]{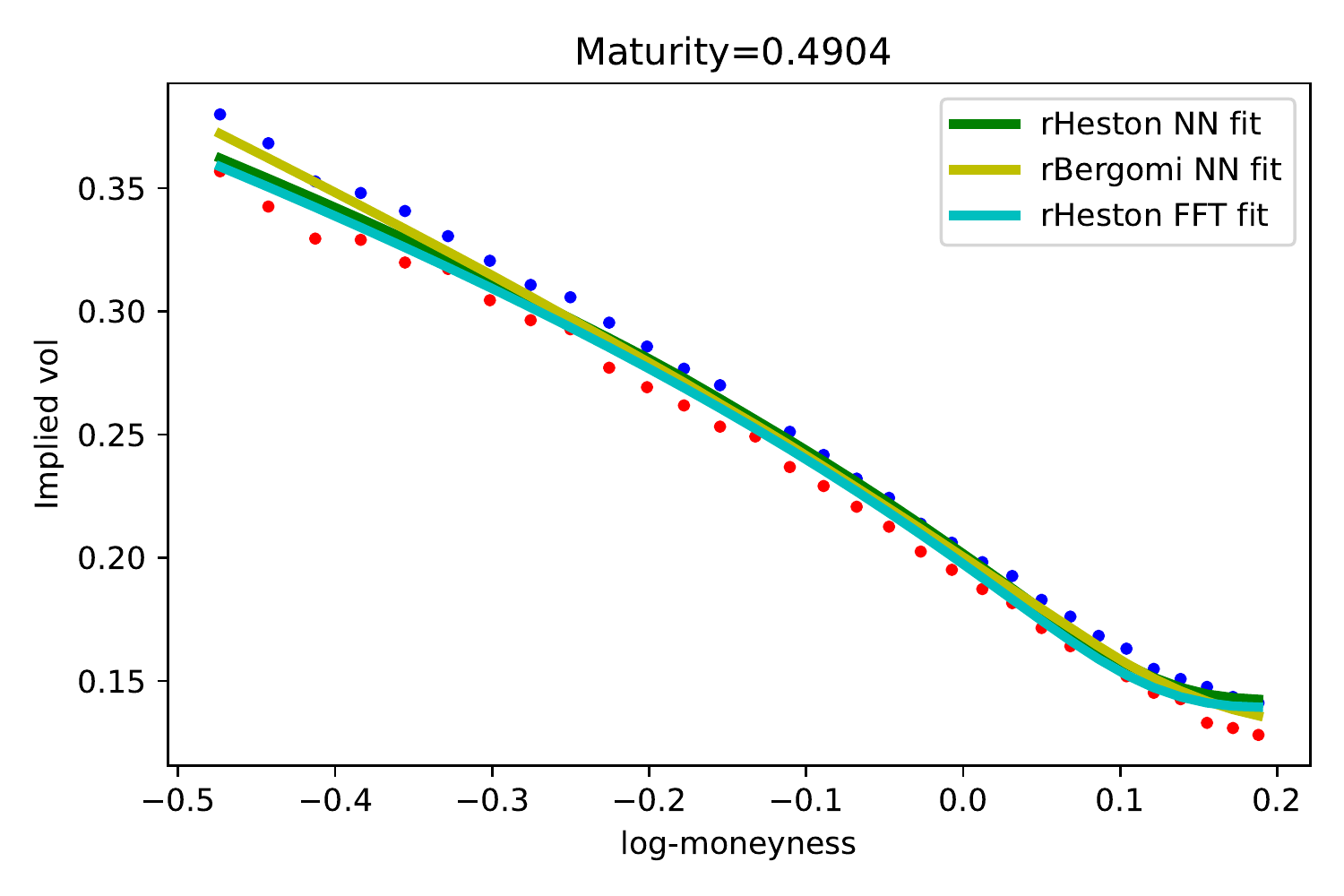}
     \end{subfigure}
     \hfill
     \begin{subfigure}[]{0.49\textwidth}
         \centering
         \includegraphics[width=\textwidth]{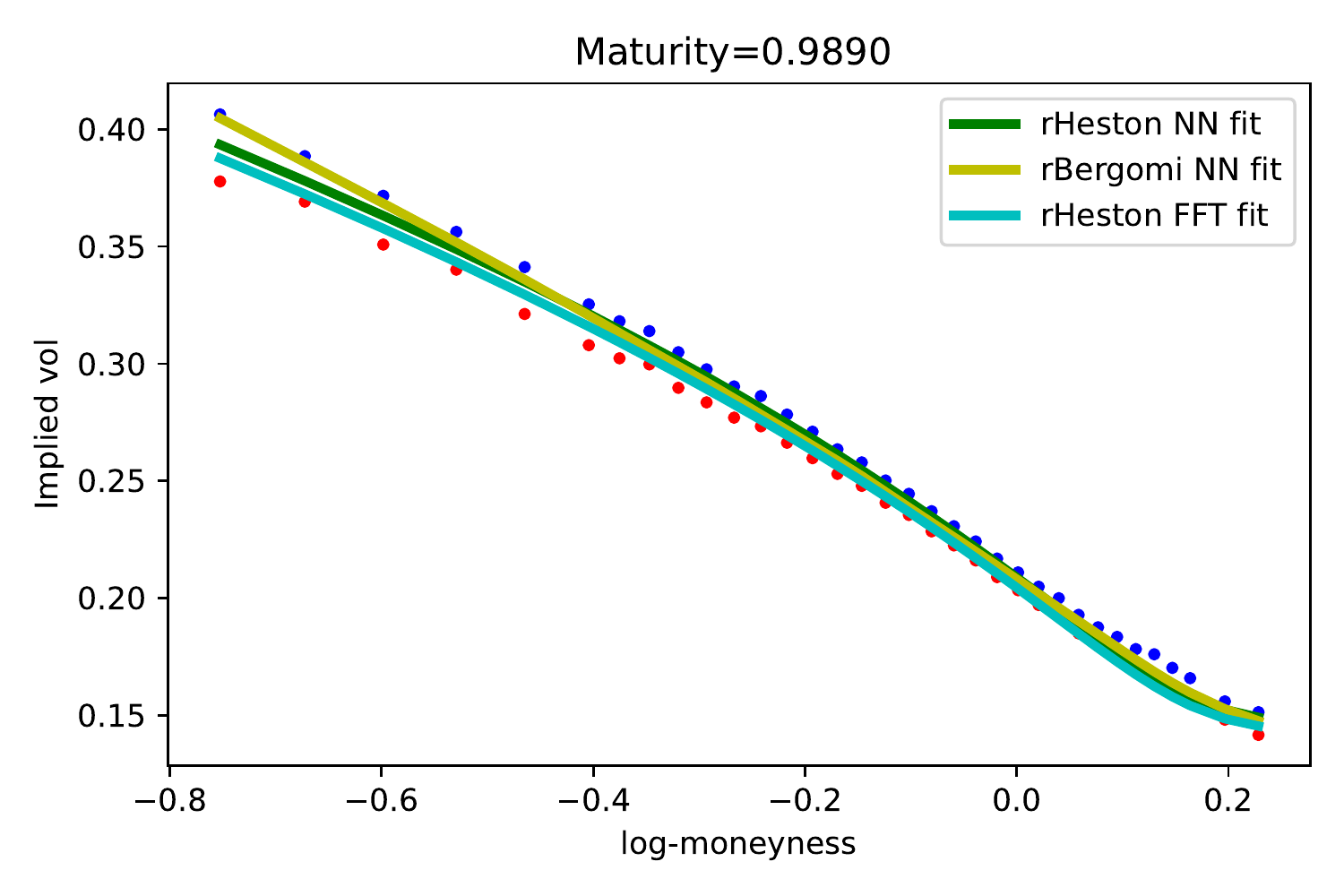}
     \end{subfigure}
     \caption{rBergomi (yellow) and rHeston fits as of March 22, 2011. Calibration via neural networks prescribes a parametric forward variance curve as in Equation \eqref{fvc_par} to be calibrated together with model parameters. FFT pricing pre-computes a piecewise constant forward variance curve and makes it a state variable. Those correspond to the green curve and the cyan curve respectively, for the rHeston model.} 
     \label{COvsSV_20110322}
\end{figure}

\noindent Still, the methodology by \cite{el2019roughening} may occasionally fail to provide a good description of the market data at very short times (say the first maturity in the market, $T_1$). This is indeed a consequence of the fact that what one is estimating with the log-strip is the value of the forward variance curve at time $T_1$, but no information is gathered about its previous history. Extrapolating the curve flat may therefore result in an over/under-estimation. In this sense, the advantage of a neural network approach is that one is encoding a whole piece of the curve over the period $[0,T_1]$ in the option price for maturity $T_1$. That's why the level of the smile in the top left corner of Figure \ref{COvsSV_20110322} is correct when using the pointwise approach with a parametric forward variance curve and a little bit biased when the curve is taken as a state variable. \\

\noindent Because extrapolation methods know nothing about the market prior to maturity $T_1$, encapsulating the primitive history of the curve into a neural network is the only way to guarantee a sensible fit to the shortest maturity. \\

\begin{figure}[h!]
     \centering
     \begin{subfigure}[]{0.49\textwidth}
         \centering
         \includegraphics[width=\textwidth]{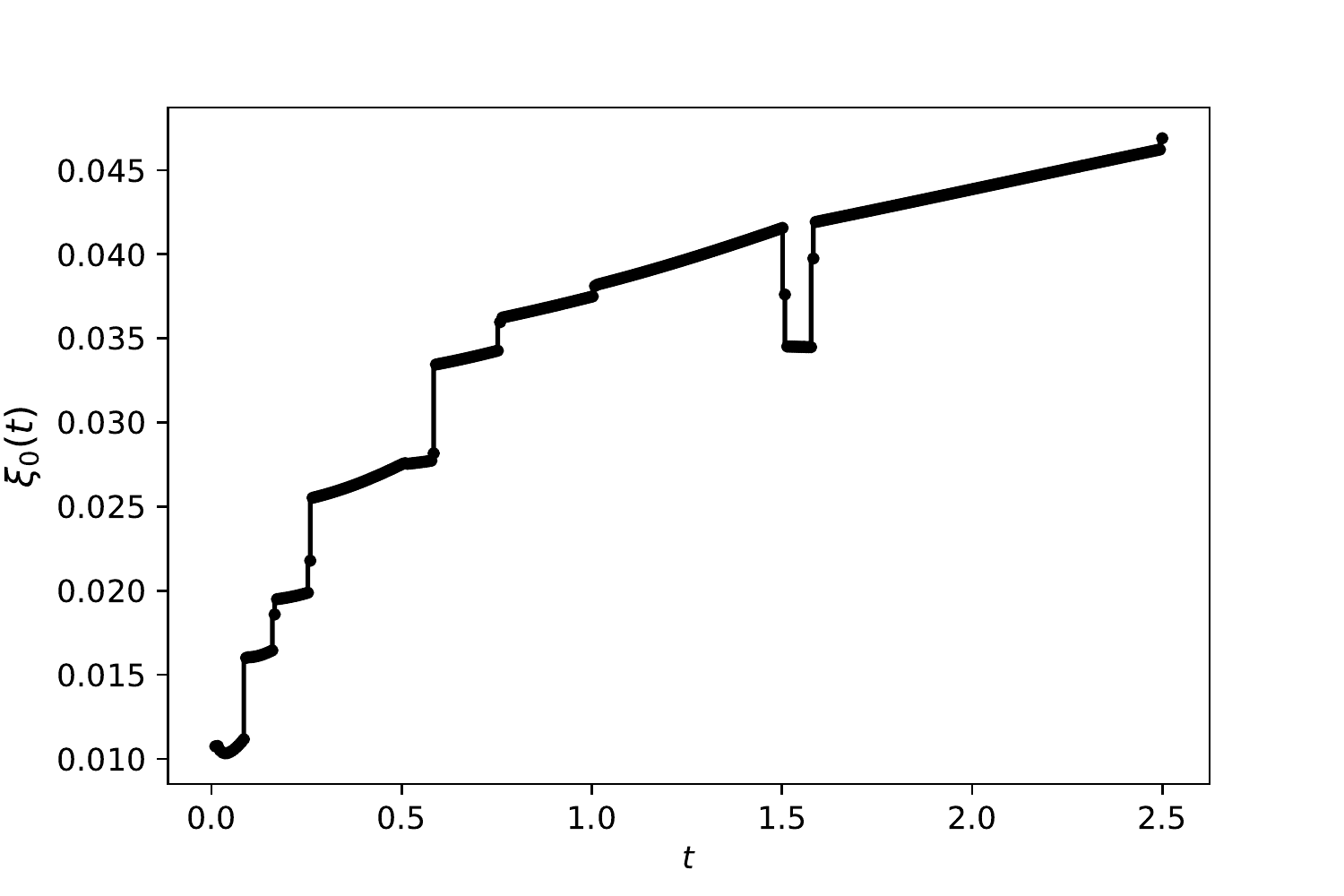}
         \caption{June 18, 2014}
     \end{subfigure}
     \hfill
     \begin{subfigure}[]{0.49\textwidth}
         \centering
         \includegraphics[width=\textwidth]{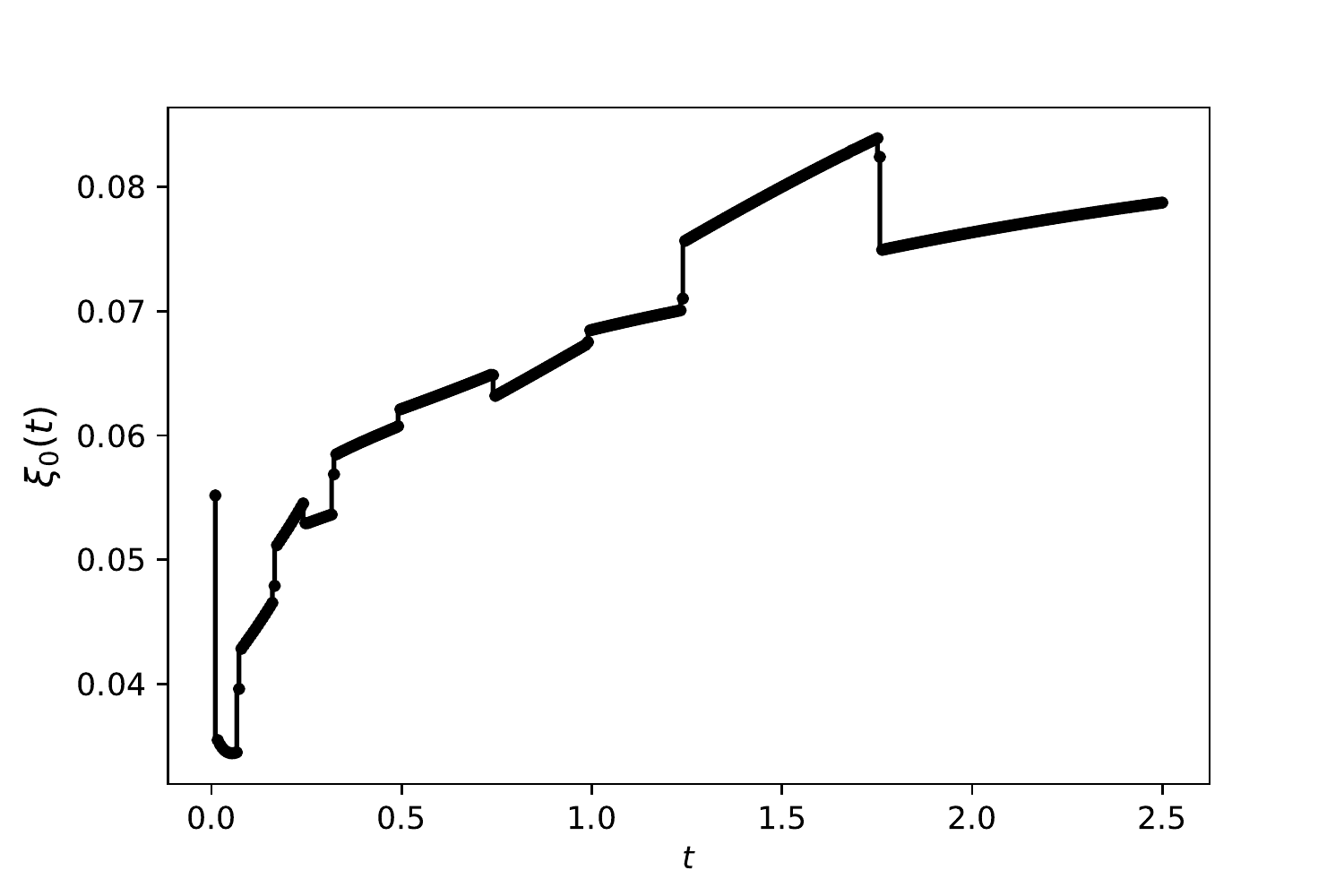}
         \caption{March 22, 2011}
     \end{subfigure}
     \begin{subfigure}[]{0.49\textwidth}
         \centering
         \includegraphics[width=\textwidth]{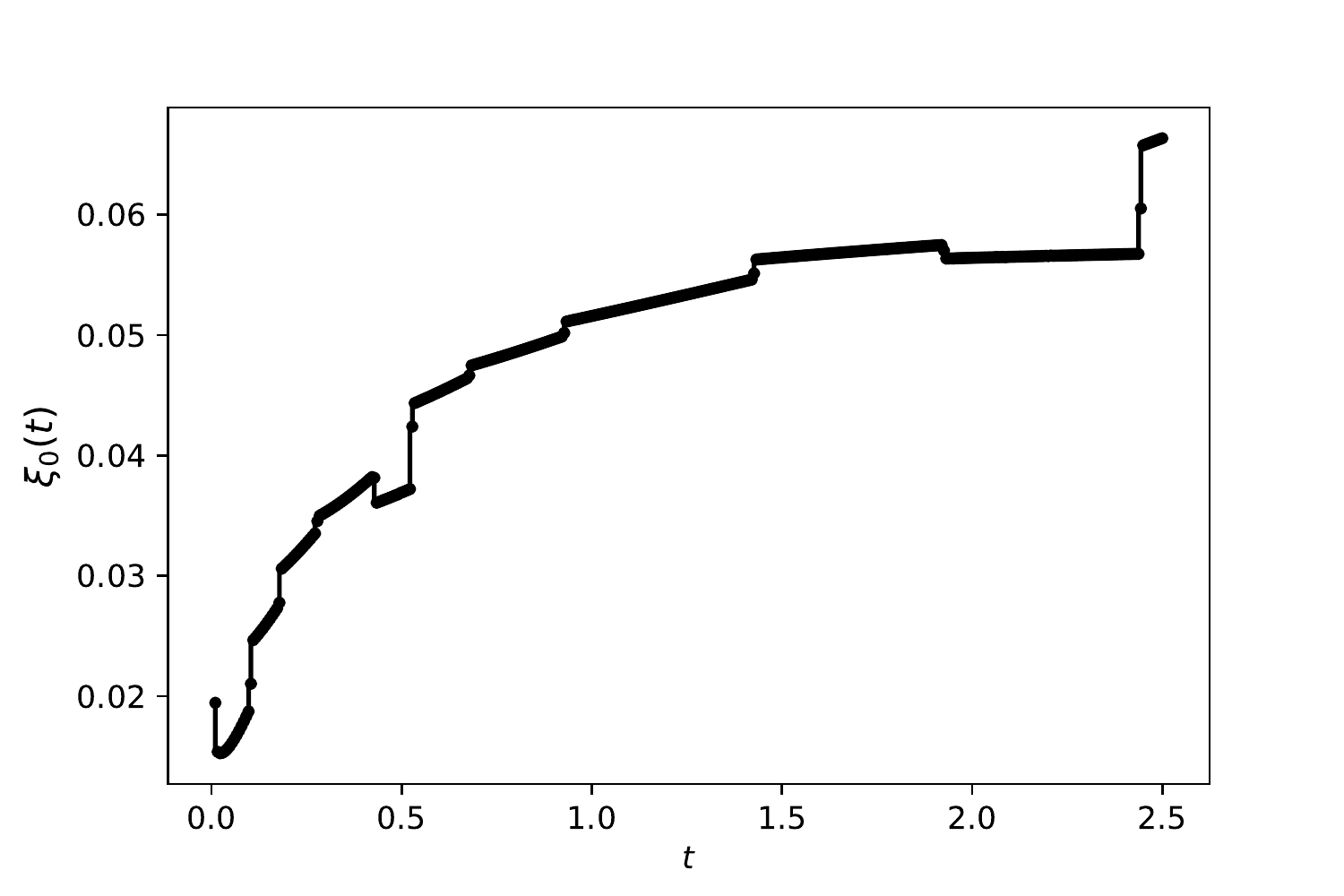}
         \caption{July 13, 2016}
     \end{subfigure}
     \hfill
     \begin{subfigure}[]{0.49\textwidth}
         \centering
         \includegraphics[width=\textwidth]{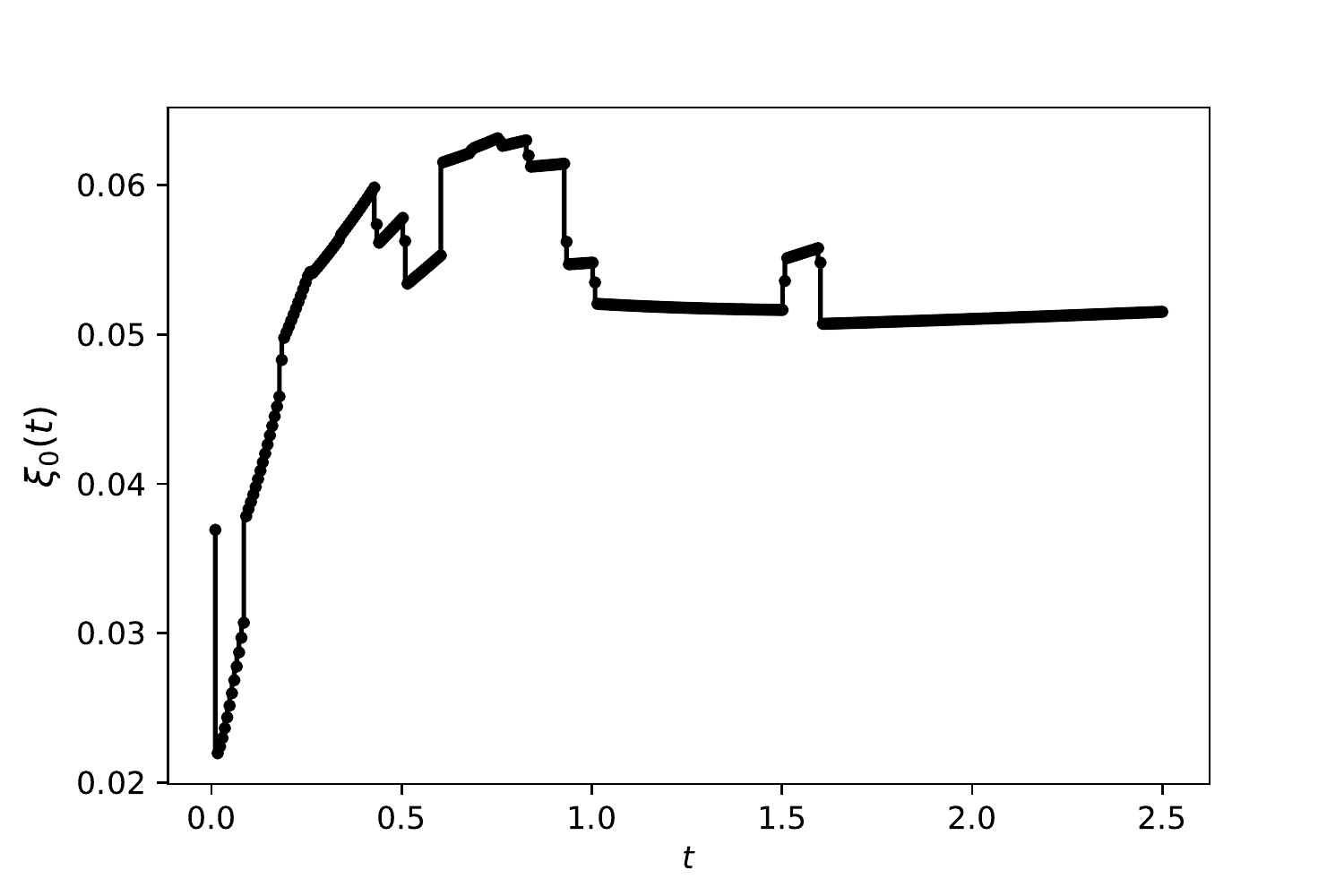}
         \caption{June 15, 2021}
     \end{subfigure}
     \caption{Market forward variance curve from applying log-strip techniques on a fine grid of market maturities as interpolated with Vola Dynamics.}
     \label{xi_lim_logstrip}
\end{figure}

\noindent Interpolation being safe, however, we can evaluate variance swaps on a much finer grid than the one imposed by the market and make the steps in the forward variance curve so short that we get a sense of it in continuous time. We do so for a few surfaces in Figure \ref{xi_lim_logstrip} and recognize that this is indeed going in the direction of the parametric form of Equation \eqref{fvc_par}. \\

\noindent As for calibration times under the pointwise approach, we had maybe better look at a piecewise constant specification of the forward variance curve first. Valuation of one single option by the neural network takes about 7e-06 seconds under this setup, which fact results in the following summary statistics in Table \ref{calib_time} for calibration of entire volatility surfaces in our sample. \\

\begin{table}[h!]
    \centering
    \begin{tabular}{|c|c|c|c|c|}
        \hline
        & rHeston calib time &  & rBergomi calib time & \\
        \hline
        piecewise constant fvc & avg & 0.5852s & 
        avg & 0.7698s \\
        &  min & 0.2266s &
        min & 0.3836s \\
        &  max & 1.4989s & 
        max & 1.7924s \\
        \hline
        parametric fvc [Eq.\eqref{fvc_par}] & avg & 0.8673s & 
        avg & 1.1438s \\
        & min & 0.2251s &
        min & 0.2844s \\
        &  max & 3.2470s & 
        max & 4.8668s \\
        \hline
    \end{tabular}
    \caption{Summary statistics for calibration time (in seconds) in our sample data, both under rHeston and rBergomi with piecewise constant and parametric forward variance curve.}
    \label{calib_time}
\end{table}

\noindent Making the forward variance curve parametric does not necessarily reduce calibration time. It will certainly be true that evaluation of option prices is slightly faster as the neural network is smaller (because the input layer is) but the overall calibration time depends on how many evaluations are needed for convergence. In particular, the problem with Equation \eqref{fvc_par} is that there are regions in parameter space where the curve experiences very little change in response to changing the input. Then, the optimizer is likely to spend some time before exiting these regions if it ever visits them because the objective function is very flat here and steps very short. Luckily enough those regions are small and they correspond to values of $\beta_1$ and/or $\beta_2$ which are close to zero. That is why we say that numerical aspects are also to be taken into account when conceiving some parametrization of the forward variance curve. Nevertheless, calibration times with a parametric forward variance curve are very much in line with what \cite{romer2022empirical} achieves by working on dense adapted grids plus interpolation/extrapolation. \\

\noindent Furthermore, reducing the number of parameters makes it easier for the network to learn the underlying pricing function and helps in producing those very nice plots that we see in Figures \ref{rHes_45} and \ref{rBer_45}.

\section{Conclusions}
We propose a pointwise approach which is trained on volatility points coming from random grids. Generating entire surfaces in the form of random grids is indeed very convenient from a computational viewpoint, because the cost for one entire smile at time $T$ is the same as a single strike for the same maturity. Most importantly, then, the method proves to be very competitive. It learns the underlying pricing function extremely well and it allows for calibration to the market in short time (less than one second on average). The random grid being adaptive in nature, our pointwise approach is financially sound and does not depend on any interpolation/extrapolation algorithm. Its use as a pricer is also possible, but requires some care. After performing the calibration step, the optimal parameters can be used by the trader to instantaneously recover via NN market consistent option prices for any desired contract specifications -- as long as interior to the training set. Possible arbitrage opportunities may arise if the quality of the NN approximation of the true pricing function is poor. However, it is worth to say that we never experienced such a situation in all explorations presented in this paper.\\

\noindent Our work fits into the flourishing literature about neural network calibration of stochastic volatility models and provides a fresh perspective on the subject. We give new life to the pointwise approach and show how this can easily compete with grid-based methods that are now more widespread in the literature. Actually, we believe (and we show in the paper) that independence on whatever interpolation technique is a valuable aspect of the pointwise approach, that neural networks should retain. \\

In addition to this, we recognize that - once the optimal parameters have been calibrated to market data - extrapolation of the volatility surface to very short maturities can be achieved with the neural network in a way which is naturally consistent with the dynamical properties of the model.\\

\noindent We prove that our solution is providing robust estimation of model parameters in a controlled environment and test it against the market as well. \\

\noindent As far as rough volatility models are concerned, our contribution also embraces a novel parametrization of the forward variance curve which guarantees very good fits to the market while reducing the dimensionality of the calibration process and easing the learning task. We also come back to a treatment of the forward variance curve as a state variable and make it 'continuous' via interpolation with the tools from Vola Dynamics LCC. When we do so, we observe that the shapes that we recover are very much in line with our parametric form, thus further corroborating it.\\

\noindent The domain of our neural network pricer
obviously extends to non-rough models. More specifically, we have in mind potential application of the method to PDV models for which option prices are only known after simulation. What is particularly interesting with those model is in fact the possibility for pricing options on the index and the VIX, thus using neural networks for the joint calibration problem. Also, because the neural network seems to have learnt the true pricing function very well, we can use it for construction of the Dupire function to be used with local stochastic volatility models.

\appendix

\section{Appendix A}\label{tables}

We report summary statistics for the estimation of rHeston model parameters under different specifications of the network used for calibration (fxd, ada, pnt for fixed grid, adapted grid, and pointwise approach, respectively). Each of the parameters has its own dedicated table, where we keep track of the following quantities: largest negative error (min), average error (mean), median error (med), $95^{th}$ percentile of the distribution of the errors (q95), largest positive error (max), standard deviation of the distribution of the errors (std), fraction of the runs which result in the true parameter being overestimated by the network (overestimation ratio), fraction of the runs which result in an absolute error $|e_\cdot| > x$ (2-tail mass at $x$). \\

\noindent Information in the tables below provides numerical support for proper interpretation of the figures in Section \ref{estH_rHes_fxdVSada}.  \\

\begin{table}[h!]
    \centering
    \begin{tabular}{|c|c|c|c|}
        \hline
        summary statistics $e_H$ true vs \dots & fxd & ada & pnt \\
        \hline
        min  & -0.1219 & -0.0353 & -0.0200 \\
        mean &  0.0031 & -0.0003 & 0.0015 \\
        med  &  0.0020 & -0.0009 & 0.0004 \\
        q95  &  0.0266 &  0.0089 & 0.0075 \\
        max  &  0.0697 &  0.0386 & 0.0622 \\
        std  &  0.0142 &  0.0057 & 0.0050 \\
        overestimation ratio & 0.225 & 0.614 & 0.406 \\
        2-tail mass at 0.02  & 0.106 & 0.012 & 0.013 \\
        2-tail mass at 0.03  & 0.054 & 0.005 & 0.006 \\
        2-tail mass at 0.05  & 0.019 & 0.000 & 0.002\\
        \hline
    \end{tabular}
    \caption{Summary statistics for the error associated with the estimation of the roughness parameter $H$ in a controlled environment using different network specifications: fixed grid (fxd), adapted grid (ada) and pointwise (pnt) [with points on the adapted grid]. Networks have piecewise constant forward variance curve, the calibration data has flat.}
    \label{eH_tab}
\end{table}

\begin{table}[h!]
    \centering
    \begin{tabular}{|c|c|c|c|}
        \hline
        summary statistics $e_{\nu}$ true vs \dots & fxd & ada & pnt \\
        \hline
        min  & -0.0302 & -0.0413 & -0.0049 \\
        mean &  0.0065 & -0.0008 &  0.0075 \\
        med  &  0.0016 & -0.0038 &  0.0047 \\
        q95  &  0.0419 &  0.0196 &  0.0201 \\
        max  &  0.1625 &  0.2542 &  0.2119 \\
        std  &  0.0176 &  0.0174 &  0.0126 \\
        overestimation ratio & 0.260 & 0.808 & 0.007 \\
        2-tail mass at 0.02  & 0.118 & 0.066 & 0.051 \\
        2-tail mass at 0.03  & 0.079 & 0.043 & 0.035 \\
        2-tail mass at 0.05  & 0.035 & 0.028 & 0.017 \\
        \hline
    \end{tabular}
    \caption{Summary statistics for the error associated with the estimation of the VOV $\nu$ in a controlled environment using different network specifications: fixed grid (fxd), adapted grid (ada) and pointwise (pnt) [with points on the adapted grid]. Networks have piecewise constant forward variance curve, the calibration data has flat.}
    \label{eNU_tab}
\end{table}

\begin{table}[h!]
    \centering
    \begin{tabular}{|c|c|c|c|}
        \hline
        summary statistics $e_{\rho}$ true vs \dots & fxd & ada & pnt \\
        \hline
        min  & -0.0301 & -0.1023 & -0.0138 \\
        mean &  0.0025 & -0.0013 &  0.0023 \\
        med  &  0.0003 & -0.0029 &  0.0007 \\
        q95  &  0.0120 &  0.0194 &  0.0140 \\
        max  &  0.1870 &  0.1779 &  0.1041 \\
        std  &  0.0139 &  0.0201 &  0.0076 \\
        overestimation ratio & 0.406 & 0.777 & 0.343 \\
        2-tail mass at 0.02  & 0.039 & 0.104 & 0.034 \\
        2-tail mass at 0.03  & 0.023 & 0.066 & 0.016 \\
        2-tail mass at 0.05  & 0.016 & 0.037 & 0.002 \\
        \hline
    \end{tabular}
    \caption{Summary statistics for the error associated with the estimation of the correlation $\rho$ in a controlled environment using different network specifications: fixed grid (fxd), adapted grid (ada) and pointwise (pnt) [with points on the adapted grid]. Networks have piecewise constant forward variance curve, the calibration data has flat.}
    \label{eRHO_tab}
\end{table}

\section{Appendix B - An additional robustness check}\label{Absence of arbitrage}

In the main text, we suggest a methodology to replace any model pricing function, semi-analytical (e.g., rHeston) or entirely Monte Carlo (e.g., rBergomi),  with a Neural Network approximation. The issue of possible absence of arbitrage violations pertains to the model specification. If the pricing model allows for arbitrages, then the NN will learn a badly specified model and will suffer the same shortcomings. If the pricing model is an arbitrage-free model -- as for the rHeston and rBergomi models -- the NN will learn arbitrage-free pricing functions. This Appendix provides an additional robustness check that the NN properly learns the arbitrage-free pricing functions. This is purely a sanity check provided in developing the model. In no way we do suggest that one would need to run it after any calibration.

Sufficient conditions for absence of arbitrage in call option prices $C=C(F,K,T)$ are well known to be given as follows
\begin{align}\label{AOA}
    \frac{\partial C}{\partial T} > 0, \qquad \frac{\partial C}{\partial K} < 0, \qquad \frac{\partial^2 C}{\partial K^2} > 0.
\end{align}
A discrete version of these is straightforward to derive and can be found in \cite{itkin2019deep}. The idea is that one fixes the option maturity $T$ and the forward price $F$, and looks at call options $C(K_1),C(K_2),C(K_3)$ for strikes $K_1<K_2<K_3$. Then 
\begin{align*}
    C(K_3) > 0, \qquad C(K_2) > C(K_3) 
\end{align*}
imposes no vertical spread, and
\begin{align*}
    (K_3-K_2)C(K_1) - (K_3-K_1)C(K_2) + (K_2-K_1)C(K_3) > 0
\end{align*}
guarantees the absence of any butterfly spread arbitrage. Fixing the strike and allowing the option maturity to move, we verify the absence of calendar spread arbitrage as
\begin{align*}
    C(T_2) > C(T_1)
\end{align*}
for maturities $T_1<T_2$. \\

We calibrate to about 50 daily market volatility surfaces and store the optimal parameter values. Using the NN, we compute the implied volatility surface for each day -- and corresponding parameter set -- over an extremely dense strike and time-to-maturity grid. We consider time-to-maturities running from 2 days up to 2.5 years ($dt=\frac{1}{365}$) with a daily resolution and evenly sampled strikes ($dK=0.01$) over the range $(S_0 (1-0.55\sqrt{t}), S_0 (1+0.30\sqrt{t}))$. On the corresponding price surfaces, computed via the Black-Scholes formula, we look for violations of the arbitrage conditions. With a parametric forward variance curve -- and the dimensionality of the learning problem consequently reduced -- we did not detect any violation out of more than 80.000 sampled points.

\end{document}